\documentclass[12pt]{article}
\usepackage{xr-hyper}
\usepackage[top=1in, bottom=1in, left=1in, right=1in]{geometry}
\usepackage{setspace}
\usepackage{booktabs}
\usepackage{rotating}
\usepackage{graphicx}
\usepackage[section]{placeins} 
\usepackage[small, bf]{caption}
\usepackage{subcaption}
\usepackage{pdfpages}
\usepackage{pdflscape}
\usepackage{textcomp}
\usepackage{longtable}
\usepackage{nicefrac}
\usepackage{adjustbox}	
\usepackage[hyphens]{url}
\usepackage{bbm}
\usepackage{floatrow}
\usepackage{wrapfig}

\usepackage{natbib}
\usepackage{bibunits}
\bibpunct{(}{)}{;}{a}{,}{,}

\usepackage{amsmath, amsfonts, amssymb, amsthm}
\usepackage[pdftex]{hyperref}
\hypersetup{
    colorlinks=true,
    linkcolor=red,
    citecolor=blue,
    filecolor=magenta,      
    urlcolor=cyan,
}

\newtheorem{lemma}{Lemma}
\newtheorem{proposition}{Proposition}[section]

\theoremstyle{definition}
\newtheorem{definition}{Definition}
\newtheorem{assumption}{Assumption}

\newtheorem{observation}{Observation}[section]

\usepackage{titlesec}
\titlespacing*{\section}{0pt}{1.5ex plus 1ex minus .2ex}{0.8ex plus .2ex}
\titlespacing*{\subsection}{0pt}{1.2ex plus 1ex minus .2ex}{0.8ex plus .2ex}

\newcommand{\cD}{\mathcal{D}}
\newcommand{\cE}{\mathcal{E}}
\newcommand{\cF}{\mathcal{F}}

\newcommand{\cM}{\mathcal{M}}
\newcommand{\cN}{\mathcal{N}}

\newcommand{\cX}{\mathcal{X}}
\newcommand{\cY}{\mathcal{Y}}

\newcommand{\tm}{\tilde{m}}

\usepackage[subtle]{savetrees}

\usepackage{titlesec}
\titlespacing\section{0pt}{0pt plus 2pt minus 2pt}{0pt plus 2pt minus 2pt}
\titlespacing\subsection{0pt}{0pt plus 2pt minus 2pt}{0pt plus 2pt minus 2pt}
\titlespacing\subsubsection{0pt}{0pt plus 2pt minus 2pt}{0pt plus 2pt minus 2pt}

\defaultbibliographystyle{aea}
\defaultbibliography{Bibliography}

\usepackage{dcolumn}
\newcolumntype{d}[0]{D{.}{.}{5}}

\usepackage{graphicx}
\usepackage[space]{grffile}
\usepackage{colortbl}

\usepackage{amssymb} 

\usepackage{tikz}
\usepackage{tikzpeople}
\usetikzlibrary{positioning}
\usetikzlibrary{snakes}
\usetikzlibrary{calc}
\usetikzlibrary{arrows}
\usetikzlibrary{decorations.markings}
\usetikzlibrary{shapes.misc}
\usetikzlibrary{matrix,shapes,arrows,fit,tikzmark}

\usepackage[linesnumbered,ruled,vlined]{algorithm2e}
\usepackage{algpseudocode}
\SetKwInput{KwInput}{Input}                
\SetKwInput{KwOutput}{Output}              

\usepackage{subfiles} 

\title{From Predictive Algorithms to \\ Automatic Generation of Anomalies\thanks{We thank audiences at Bristol, Georgetown, Harvard/MIT, LMU Munich, Stanford, UCLA, UCL, University of Chicago, USC, Warwick, Wisconsin, Yale, and the NBER Summer Institute Digital Economics and AI session, Nikhil Agarwal, Nicolas Barberis, Raf Batista, Alex Imas, Suproteem Sarkar, Josh Schwartzstein, Andrei Shleifer, Cassidy Shubatt, Richard Thaler, Keyon Vafa, and especially our discussant Colin F. Camerer for helpful comments. We also thank Peter Chang for exceptional research assistance.
We are grateful to the Center for Applied Artificial Intelligence at the Booth School of Business for generous funding. 
All errors are our own.}}
\author{Sendhil Mullainathan \and Ashesh Rambachan\thanks{Mullainathan: Massachusetts Institute of Technology and NBER (\texttt{Sendhil.Mullainathan@uchicago.edu}). Rambachan: Massachusetts Institute of Technology (\texttt{asheshr@mit.edu}).}}	
\date{\today}

\begin{document}

{\singlespacing \maketitle}
\thispagestyle{empty} 
\setcounter{page}{0}

{\singlespacing
\begin{abstract}
How can we extract theoretical insights from machine learning algorithms? 
We take a familiar lesson: researchers often turn their intuitions into theoretical insights by constructing ``anomalies'' --- specific examples highlighting hypothesized flaws in a theory, such as the Allais paradox and the Kahneman-Tversky choice experiments for expected utility.
We develop procedures that replace researchers' intuitions with predictive algorithms: given a predictive algorithm and a theory, our procedures automatically generate anomalies for that theory.
We illustrate our procedures with a concrete application: generating anomalies for expected utility theory.
Based on a neural network that accurately predicts lottery choices, our procedures recover known anomalies for expected utility theory and discover new ones absent from existing work.
In incentivized experiments, subjects violate expected utility theory on these algorithmically generated anomalies at rates similar to the Allais paradox and common ratio effect.
\end{abstract}}

\newpage
\section{Introduction}
How do we improve economic theories? 
While there is no single answer, one pattern stands out. 
Consider the ``Allais paradox.''
To highlight a conjectured inconsistency between expected utility theory and how people make risky choices, \citet[][]{Allais(53)} crafted two menus of lotteries (see Table \ref{tab: the allais paradox}). 
He hypothesized that if people were presented this pair, their choices would violate expected utility theory.
When collected data confirmed Allais' intuition, it led to a reappraisal of expected utility theory.
More examples like the Allais paradox were constructed, such as the certainty effect and the Kahneman-Tversky choice experiments \citep[][]{KahnemanTversky(79)}.
These examples eventually led to a new theory of risky choice: cumulative prospect theory \citep[][]{TverskyKahneman(92)}.

\begin{table}[!h]
\caption{Menus of lotteries in the Allais paradox \citep{Allais(53)}.}
\begin{subtable}{.5\linewidth}
    \centering
    \caption{Menu A}
    \begin{tabular}{c | c c c}
            \cellcolor{green!25}{Lottery 0} & \$1 million \\
            & 100\% \\
            \hline 
            Lottery 1 & \$1 million & \$0 & \$5 million \\ 
            & 89\% & 1\% & 10\%
        \end{tabular}
    \end{subtable}%
    \begin{subtable}{.5\linewidth}
      \centering
        \caption{Menu B}
        \begin{tabular}{c | c c c}
            Lottery 0 & \$0 & \$1 million \\
            & 89\% & 11\% \\
            \hline 
            \cellcolor{green!25}{Lottery 1} & \$0 & \$5 million \\
            & 90\% & 10\%
        \end{tabular}
    \end{subtable} 
    \label{tab: the allais paradox}
    \floatfoot{\textit{Notes}: The hypothesized choices are highlighted in green on these two menus.
    \cite{Allais(53)} originally denominated the payoffs in French Francs, and we reproduce the version of the Allais paradox used in \cite{SlovicTversky(74)}.} 
\end{table}

The Allais paradox does not have a natural place in our toolkit. 
It is neither an innovation in data collection on risky choice, nor is it a novel test statistic for whether expected utility theory is misspecified.
Rather the Allais paradox is a hypothesis about where expected utility theory might fail, instantiated by a concrete pair of menus that guided subsequent empirical analysis. 
The Allais paradox is a specific instance of what we call an ``anomaly'' in this paper. 

This is a recurring pattern of how economic theories are improved across fields: researchers construct anomalies that hypothesize where an existing theory might fail, and then researchers test the theory in newly collected data on those hypothesized anomalies.\footnote{See for example \citet[][]{CamererThaler(95)-Anomalies-Ultimatums} in game theory, \citet[][]{KahnemanKnetschThaler(91)-Anomalies-EndowmentEffectLossAversionStatusQuo} in decision-making under uncertainty, and \citet[][]{LoewensteinPrelec(92)-AnomaliesIntertemporal} in intertemporal choice.}
The second step meticulously applies statistical tests to collected data, bringing to bear a foundational literature in econometrics and economic theory on specification testing and rationalizability tests \citep[e.g.,][]{Afriat(67), Varian(82)}.
By contrast, the first step of anomaly generation largely happens offstage in a creative process.
Researchers construct anomalies by contrasting their intuition with an existing theory and crafting examples where the theory differs from their hypothesized empirical patterns. 

We are interested in anomalies because they offer a familiar solution to a new problem: how can we extract theoretical insights from machine learning algorithms?  
We propose procedures to automatically construct anomalies for an existing theory from predictive algorithms. 
We use supervised machine learning algorithms to build a black box predictive model, which serves as our procedures' empirical intuition. 
Our procedures then contrast the theory with the predictive model, searching for minimal examples on which the theory cannot explain the model's predictions.
Like the Allais paradox, the resulting generated anomalies can be examined by researchers to interpret the predicted violation, and they provide natural targets for subsequent data collection. 
We illustrate these procedures by generating anomalies for expected utility theory based on a neural network that accurately predicts lottery choices. 

We focus on settings summarized by input features $x$ and a modeled outcome $y^*$. 
Following influential work that views economic theories as tools for prediction \citep[e.g.,][]{Selten(91), FudenbergEtAl(22)-Completeness, FudenbergEtAl(20)-Restrictiveness}, we model a theory as a restrictive class of functions that map features to the modeled outcome, consistent with its underlying structure. 
We call this the theory’s allowable function class. 
In choice under risk, $x$ is a menu of lotteries, $y^*$ a choice probability over the menu, and the allowable function class is the set of choice-probability functions consistent with expected utility theory. 
Anomalies are minimal collections of examples that are inconsistent with the theory’s allowable function class. Put simply: given any examples, the theory searches for an allowable function that fits them; an anomaly is a collection crafted to foil this search.

In this framework, constructing anomalies implied by an estimated prediction function is an adversarial game between a falsifier and the theory.  
The falsifier proposes collections of features and the estimated prediction function evaluated on those features, and the theory attempts to explain them by fitting an allowable function. 
The falsifier's payoff is increasing in the theory's loss on the proposed collection.
An anomaly arises if the falsifier finds a collection of examples that the theory cannot fit. 
In other words, anomalies are \textit{adversarial examples} for the theory in such a game.

Our first anomaly generation procedure solves the falsifier’s problem by directly optimizing a max-min program over the theory’s allowable functions. 
We analyze feasible implementation, establishing a finite‐sample bound on how well it approximates its population analog.
Optimizing this max-min program may be challenging as the falsifier's maximization will typically be non-concave.
An active literature in adversarial learning proposes a variety of gradient‐based methods for problems \citep[e.g.,][]{RazaviyaynEtAl(20)};
we adapt a particularly simple gradient descent--ascent scheme and describe its properties.
By iteratively updating the falsifier's proposed collection of examples to maximize the loss of the theory's best-responding allowable function,  this procedure generates anomalies implied by the predictive model.

There may, however, exist additional structure in theories that this procedure does not exploit. 
Theories often behave as if they have a lower-dimensional representation of the features (i.e., there may exist some pair of feature values that all allowable functions assign the same modeled outcome value).
Some anomalies, like the Allais paradox, reveal a possible dimension that is relevant for the modeled outcome but which the theory's representation fails to capture. 
Our second procedure generates these representational anomalies for a theory. 
Given an initial feature value, it uses projected gradient descent to search for nearby feature values across which the theory's allowable functions do not vary but across which the estimated prediction function varies.

Our results so far provide conditions under which our procedures will generate anomalies implied by the predictive algorithm.
How should we practically evaluate anomaly generation procedures? 

We suggest two complementary evaluations that mirror the two-step workflow already used by researchers.
First, logical verification: how often does the anomaly generation procedure return collections on which the theory is inconsistent with the predictive algorithm? 
This step assesses how effectively the procedure contrasts the predictive algorithm with the theory; better optimization should yield more generated anomalies implied by the predictive algorithm.
Second, empirical testing: treating the algorithmically generated anomaly as a fixed design, collect independent data via new experiments on these generated anomalies and formally test for misspecification.
This decouples anomaly generation from downstream inference, permitting the use of any supervised learner or optimizer in the procedure.
Better prediction and better optimization should lead to generated anomalies that hold in new experiments. 
Together, these provide evaluation metrics for comparing anomaly generation procedures, and they enable researchers to reuse existing experimental datasets as ``common tasks'' for developing anomaly generation \citep[][]{Donoho(24)}.

With this framework in hand, we turn to a concrete illustration: generating anomalies for expected utility theory. 
As a first step, we generate anomalies for expected utility theory that are implied by prospect theory in simulated lottery choice data.  
Since prospect theory has been well studied, we compare our algorithmically generated anomalies against known anomalies for expected utility theory constructed by researchers, such as those produced in \citet{Allais(53)}, \citet{KahnemanTversky(79)}, and many others.
In this first exercise, we therefore have a floor: do our procedures reproduce known anomalies for expected utility theory implied by prospect theory?
Our procedures successfully reach this floor, recovering known anomalies for expected utility theory over menus of two-payoff lotteries.

Motivated by their performance in simulation, we next apply our anomaly-generation procedures to \emph{real} lottery-choice data. 
Using a publicly available dataset of human decisions on nearly 10{,}000 lottery menus collected by \citet{PetersonEtAl(21)-MLforChoiceTheory}, we train a neural network to predict choice rates from the menus’ payoffs and probabilities.
On a held-out set of menus, the predictive algorithm meaningfully improves on the accuracy of leading theories of decision-making under risk.

Using the predictive algorithm as its empirical intuition, our anomaly generation procedures again reconstruct known anomalies for expected utility theory, such as those produced in \citet{Allais(53)} and \citet{KahnemanTversky(79)}. 
Our procedures go further and generate novel anomalies for expected utility theory that are implied by the predictive algorithm, which to our knowledge are nowhere in the existing literature.
These anomalies fall into distinct categories that illustrate new choice reversals could be produced across menus of lotteries \textit{if} people behave as predicted.
Even in this well-trodden domain, our procedures suggest there might be novel patterns in lottery choice behavior implied by neural network's predictions.

Having generated these anomalies, we proceed to preliminary empirical testing: collect new data on these generated anomalies and test whether expected utility theory is violated. 
We present promising evidence. 
We recruit participants to make incentivized choices on our algorithmically generated anomalies.
Our experimental design mirrors existing tests of the Allais paradox and the common ratio effect \citep[e.g.,][]{HarlessCamerer(94), BlavatskyyEtAl(22)-RobustnessOfAllais, JainNielsen(23)}, enabling us to evaluate how ``significant'' these novel anomalies are. 
On our algorithmically generated anomalies, participants exhibit behavior inconsistent with expected utility theory at rates comparable to those for well-established anomalies in behavioral economics. 

Ultimately, our goal was not to revisit choice under risk or expected utility theory. 
Rather we hope to illustrate that anomaly generation is a distinct and valuable step in improving theories: it suggests \textit{where} to look for violations and delivers concrete examples that guide subsequent empirical work. 
We show that procedures can be built to accelerate this step.

Substantial progress has been made in exploring how machine learning interacts with economic theories. 
Recent work compares the out-of-sample predictive performance of machine learning models against economic theories, measuring the ``completeness'' of economic theories \citep[][]{FudenbergEtAl(22)-Completeness}. 
\cite{AndrewsEtAl(22)-Transfer} measure the out-of-distribution predictive performance of economic theories.
When a supervised machine learning model predicts some outcome of interest accurately out-of-sample, researchers often attempt to open the prediction function and investigate particular behavioral hypotheses.
See, for example, \cite{PeysakhovichNaecker(17)} and \cite{PetersonEtAl(21)-MLforChoiceTheory} for choice under risk, \cite{WrightLeytonBrown(17), HirasawaEtAl(22)-GameTheoryPred} for strategic behavior in normal-form games, and \cite{MullainathanObermeyer(21)} for medical decision-making.
We use supervised machine learning algorithms as stepping stones to automatically generate anomalies, rather than relying on researchers to inspect the prediction function or generate hypotheses themselves. 
\cite{FudenbergLiang(19)-InitialPlay} use supervised machine learning algorithms to predict on which normal-form games will observed play differ from alternative theories of strategic behavior.
They use the resulting prediction function to sample normal-form games where a particular theory will predict poorly.
This intuitive procedure can be reinterpreted as a heuristic solution to our adversarial characterization of anomalies tailored to the models of strategic behavior they study.
\cite{LudwigMullainathan(23)-Faces} develop a morphing procedure for images based on generative adversarial networks in order to uncover implicit characteristics of defendant mug-shots that affect pretrial release decisions. 

Finally, a large machine learning literature seeks to explain or interpret predictive algorithms \citep[][]{Molnar(25)}.
These intepretability techniques often have different goals than anomaly generation for an existing theory. 
For example, counterfactual explanations search for nearby inputs that would yield alternative predictions given an input value and its prediction \citep[][]{MothilalEtAl(20)}.
Other work examines the sensitivity of predictions to alternative training examples \citep[][]{KohLiang(17)}, and others seek to locally approximate a complex, predictive algorithm around some input value using simpler functions \citep[][]{RibeiroEtAl(16)}.
By contrast, anomaly generation is an adversarial problem: we search for examples on which the predictions are inconsistent with the theory's allowable function class. 
Nonetheless, exploring how these interpretability techniques could be adapted for anomaly generation is a valuable direction, and our evaluation metrics provide a common yardstick for future comparisons.

\section{The anomaly generation problem}\label{section: anomaly generation problem}

In this section, we develop a framework that analyzes an economic theory as a tool for prediction.
This serves as the basis for developing anomaly generation procedures.
Let $x \in \cX$ be some features and $y^* \in \cY^*$ be some modeled outcome in an economic domain. 
Any pair $(x, y^*) \in \cX \times \cY^*$ is an example, and $D := \{ (x_1, y_1^*), \hdots, (x_n, y_n^*) \}$ is a finite collection of examples. 
Let $\cD$ denote all collections of examples and $\cF$ the collection of all mappings $f \colon \cX \rightarrow \cY^*$. 

For our purposes, many economic theories can be summarized as a  restrictive collection of mappings used to make novel predictions given examples.
To state this precisely, we say a mapping $f \in \cF$ is \textit{consistent} with $D \in \cD$ if $f(x) = y^*$ for all $(x, y^*) \in D$.  
A collection $D$ is \textit{inconsistent} with function class $\widetilde{\cF} \subseteq \cF$ if there exists no $f \in \widetilde{\cF}$ consistent with $D$.  

\begin{definition}\label{defn: theory as allowable functions}
    A \textit{theory} consists of a function class $\mathcal{F}^T \subseteq \mathcal{F}$ and a predictive algorithm $T$ satisfying, for all $x \in \mathcal{X}$ and $D \in \mathcal{D}$, $T(x; D) := \{ f(x) \colon f \in \mathcal{F}^T \mbox{ and } f \mbox{ is consistent with } D \}$.
\end{definition}

\noindent We write $T(x; D) \subseteq \cY^*$ as the theory's predictions at feature $x$ when applied to examples $D$, and $\cF^{T}$ is the \textit{allowable function class}. 
The allowable function class $\cF^{T}$ summarizes all mappings consistent with the theory's underlying structure, however that may be modeled.
To make new predictions, the theory searches for allowable functions $f \in \cF^{T}$ that are consistent with the given examples $D \in \cD$.
When the allowable functions $\mathcal{F}^T$ are inconsistent with $D$, then $T(\cdot; D)$ is empty as the theory is unable to make new predictions.

We illustrate how two popular economic domains map into this framework. 
We provide additional examples in Appendix \ref{section: additional examples}.

\vspace{-1em}
\paragraph{Example: choice under risk} 
Consider individuals making choices from menus of two lotteries over $J > 1$ monetary payoffs.\footnote{\cite{ErevEtAl(17)-FromAnomaliesToForecasts, PeysakhovichNaecker(17), PetersonEtAl(21)-MLforChoiceTheory} collect experimental datasets of individuals making choices from menus of risky lotteries and compare the choice predictions made by economic theories of risky choice against those made by black-box predictive algorithms.}
The features describe the menu $x = (z_0, p_0, z_1, p_1)$, where $z_0, z_1 \in \mathbb{R}^{J}$ are the payoffs and $p_0, p_1 \in \Delta^{J-1}$ are the probabilities associated with lottery 0 and lottery 1 respectively.
The features may also, for example, include information about how each lottery is presented or the ordering of lotteries. 
The modeled outcome is the choice probability $y^* \in [0,1]$ for lottery 1.

Expected utility theory searches for utility functions in some researcher-chosen class $\mathcal{U}$ that rationalize lottery choice probabilities. 
This yields the allowable function class $\mathcal{F}^T = \{ f \colon \exists u \in \mathcal{U} \mbox{ s.t. } f(x) \in \arg \max_{k \in \{0, 1\}} \sum_{j=1}^{J} p_k(j) u(z_k(j)) \mbox{ for all } x \in \mathcal{X}\}$; in other words, each allowable function summarizes the choices consistent with some utility function with any tie-breaking rule for indifferences. 
Given a collection $D$, expected utility theory returns all utility functions $u \in \mathcal{U}$ satisfying $y^* = \arg \max_{k \in \{0, 1\}} \sum_{j=1}^{J} p_k(j) u(z_k(j))$ for all $(x, y^*) \in D$.
On any new menu $x$, it returns predictions $T(x; D)$, where $y^* \in T(x; D)$ if and only if $y^* \in \arg \max_{k \in \{0, 1\}} \sum_{j=1}^{J} p_k(j) u(z_k(j))$ for some utility function $u \in \mathcal{U}$ rationalizing $D$. 

Incorporating noise yields an alternative theory.
For instance, expected utility theory with idiosyncratic errors allows the individual to mistakenly select the wrong lottery with some probability $\epsilon \in [0, 0.5]$ \citep[e.g.,][]{HarlessCamerer(94)}. 
Each utility function $u \in \mathcal{U}$ and idiosyncratic error $\epsilon$ is now associated with an allowable function $f \in \mathcal{F}^T$, and the pair is chosen to satisfy $y^* = (1 - \epsilon) 1\{ \sum_{j=1}^{J} p_1(j) u(z_1(j)) \geq  \sum_{j=1}^{J} p_0(j) u(z_0(j)) \} + \epsilon 1\{ \sum_{j=1}^{J} p_1(j) u(z_1(j)) < \sum_{j=1}^{J} p_0(j) u(z_0(j)) \}$ for all $(x, y^*) \in D$.
Other models of noisy choices can be captured by our framework, such as i.i.d. additive noise \citep[e.g.,][]{BallingerWilcox(97)} or more general noise models \citep[][]{McGranahanEtAl(23)-CommonRatio}.
$\blacktriangle$

\vspace{-1em}
\paragraph{Example: play in normal-form games}
Consider individuals playing $J \times J$ normal-form games.\footnote{\citet{HartfordEtAl(16), WrightLeytonBrown(17), FudenbergLiang(19)-InitialPlay, HirasawaEtAl(22)-GameTheoryPred} collect datasets of individuals selecting actions in normal-form games and compare the action predictions made by game-theoretic models of behavior against those made by black-box predictive algorithms.}
Let $\{1, \hdots, J\}$ denote the actions available to the row and column players, and $\pi_{row}(j, k)$, $\pi_{col}(j, k)$ denote the payoff to the row and column players respectively from action profile $(j, k)$.
The features describe the payoff matrix with $x = (\pi_{row}(1,1), \pi_{col}(1,1), \hdots, \pi_{row}(J,J), \pi_{col}(J, J))^\prime$.
The modeled outcome is the row player's strategy profile, which is a probability distribution over actions $y^* \in \Delta^{J-1}$. 
For Nash equilibrium, its set-valued predictions on any payoff matrix $x$ are $\mathcal{Y}^{NE}(x) := \{ y^* \colon \exists y_{col}^* \in \Delta^{J-1} \mbox{ s.t. } y^* \in \arg \max_{\tilde{y}} \ \tilde{y}^\intercal \pi_{row} y_{col}^* \mbox{ and } y_{col}^* \in \arg \max_{\tilde{y}_{col}} \ (y^*)^\intercal \pi_{col} \tilde{y}_{col} \}$. 
This yields the allowable function class $\mathcal{F}^{T} = \{ f \colon f(x) \in \mathcal{Y}^{NE}(x) \mbox{ for all } x \in \mathcal{X}\}$; each allowable function corresponds to a particular choice among Nash equilibria in each game. 
Given the collection $D$, it returns the predictions $T(x; D)$ satisfying $T(x; D) = \{y^*\}$ for all $(x, y^*) \in D$ and $y^* \in T(x; D)$ for any $x \notin D$ if and only if $y^* = f(x)$ for some allowable function $f \in \mathcal{F}^T$ consistent with the collection.
Alternatively, for example, ``level-0'' strategy behavior follows the same construction except instead it is associated with the point predictions $T^{level-0}(x) := \{(1/J, \hdots, 1/J)\}$ for all payoff matrices $x$. $\blacktriangle$

\medskip

Definition \ref{defn: theory as allowable functions} builds on an influential literature that views economic theories as tools for prediction. 
\cite{Selten(91)} measures the ``predictive success'' of a theory as the comparison between the fraction of correct predictions it makes and the fraction of outcomes it deems possible \citep[see also][]{HarlessCamerer(94)}.
\cite{FudenbergEtAl(20)-Restrictiveness} measure the ``restrictiveness'' of economic theories, generalizing Selten's definition.
Recent work bridging economic theory and supervised learning take the same perspective.
\citet[][]{FudenbergEtAl(22)-Completeness} formalize the predictive completeness of an economic theory, and \citet[][]{AndrewsEtAl(22)-Transfer} study their transfer performance across domains. 

\subsection{Anomalies}\label{section: inconsistent datasets and anomalies}

On any collection inconsistent with a theory $\mathcal{F}^T$, it may be difficult for researchers to understand what drives the theory's failure. 
Consequently, researchers like Allais do not simply characterize all possible collections that are inconsistent with a theory; rather they construct minimally inconsistent collections, which we refer to as ``anomalies.''

\begin{definition}\label{defn: anomaly}
A collection $D \in \mathcal{D}$ is an \textit{anomaly} for theory $\mathcal{F}^T$ if $D$ is inconsistent with $\mathcal{F}^T$ and $\widetilde{D}$ is consistent with $\mathcal{F}^T$ for all $\widetilde{D} \subset D$. 
\end{definition} 

\vspace{-1em}
\paragraph{Example: the Allais paradox} 
Consider the Allais paradox (Table \ref{tab: the allais paradox}), which is a pair of examples consisting of the menus $x_A, x_B$ and the associated modeled outcomes $y_A^* = 0, y_B^* = 1$. 
The independence axiom implies that the choice on $x_A$ must be the same as the choice on $x_B$, provided the considered utility function class $\mathcal{U}$ rules out indifferences; that is, for any $f \in \mathcal{F}^T$, $f(x_A) = f(x_B)$.
The Allais paradox is inconsistent with expected utility theory. 
Since any single choice $(x_A, y_A^*)$ or $(x_B, y_B^*)$ is consistent with expected utility theory, the Allais paradox satisfies Definition \ref{defn: anomaly}. $\blacktriangle$

\medskip

\noindent An anomaly is a minimally inconsistent collection of examples that $\mathcal{F}^T$ is consistent with any of its subsets. 
In Appendix \ref{section: additional examples}, we illustrate additional anomalies, such as the certainty effect for expected utility theory and an anomaly for Nash equilibrium in normal-form games.

We next observe that inconsistent collections and anomalies have a simple characterization in terms of a theory's allowable functions $\cF^T$ and a loss function.

\begin{observation}\label{observation: inconsistent/anomalies in terms of allowable functions}
Consider a loss function $\ell \colon \cY^* \times \cY^* \rightarrow \mathbb{R}_{+}$ satisfying $\ell(y, y^\prime) = 0$ if and only if $y = y^\prime$. 
Then, (i) a collection $D \in \cD$ is inconsistent with theory $\mathcal{F}^T$ if and only if $\min_{f \in \cF^{T}} |D|^{-1} \sum_{(x, y^*) \in D} \ell\left( f(x), y^* \right) > 0$; and (ii) if there exists no inconsistent collection with fewer than $n > 1$ examples, then any inconsistent collection with $n$ examples is an anomaly.
\end{observation}

\noindent This is immediate from Definitions \ref{defn: theory as allowable functions}-\ref{defn: anomaly}. 
Searching for inconsistent collections is equivalent to searching for collections that induce a positive loss for the theory's allowable functions. 
Furthermore, we can search for anomalies by iteratively searching for larger inconsistent collections. 
Observation \ref{observation: inconsistent/anomalies in terms of allowable functions}(i) can be reinterpreted as an adversarial game between the theory (the min-player) and a falsifier. 
The falsifier proposes examples to the theory, and the theory attempts to explain them by fitting its allowable functions. 
The theory's payoffs are decreasing in its loss, and the falsifier wishes to search for examples that induce a positive loss for the theory's best-responding allowable function.

\subsection{Connecting anomalies to observable data}\label{section: observable data and SL}

To this point, we analyzed the behavior of a theory on collections of hypothetical examples. 
Researchers construct anomalies by contrasting their intuition with a theory $\mathcal{F}^T$ and crafting examples where the theory's predictions differ from their hypothesized outcomes. 
We will instead use predictive algorithms as our procedures' empirical intuition. 
This step requires some assumption on how observable data map onto the theory's examples. 

We will assume data are drawn from some joint distribution over $(X_i, Y_i) \sim P$, where $Y_i \in \cY$ is an observed outcome and $P(X_i = x) > 0$ for all $x \in \cX$.
The observed outcome is statistically related to the modeled outcome through $f^*(x) := \mathbb{E}\left[ g(Y_i) \mid X_i = x \right]$ for some researcher-specified function $g$, where $\mathbb{E}[\cdot]$ denotes the expectation under $P$.
An example is now a pair $(x, f^*(x))$.
The theory's modeled outcome is simply the conditional expectation of some known transformation of the observed outcome.
For example, researchers often first estimate choice probabilities from data on discrete choices and strategy profiles in normal-form games from data on actions.

We would like to generate anomalies for a theory $\mathcal{F}^{T}$ of the form $D = \{ (x_1, f^*(x_1)), \hdots, (x_n, f^*(x_n)) \}$.
In this light, we can reinterpret the largely informal process used by researchers to generate anomalies themselves: researchers ``estimate'' $f^*$ using intuition about the economic domain and then ``contrast'' it against the theory $\mathcal{F}^T$ using their creativity.
We instead propose two procedures to generate anomalies that will first use data and supervised learning algorithms, rather than intuition, to estimate $f^*$ and second use optimization techniques, instead of creativity, to contrast the resulting estimate against the theory $\mathcal{F}^T$.

\section{Two procedures for anomaly generation}\label{section: two algorithms for anomaly generation}

In this section, we describe our two specific procedures for anomaly generation. 
Both procedures contrast the theory $\mathcal{F}^T$ with a predictive algorithm, searching for minimal collections on which the theory cannot explain the model's predictions. 
The quality of the resulting generated anomalies depend on both the model's predictive fit and the optimization performance of the search procedure. 
We discuss two complementary evaluations for algorithmically generated anomalies that mirror how we evaluate hypothesized anomalies from researchers: logical verification and empirical testing. 

\subsection{An adversarial procedure}\label{section: adversarial algorithm, main text}

For $x_{1:n} := (x_1, \hdots, x_n)$, let $\cE(x_{1:n}) := \min_{f \in \cF^{T}} n^{-1} \sum_{i=1}^{n} \ell(f(x_i), f^*(x_i))$ be the theory's loss on $D = \{ (x_1, f^*(x_1)), \hdots, (x_n, f^*(x_n)) \}$. 
Following Observation \ref{observation: inconsistent/anomalies in terms of allowable functions}, the falsifier's adversarial problem is given by 
\begin{equation}\label{eqn: max-min program}
    \max_{x_{1:n}} \ \cE(x_{1:n}),
\end{equation}
which searches for collections that generate large positive loss for the theory's best responding function (if they exist). 
Our first procedure for generating anomalies is based on this program. 
We iterate over collection sizes $n$ and solve the falsifier's problem \eqref{eqn: max-min program}. 
The collection associated with feature vector $x_{1:n}$ is inconsistent with $\mathcal{F}^T$ if and only if $\cE(x_{1:n}) > 0$  following Observation \ref{observation: inconsistent/anomalies in terms of allowable functions}. 
We verify it is an anomaly by checking whether $\mathcal{F}^T$ is consistent with its subsets. 

This search procedure is not directly feasible for two reasons: $f^*$ is not known; and the falsifier's problem \eqref{eqn: max-min program} is a possibly difficult max-min optimization. 
Our first procedure tackles these challenges by estimating $f^*$ from data using supervised learning algorithms and practically optimizing via a simple gradient descent ascent (GDA) routine. 

\subsubsection{Analysis of plug-in max-min optimization}\label{section: finite sample bound}

Recall $f^*(x) = \mathbb{E}[g(Y_i) \mid X_i = x]$ for some researcher-specified function $g$. 
Suppose we observe $(X_i, Y_i) \sim P$ for $i = 1, \hdots, N$ and construct an estimate $\widehat{f}^*$. 
This might be constructed using supervised machine learning algorithms that predict $g(Y_i)$ based on the features $X_i$ such as deep neural networks or classic nonparametric regression techniques. 
Consider the \textit{plug-in} max-min program 
\begin{equation}\label{equation: plug-in max min}
    \max_{x_{1:n}} \min_{f \in \cF^{T}} n^{-1} \sum_{i=1}^{n} \ell\left( f(x_i), \widehat{f}^*(x_i) \right).
\end{equation}
We analyze the plug-in program's error for the infeasible program \eqref{eqn: max-min program} assuming access to approximate optimization routines that solve the inner minimization and outer maximization problems possibly up to some errors. 

\begin{assumption}[Approximate optimization]\label{assumption: optimization oracles}
\hfill
\begin{enumerate}
\item[i.] For any $x_{1:n}$ and $\widehat{f}^* \in \cF$, the minimization routine returns $\widetilde{f}(\cdot; x_{1:n}) \in \cF^{T}$ satisfying, for some $\delta \geq 0$, $n^{-1} \sum_{i=1}^{n} \ell\left( \widetilde{f}(x_i; x_{1:n}), \widehat{f}^*(x_i) \right) \leq \, \min_{f \in \cF^{T}} n^{-1} \sum_{i=1}^{n} \ell\left(f(x_i), \widehat{f}^*(x_i)\right) + \delta$.

\item[ii.] For any $f(\cdot; x_{1:n}) \in \cF^{T}$ and $\widehat{f}^* \in \cF$, the maximization routine returns $\widetilde{x}_{1:n}$ satisfying, for some $\nu \geq 0$, $n^{-1} \sum_{i=1}^{n} \ell\left( f(\widetilde{x}_i; \widetilde{x}_{1:n}), \widehat{f}^*(\widetilde{x}_i) \right) \geq \max_{x_{1:n}} \, n^{-1} \sum_{i=1}^{n} \ell\left( f(x_i, x_{1:n}), \widehat{f}^*(x_i) \right) - \nu$.
\end{enumerate}
\end{assumption}

\noindent Our analysis will yield an inequality that decomposes the total error into a statistical component (from estimating $f^*$) and an optimization component (from the routines' optimality gaps). 
Alternative estimators or optimization routines may yield better generated anomalies by reducing their corresponding terms. 

Define $\widetilde{f}^{T}(\cdot; x_{1:n})$ to be the allowable function returned when the minimization routine solves $\min_{f \in \cF^{T}} n^{-1} \sum_{i=1}^{n} \ell\left( f(x_i), \widehat{f}^*(x_i) \right)$ at any $x_{1:n}$, and $\widetilde{x}_{1:n}$ to be the feature values returned when the maximization routine solves $\max_{x_{1:n}} n^{-1} \sum_{i=1}^{n} \ell\left( \widetilde{f}^{T}(x_i; x_{1:n}), \widehat{f}^*(x_i) \right)$.
The optimal values of the plug-in and population programs are
\begin{equation}
    \widehat{\cE} := n^{-1} \sum_{i=1}^{n} \ell\left( \widetilde{f}^T(\widetilde{x}_i, \widetilde{x}_{1:n}), \widehat{f}^*(\widetilde{x}_i) \right) \mbox{ and } \cE = \max_{x_{1:n}} \min_{f \in \cF^{T}} n^{-1} \sum_{i=1}^{n} \ell\left( f(x_i), f^*(x_i) \right).
\end{equation}

\begin{proposition}\label{prop: plug in vs. pop max-min program}
Suppose the loss function $\ell(\cdot, \cdot)$ is differentiable with gradients bounded by some $K < \infty$ and convex in its second argument. 
Then, for any $n \geq 1$, 
\begin{equation}
    \left\| \widehat{\cE} - \cE \right\| \leq (\delta + \nu) + 3 K \| \widehat{f}^*(\cdot) - f^*(\cdot) \|_{\infty},
\end{equation}
where $\| f_1(\cdot) - f_2(\cdot) \|_{\infty} = \sup_{x \in \cX} |f_1(x) - f_2(x) |$ is the supremum norm between two functions $f_1(\cdot), f_2(\cdot) \in \cF$.
\end{proposition}

\noindent The error is bounded by the optimality gaps of the optimization routines and the estimation error of $\widehat{f}^*$.
The estimation error contributes through the worst-case (supremum norm) error of $\widehat{f}^*$ for $f^*$.
If we could exactly optimize and set $\delta, \nu = 0$, the rate at which the plug-in optimal value converges to the population optimal value is bounded by the rate at which $\widehat{f}^*$ converges uniformly to $f^*$.
While strong, it is unsurprising that this form of convergence is sufficient to control the plug-in's error as the max-min optimization program possibly extrapolates to unseen features in searching for inconsistent collections.
While Proposition \ref{prop: plug in vs. pop max-min program} asks for error control at the realized collection, one could instead target bounds on the average error over realized collections produced across optimization runs; this might permit $L^2$-type errors at the cost of assumptions on the distribution over collections induced by the optimization routines. This is an interesting direction. 

Proposition \ref{prop: plug in vs. pop max-min program} is agnostic, applying to any choice of estimator $\widehat{f}^*$ and optimization routines.
First, this is valuable as there is tremendous appetite for comparing economic theories against a wide range of supervised learning algorithms; in studies of choice under risk and strategic behavior alone, researchers already estimate regularized linear regression \citep[][]{PeysakhovichNaecker(17)}, kernel ridge regression \citep[][]{AndrewsEtAl(22)-Transfer}, random forests \citep[][]{FudenbergLiang(19)-InitialPlay}, deep neural networks \citep[][]{HartfordEtAl(16), PetersonEtAl(21)-MLforChoiceTheory} and recurrent neural networks \citep[][]{HirasawaEtAl(22)-GameTheoryPred}. 
We should expect new innovations in machine learning algorithms to quickly enter these comparisons.
The estimation error term also connects anomaly generation to theoretical innovations that provide high-probability bounds on worst-case error for modern nonparametric estimators --- see, for example, \citet[][]{BelloniEtAl(15)-Series, ChenChristensen(15)-Series} on series based estimators, reproducing kernel Hilbert space methods \citep[e.g.,][]{FischerEtAL(20)-RKHS}, and deep neural networks \citep[e.g.,][]{Imaizumi(23)-DeepNetsSupNorm}.
As such results are developed for novel supervised learning algorithms, they can be combined with Proposition \ref{prop: plug in vs. pop max-min program} to provide high probability bounds on the error.

Second, it is also valuable to remain agnostic about the choice of optimization routines. 
A wide array of optimization routines are available for solving max-min programs, and different problems might reward different routines \citep[e.g., see][for overviews]{RazaviyaynEtAl(20)}.
The optimality gap term creates an interface between anomaly generation and improvements in optimization --- better optimization routines in a particular problem improve the resulting generated anomalies.
In the next section, we discuss one concrete optimization routine that we use in our application to generating anomalies for expected utility theory.

\subsubsection{Optimization via gradient descent ascent}\label{section: main text, GDA}

The plug-in max-min program \eqref{equation: plug-in max min} has connections to a computer science literature on adversarial learning \citep[e.g.,][]{AkhtarMian(18)-AdversarialAttacks}.
In adversarial learning, ``data-poisoning attacks'' are studied to understand the robustness of black box predictive algorithms: Given an estimated neural network for image classification, for example, we search for perturbations to pixel values that would lead the neural network to (humorously) classify a picture of a pig as an airliner or (more dangerously) fail to notice a stop sign in a self-driving car.
The resulting perturbed images are known as ``adversarial examples.'' 

The plug-in max–min program’s search for anomalies can be viewed as a data-poisoning attack on the theory’s allowable class $\cF^{T}$. The falsifier seeks collections $D$ that degrade the performance of every $f\in\cF^{T}$; the resulting generated anomalies are adversarial examples for the theory. 
Motivated by this connection, we solve the plug-in max–min problem with a gradient descent–ascent (GDA) routine. 
An active literature in adversarial learning offers many gradient-based methods for analogous problems \citep[e.g.,][]{RazaviyaynEtAl(20), MokhtariEtAl(20)}, and since these techniques have been used to construct adversarial examples in machine learning, they are a natural starting point for anomaly generation. 
We use a particularly simple GDA implementation and describe its properties.

To describe our specific implementation, we first simplify the inner minimization over the theory's allowable functions. 
We assume the theory's allowable functions can be parametrized, meaning $\cF^{T} = \{ f_{\theta}(\cdot) \colon \theta \in \Theta \}$ for some parameter vector $\theta$ and compact parameter space $\Theta$. 
In expected utility theory, for example, such a parametrization may involve a flexible sieve basis or a class of neural networks for the possible utility functions. 
The inner minimization over the theory's allowable functions then becomes 
\begin{equation}
    \min_{\theta \in \Theta} \ n^{-1} \sum_{i=1}^{n} \ell\left( f_{\theta}(x_i), \widehat{f}^*(x_i) \right).
\end{equation}
This is equivalent to an empirical risk minimization problem, and so we can apply standard optimization techniques (e.g., convex methods for particular parametrizations and loss functions or gradient descent with random initializations). 
The outer maximization will often be non-concave; varying the features induces variation in the estimated function $\widehat{f}^*$, the theory's allowable function $f_{\theta}$ and the theory's best-fitting parameter vector $\theta \in \Theta$.
We nonetheless use a gradient-based optimization procedure. 
Let $\widehat{\cE}(x_{1:n}, \theta) := n^{-1} \sum_{i=1}^{n} \ell\left( f_{\theta}(x_i), \widehat{f}^*(x_i) \right)$ and we assume $\widehat{\cE}(x_{1:n}, \theta)$ is differentiable in $x_{1:n}$ for all $\theta \in \Theta$. 
For a collection of initial feature values $x_{1:n}^0$, maximum number of iterations $S > 0$, and some chosen step size sequence $\{\eta_s\}_{s=0}^{S} > 0$, we iterate over $s = 0, \hdots, S$ and calculate at each iteration
\begin{align}
    & \theta^{s+1} = \arg \min_{\theta \in \Theta} \ \widehat{\cE}(x_{1:n}^{s}; \theta) \label{eqn: GDA inner minimization} \\
    & x_{1:n}^{s+1} = x_{1:n}^{s} + \eta_{s} \nabla \widehat{\cE}(x_{1:n}^{s}; \theta^{s+1}). \label{eqn: GDA outer maximization step}
\end{align}
At each iteration $s$, we construct an approximate solution to the inner minimization problem $\theta^{s+1}$ and take a gradient ascent step on the feature values plugging in $\theta^{s+1}$. 
As we discuss in Appendix \ref{section: GDA over allowable functions}, such simple fixed step-size GDA routines converge to an approximate stationary point of the outer maximization problem \citep[e.g.,][]{JinEtAl(19)}, loosely meaning that $\nabla \widehat{\cE}(x_{1:n}, \theta) \approx 0$ at the returned feature and parameter vectors.

The family of GDA routines is a reasonable choice for anomaly generation when the theory's allowable functions can be parametrized and the theory's average loss on a given collection is differentiable. 
While we consider a fixed step size routine here, there exists a variety of GDA algorithms as mentioned earlier; and a fruitful direction is to explore alternative optimization routines depending on the domain. 
We will return to this point in Section \ref{section: evaluating algorithmically generated anomalies} below.

\subsection{Representational anomalies and an example morphing procedure}\label{section: example algorithm, main text}

Our adversarial procedure exploits no structure beyond the theory's allowable functions $\mathcal{F}^{T}$. 
Yet many economic theories operate on a lower-dimensional representation of the domain, meaning each allowable function $f \in \mathcal{F}^{T}$ behaves as if it pools together some distinct feature values. 
We next propose an example morphing algorithm to search for ``representational anomalies'' that highlight how the allowable functions fail to capture some dimension along which predicted outcomes vary.

\subsubsection{Representational equivalence and anomalies}

Many economic theories draw the same implications at distinct features $x, x^\prime$, which we formalize in the following definition.

\begin{definition}\label{defn: representational equivalence in terms of allowable functions}
$x_1, x_2 \in \mathcal{X}$ are \textit{representationally equivalent} if and only if $f(x_1) = f(x_2)$ for all $f(\cdot) \in \cF^{T}$.
\end{definition}

To build intuition, a theory $\mathcal{F}^{T}$ has representationally equivalent features whenever it ignores some aspect of the economic domain. 
In choice under risk, consider presenting a menu of risky lotteries as either a two-stage lottery or as a simple lottery over final payoffs.  
Expected utility theory is silent on whether this presentational choice, which could be encoded in the features $x \in \mathcal{X}$, would influence an individual's decision. 
Any pair of menus of $x_1, x_2$ whose constituent lotteries have the same final payoffs and probabilities over those final payoffs yet differ in their presentation are representationally equivalent.

While ignoring an aspect of the economic domain is sufficient for the existence of representationally equivalent features, it is not necessary. 
In choice under risk, suppose we defined the features $x \in \mathcal{X}$ to only consist of the final payoffs and probabilities associated with the lotteries in the menu. 
Expected utility theory of course does not ignore any of these features in modeling risky choices.
Nonetheless, there exists a representationally equivalent menu $x_2$ consisting of the compound lotteries $\alpha (p_0, z_0) + (1 - \alpha) (\widetilde{p}, \widetilde{z})$ and $\alpha (p_1, z_1) + (1 - \alpha) (\widetilde{p}, \widetilde{z})$. 
The pair satisfies $f(x_1) = f(x_2)$ due to the independence axiom.

Representationally equivalent features, if they exist, provide more structure that can be exploited for anomaly generation.
Particular anomalies highlight failures in the theory's representation, as we define next.

\begin{definition}\label{definition: taxonomy of anomalies}   
$D = \{ (x_1, y^*_1), (x_2, y^*_2) \}$ is a \textit{representational anomaly} for theory $\mathcal{F}^{T}$ if $x_1, x_2$ are representationally equivalent under $\mathcal{F}^{T}$ but $y^*_1 \neq y^*_2$.
\end{definition} 

\noindent Researchers often generate representational anomalies for theories themselves. 
Many classic examples of anomalies for expected utility theory are, in fact, representational anomalies.
Consider again the Allais paradox (Table \ref{tab: the allais paradox}). 
Due to the independence axiom, expected utility theory requires $f(x_A) = f(x_B)$ for all $f \in \mathcal{F}^{T}$, and so the menus $x_A, x_B$ are representationally equivalent. 
Yet the Allais paradox highlights that choices may vary across these two menus, and it is therefore a representational anomaly. 
By the same reasoning, the certainty effect (Table \ref{tab: the certainty effect}) is also a representational anomaly.
\cite{TverskyKahneman(81)} constructed other representational anomalies for expected utility theory that highlight whether lotteries are presented as two-stage lotteries versus simple lotteries may affect individuals' risky choices.

\subsubsection{An example morphing procedure}

Given $f^*(x) \equiv \mathbb{E}[Y_i \mid X_i=x]$, our goal is to search for representational anomalies ${(x_1,f^*(x_1)),(x_2, f^*(x_2))}$ using only the theory’s allowable class $\mathcal{F}^T$. 
Our procedure attempts to first characterize the representational equivalences implied by $\mathcal{F}^T$ and then search adversarially within them. 
For the implementation below, we assume that $f^*$ and all $f\in\mathcal{F}^T$ are differentiable (this is satisfied in our illustration with expected utility theory and additive logit noise) and that the theory admits a local representation.

\begin{assumption}[Local representational equivalence]\label{asm: local representational equivalence}
If $x_1, x_2 \in \cX$ are representationally equivalent under $\mathcal{F}^{T}$, then so are $\lambda x_1 + (1 - \lambda) x_2$ for any $\lambda \in (0, 1)$.
\end{assumption}

\noindent Representations are \textit{local} in the sense that there exists a small deviation from $x_1$ or $x_2$ that is also representationally equivalent. 
Under Assumption \ref{asm: local representational equivalence}, we could search for representational anomalies by taking small gradient steps. 
Given $f^*$ and an initial value $x^0$, we could search for directions $v \in \mathbb{R}^{dim(x)}$ along which no allowable function $f \in \cF^{T}$ changes but $f^*$ changes substantially, and we then update or \textit{morph} $x^0$ in the direction $v$. 
Specifically, let $\cN(x) = \{ v \in \mathbb{R}^{dim(x)} \colon \nabla f(x)^\prime v = 0 \mbox{ for all } f(\cdot) \in \cF^{T} \}$ denote the subspace of directions that are orthogonal to the gradient of each allowable function. 
Under Assumption \ref{asm: local representational equivalence}, $\cN(x)$ is non-empty at any $x$ for which there exists some representationally equivalent $x^\prime$. 
For an initial feature value $x^0$, maximum number of iterations $S$, and step size sequence $\{\eta_s\}_{s=0}^{S}$, we iterate over $s = 0, \hdots, S$ and compute the update step 
\begin{equation}
    x^{s+1} = x^{s} - \eta_s \mbox{Proj}\left( \nabla f^*(x^s) \mid \cN(x^s) \right),
\end{equation}
where $\mbox{Proj}\left( \cdot \right)$ is the projection operator and $\mbox{Proj}\left( \nabla f^*(x) \mid \cN(x)\right)$ is the projection of the gradient of $f^*(\cdot)$ onto the null space of the allowable functions. 
We therefore move in directions of $f^*$ that hold fixed the value of any allowable function $f \in \cF^{T}$.

Since we do not observe $f^*$, we again construct an estimator $\nabla \widehat{f}^*$ based on a random sample $(X_i, Y_i) \sim P(\cdot)$ i.i.d. for $i = 1, \hdots, n$. 
We plug the estimator into the morphing procedure and apply the update step $x^{s+1} = x^{s} - \eta_s \mbox{Proj}\left( \nabla \widehat{f}^*(x^s) \mid \cN(x^s)  \right)$ at each iteration $s = 0, \hdots, S$.
Provided the error $\nabla \widehat{f}^* - \nabla f^*$ is small, $\mbox{Proj}\left( \nabla \widehat{f}^*(x) \mid \cN(x) \right)$ remains a descent direction for $f^*(\cdot)$, 

\begin{proposition}\label{prop: descent direction for morphing}
    Under Assumption \ref{asm: local representational equivalence}, $-\mbox{Proj}\left( \nabla f^*(x) \mid \cN(x) \right)$ is a descent direction for $f^*$. 
    Furthermore, $-\mbox{Proj}\left( \nabla \widehat{f}^*(x) \mid \cN(x)  \right)$ is also a descent direction for $f^*$ provided $\|\nabla \widehat{f}^*(x) - \nabla f^*(x)\|_2 \leq \| \mbox{Proj}\left( \nabla f^*(x) \mid \cN(x) \right) \|_2$ is satisfied.
\end{proposition}

While Proposition \ref{prop: descent direction for morphing} analyzes the properties of plugging the estimated gradient of the true function into the morphing procedure, it still leaves open the question of how to practically characterize the subspace $\mathcal{N}(x)$ and implement the projection operator. 

In our implementation, we again parametrize the allowable functions, meaning $\cF^{T} = \{ f_{\theta} \colon \theta \in \Theta\}$ for some $\theta \in \Theta$ as in Section \ref{section: main text, GDA}. 
We implement the projection operator by sampling $B > 0$ parameter values $\theta \in \Theta$ at each update step and directly orthogonalizing the gradient $\nabla \widehat{f}^*(x)$ with respect to each of the sampled gradients $\nabla f_{\theta}(x)$.
As $B$ grows large, this better approximates the null space of the allowable function $\cN(x)$.
Our implementation of example morphing only requires the access to the allowable functions. 
Of course, if there are known directions $v^s$ along which no allowable functions vary, then this sampling step is not needed and we can directly orthogonalize the gradient $\nabla \widehat{f}^*(x)$ with respect to the known directions. In this case, example morphing uses gradient steps to adversarially search within the representational equivalences of the allowable functions.

\subsection{Evaluating algorithmically generated anomalies}\label{section: evaluating algorithmically generated anomalies}

Having introduced two specific anomaly generation procedures, we next suggest two complementary evaluations: logical verification and empirical testing.
These mirror the two-step workflow already used by researchers: researchers first assess whether a hypothesized anomaly would be inconsistent with the theory if nature behaved as they postulated; and second, researchers then collect new data on the hypothesized anomaly and test whether nature in fact does so. 

Let $\widehat{X}_{1:n}$ denote a collection of features returned by an anomaly generation procedure using training data $D_N = \{(X_i,Y_i)\}_{i=1}^N$, $\{f_\theta:\theta\in\Theta\}$ be a parametrization of the theory’s allowable functions and $\widehat{\cE}(x_{1:n};\theta)$ be the average loss of $f_\theta$ evaluated on $\{(x_j,\widehat f^{\,*}(x_j))\}_{j=1}^n$.
Logical verification reports two evaluation metrics, conditioning on the training dataset $D_{N}$,
\begin{equation}\label{eqn: logical verification}
    P\left(\min_{\theta\in\Theta} \ \widehat{\cE}(\widehat{X}_{1:n};\theta)>0\mid D_N\right) \mbox{ and } P\left(\widehat{\cE}(\widehat{X}_{1:n})>0\mid D_N\right), 
\end{equation}
where the first and second term verify whether $\widehat{X}_{1:n}$ is an anomaly for parametrized allowable function class $\{f_\theta:\theta\in\Theta\}$ and the allowable function class $\mathcal{F}^T$ respectively.
These two can be different since $\{f_\theta:\theta\in\Theta\}$ may only approximate $\mathcal{F}^T$.
Because $\widehat{X}_{1:n}$ may depend on random initializations and other optimization choices (e.g., number of update steps), these probabilities summarize the fraction of optimization runs that yield an inconsistent collection taking $\widehat{f}^*$ as given (i.e., assuming nature behaves as postulated by the predictive algorithm). 

Logical verification evaluates the quality of the optimization used by an anomaly generation procedure --- how well does a specific procedure contrast the predictive algorithm against the theory? 
Better optimization should yield more generated anomalies taking $\widehat{f}^*$ as given. 
This is valuable as researchers should explore diverse optimization techniques in anomaly generation. 
For example, the adversarial formulation given in Section \ref{section: adversarial algorithm, main text} admits many GDA routines. Since the anomaly generation problem is often nonconvex-nonconcave, different routines may excel on different problem instances. 
The logical verification metrics provide scores that can be used for comparisons. 
Of course, logical verification only reveals inconsistencies between the predictive model and the theory; it does not by itself establish violations in nature.

For this reason, empirical testing is an essential next step: treat the algorithmically generated anomalies as experimental designs for data collection and formally test whether the theory is violated on independent data. 
More specifically, let $S(\widehat{X}_{1:n},Y_{1:n})$ denote a particular test statistic for misspecification (e.g., a violation rate with expected utility theory) computed on independent responses $Y_{1:n}^\prime \mid  \widehat{X}_{1:n}$. Empirical testing then reports, for example, 
\begin{equation}\label{equation: empirical testing}
\mathbb{E}\left[S(x_{1:n},Y_{1:n}^\prime)\;\middle|\; \widehat{X}_{1:n}=x_{1:n}\right] \mbox{ and } \mathbb{E}\!\left[S(\widehat{X}_{1:n},Y_{1:n}^\prime)\;\middle|\; D_N\right],
\end{equation}
where the first term is the average violation on a particular generated anomaly and the second term is the average violation across anomalies generated by the procedure given $D_N$ (of course, researchers might instead report a hypothesis test on a generated anomaly and a rejection rate across generated anomalies).
We propose researchers estimate these quantities with independently collected data in new experiments. 

Empirical testing is practical in many domains like risky choice and strategy behavior, where online experiments are feasible and relatively cheap.
Furthermore, this serves to statistically decouple testing for misspecification from the anomaly generation step. 
Any supervised learning algorithm and any optimization routine can be used upstream in anomaly generation to produce $\widehat{X}_{1:n}$ --- including future, more powerful tools --- without distorting downstream inferences on newly collected data.
Researchers are further free to bring to bear whatever statistic $S(\widehat{X}_{1:n},Y_{1:n})$ they prefer from rich work on specification testing and rationalizability tests in econometrics and economic theory \citep[e.g.,][]{Afriat(67),Varian(82), HarlessCamerer(94)}.
 
More broadly, logical verification and empirical testing allow researchers to reuse benchmark datasets, like large-scale datasets on risky choice \citep[e.g.,][]{PetersonEtAl(21)-MLforChoiceTheory} and strategic behavior \citep[e.g.,][]{WrightLeytonBrown(17), HirasawaEtAl(22)-GameTheoryPred}.
Anomaly generation procedures can be prototyped on these benchmarks and subsequently compared against each other based on subsequent logical verification and empirical testing.
This fosters comparable ``common tasks'' essential for the rapid development of methods \citep[][]{Donoho(24)}.

\section{Revisiting prospect theory through the lens of anomaly generation}\label{section: anomalies for choice under risk, simulation}

As a first illustration, we algorithmically generate anomalies for expected utility theory based on simulated lottery choice data consistent with prospect theory. 
We compare the anomalies generated by our procedures against known anomalies for expected utility theory that are implied by prospect theory, such as those produced in \citet{Allais(53)} and \citet{KahnemanTversky(79)}.

\subsection{Data generating process}\label{section: main text, simulation design}

We simulate lottery choice data from an individual who evaluates lotteries over $J > 1$ monetary payoffs according to the probability weighting function $\pi_j(p; \delta, \gamma) = \frac{\delta p_j^{\gamma}}{ \delta p_j^{\gamma} + \sum_{k \neq j} p_k^{\gamma} }$ for $j = 1, \hdots, J$, where $p \in \Delta^{J-1}$ and $\delta \geq 0, \gamma \geq 0$ govern the probability weighting function \citep[][]{LattimoreEtAl(92)}.
We calibrate $(\delta, \gamma)$ using estimates based on \cite{BruhinEtAl(10)} (reported in their Table V and Table IX), setting $(\delta, \gamma)$ to be equal to one of $(0.926, 0.377)$, $(0.726, 0.309)$, or $(1.063, 0.451)$. 
For these parameter values, the individual over-weights probabilities close to zero, under-weights probabilities close to one, and compresses intermediate probabilities (Appendix Figure \ref{figure: probability weighting function, visualization}).
This can generate known anomalies for expected utility theory, such as the Allais paradox and the Certainty effect.
These parameter values also introduce ``outcome pessimism'' when $\delta < 1$ (i.e., probability weights may sum to less than one), or ``outcome optimism'' when $\delta > 1$ (i.e., probability weights may sum to greater than one), which may lead the individual to select a first-order stochastically dominated lottery.

The individual has a linear utility function. 
The individual evaluates a lottery $(p, z)$ by $CPT(p, z; \delta, \gamma) := \sum_{j=1}^{J} \pi_j(p; \delta, \gamma) z_j$. 
On a menu $x = (p_0, z_0, p_1, z_1)$, we simulate the individual's probability of selecting lottery $1$ according to $f^*(x) = P\left( CPT(p_1, z_1; \delta, \gamma) - CPT(p_0, z_0; \delta, \gamma) + \xi \geq 0 \right)$, where $\xi$ is an i.i.d. logit shock.
The individual's binary choice is $Y_i \mid X_i = x \sim Bernoulli(f^*(x))$. 

\subsection{Generating anomalies for expected utility theory}\label{section: main text, simulation, implementation discussion and benchmark}

To apply our procedures, we parametrize the allowable functions of expected utility theory.
We model the utility function as a linear combination of non-linear basis functions with $u_{\theta}(z) = \sum_{k=1}^{K} \theta_k b_k(z)$ for basis functions $b_1(\cdot), \hdots, b_K(\cdot)$ (e.g., polynomial bases or monotone I-splines), $K$ finite, and parameter vector $\theta \in \Theta$.
We consider the parametrized allowable functions of expected utility theory $\{ f_{\theta} \colon \theta \in \Theta \}$ for $f_{\theta}(x) = P( \sum_{j=1}^{J} p_1(j) u_{\theta}(z_1(j)) - \sum_{j=1}^{J} p_0(j) u_{\theta}(z_0(j)) + \xi \geq 0)$ for $\xi$ an i.i.d. logit shock. 

We generate anomalies for expected utility theory over menus of two-payoff lotteries, applying our procedures to the choice probability function $f^*$.
For each parameter value $(\delta, \gamma)$, we apply our adversarial algorithm to 25,000 randomly initialized menus of lotteries $x^0$ and our example morphing algorithm to 15,000 randomly initialized menus. 
Appendix \ref{section: simulations, implementation details} provides further details on the implementation, and we generate anomalies based on an estimated choice probability function $\widehat{f}$ in Appendix \ref{section: simulations, estimated choice probabilities, binary payoffs}.

We evaluate our anomaly generation procedures by logical verification as discussed in Section \ref{section: evaluating algorithmically generated anomalies}: among the menus generated, how many are inconsistent with expected utility theory?
For the parametrized allowable function classes, our procedures generate $1,581$, $1,852$, and $2,564$ anomalies across calibrated parameter values $(\delta, \gamma)$ for the probability weighting function respectively.
Since these parametrized allowable functions are restrictive, we numerically verify whether the same menus are inconsistent with expected utility theory at any increasing utility function and without noisy choices (see Appendix \ref{section: simulations, numerical verification of EUT anomalies} for details).
Altogether our procedures generate 309 anomalies for expected utility theory across calibrated parameter values $(\delta, \gamma)$ for the probability weighting function.

As a baseline comparison, we randomly sample $25,000$ menus at random for each calibrated parameter value $(\delta, \gamma)$ and flag those on which the implied choices would violate expected utility theory. 
This corresponds to the initialization step of boosting-style procedure for anomaly generation, which would subsequently reweight toward hard instances for the theory \citep[e.g.,][]{FudenbergLiang(19)-InitialPlay}.
This baseline returns only $15$, $17$, and $7$ anomalies for the parametrized allowable function class across calibrated parameter values $(\delta, \gamma)$ respectively; furthermore, it generates zero anomalies for expected utility theory at any increasing utility function.
This comparison highlights two lessons: first, anomalies for a theory can be rare in the feature space, and so targeted search is necessary; and second, our optimization routines add value in this illustration, navigating the complicated landscape to generate anomalies that would fail to be uncovered through brute force.

\subsection{What anomalies do the algorithms generate?}\label{section: anomalies, categorization, binary payoffs}

To interpret the 309 anomalies for expected utility theory generated by our procedures, we categorize them in two steps based on our knowledge of expected utility theory and the particular violations they highlight.
We first numerically check whether the generated anomaly is a first-order stochastic dominance violation, which is a well known anomaly implied by particular parametrizations of the probability weighting functions.
We next numerically check whether the generated pair of menus can be represented as compound lotteries over the same simple lotteries and whether there is a choice reversal across the menus, which would indicate a violation of the independence axiom of expected utility theory. 
Since the space is simple enough, we enumerate the possible compounding operations.
Appendix \ref{section: categorization details, two payoff anomalies} provides further details on our numeric categorization.
Our algorithmically generated anomalies fall into distinct categories, which illustrate choice reversals across menus of lotteries predicted by the probability weighting function (Table \ref{tab: anomaly categorization for binary payoffs}). 

\begin{table}[h!]
\caption{Algorithmically generated anomalies for expected utility theory over menus of two lotteries on two monetary payoffs.}
\centering
\begin{tabular}{c c c c}
 & \multicolumn{3}{c}{Prob. Weighting Function: $(\delta, \gamma)$} \\
 & $(0.726, 0.309)$ & $(0.926, 0.377)$ & $(1.063, 0.451)$ \\
 \multicolumn{1}{r|}{Dominated Consequence Effect} & 85 & 34 & 10 \\
 \multicolumn{1}{r|}{Reverse Dominated Consequence Effect} & 17 & 15 & 14 \\
 \multicolumn{1}{r|}{Strict Dominance Effect} & 45 & 1 & 0 \\
 \multicolumn{1}{r|}{First Order Stochastic Dominance} & 81 & 0 & 2 \\
 \multicolumn{1}{r|}{Other} & 3 & 1 & 1 \\
 \hline \hline 
 \multicolumn{1}{r|}{\# of Anomalies} & 231 & 51 & 27 
\end{tabular}
\label{tab: anomaly categorization for binary payoffs}
\floatfoot{\textit{Notes}: This table summarizes the anomalies for expected utility theory over menus of two-payoff lotteries produced by the adversarial algorithm and the example morphing algorithm. 
The anomalies are organized by probability weighting parameter values $(\delta, \gamma)$ and categories. See Section \ref{section: anomalies, categorization, binary payoffs} for discussion.}
\end{table}

\vspace{-1em}
\paragraph{The dominated consequence effect:} Consider the algorithmically generated pair of menus in Table \ref{tab: generated anomalies, binary lotteries, dominated consequence, main text}.
The individual selects lottery A0.
Since lottery A0 has a lower expected value than lottery A1, expected utility theory could rationalize this choice with an appropriate degree of risk aversion. 
Yet the individual also selects the lottery B1, which has a higher expected value than lottery B0. 
It appears as if the individual's risk attitudes have reversed on menu B.

More formally, menu A and menu B in Table \ref{tab: generated anomalies, binary lotteries, dominated consequence, main text} are engineered such that the choice predictions of expected utility theory are constant across them.
Lottery B0 can be expressed as a compound lottery over lottery A0 and a degenerate lottery that yields the certain payoff $6.44$; that is, $B0 = \alpha_0 A0 + (1 - \alpha_0) \delta_{6.44}$ for some $\alpha_0 \in (0, 1)$. 
Analogously, lottery B1 can be written as $B1 = \alpha_1 A1 + (1 - \alpha_1) \delta_{5.72}$ for some $\alpha_1 < \alpha_0$. 
The individual's choices express that lottery $A0$ is preferred to lottery $A1$ but $\alpha_1 A1 + (1 - \alpha_1) \delta_{5.72}$ is preferred to $\alpha_0 A0 + (1 - \alpha_0) \delta_{6.44}$, contradicting expected utility theory since it can be shown that $A0$ being preferred to $A1$ must imply that $\alpha_0 A0 + (1 - \alpha_0) \delta_{6.44}$ must be preferred to $\alpha_1 A1 + (1 - \alpha_1) \delta_{5.72}$.
We provide a formal proof in Appendix \ref{section: simulations, binary payoffs, proofs of anomalies}.

\begin{table}[h!]
\caption{An algorithmically generated anomaly for expected utility theory illustrating the dominated consequence effect.}
\begin{subtable}{.45\linewidth}
\centering
\begin{tabular}{c c c}
 & \multicolumn{1}{l}{Menu A $(x_A, y_A^*)$}\\
 \multicolumn{1}{c|}{\cellcolor{green!25}Lottery 0} & \$6.44 & \$6.71 \\
 \multicolumn{1}{c|}{} & 0\% & 100\% \\
 \hline
 \multicolumn{1}{c|}{Lottery 1} & \$5.72 & \$8.64 \\
 \multicolumn{1}{c|}{} & 13\% & 87\% \\
 \hline
 \end{tabular}
\end{subtable} %
\begin{subtable}{.45\linewidth}
\centering
 \begin{tabular}{c c c}
 & \multicolumn{1}{l}{Menu B $(x_B, y_B^*)$} \\
 \multicolumn{1}{c|}{Lottery 0} & \$6.44 & \$6.71 \\
 \multicolumn{1}{c|}{} & 11\% & 89\% \\
 \hline
 \multicolumn{1}{c|}{\cellcolor{green!25} Lottery 1} & \$5.72 & \$8.64 \\
 \multicolumn{1}{c|}{} & 34\% & 66\% \\
 \hline
 \end{tabular}   
\end{subtable}
\floatfoot{\textit{Notes}: 
In each menu, we color the lottery that is selected with probability at least 0.50 in green. 
This anomaly was produced by the example morphing algorithm applied to the choice probability function with probability weighting parameter values $(\delta, \gamma) = (0.726, 0.309)$.
We round each payoff to the nearest cent and each probability to the nearest percentage.
See Section \ref{section: anomalies, categorization, binary payoffs} for further discussion.}
\label{tab: generated anomalies, binary lotteries, dominated consequence, main text}
\end{table}

All anomalies in the first row of Table \ref{tab: anomaly categorization for binary payoffs} have a common structure. 
We define the appropriate pair of lotteries as $\ell_0 = (p_0, z_0)$, $\ell_1 = (p_1, z_1)$ with $z_0 = (z_{0,1}, z_{0,2})$, $z_1 = (z_{1,1}, z_{1,2})$ and $\underline{z}_0 := \min_{j \in \{1, 2\}} z_{0j} < \min_{j \in \{1, 2\}} z_{1j} := \underline{z}_1$.
Each anomaly can be summarized as: for some $\alpha_0 \leq \alpha_1$, one menu consists of the choice between lottery $\ell_0$ and lottery $\ell_1$, and the other menu consists of the choice between the compound lotteries $\alpha_0 \ell_0 + (1 - \alpha_0) \delta_{\underline{z}_0}$ and $\alpha_1 \ell_1 + (1 - \alpha_1) \delta_{\underline{z}_1}$.
Selecting $\ell_1$ over $\ell_0$ implies that the individual also prefers $\alpha_1 \ell_1 + (1 - \alpha_1) \delta_{\underline{z}_1}$ over $\alpha_0 \ell_0 + (1 - \alpha_0) \delta_{\underline{z}_0}$ under expected utility theory.

We refer to this category as a ``dominated consequence effect'' as the pair of menus highlight a violation of the expected utility theory based on mixing each lottery with dominated, certain consequences (Appendix Table \ref{tab: generated anomalies, binary lotteries, dominated consequence} provides additional examples). 
The Common ratio effect (itself a generalization of the Certainty effect and Bergen paradox) is a special case of the dominated consequence effect; it can be recovered by setting $\alpha_0 = \alpha_1$ and placing additional restrictions on how the probabilities $p_0, p_1$ relate to one another. 
The dominated consequence effect therefore nests the well-known anomalies for expected utility theory over pairs of menus of two lotteries over two monetary payoffs. 

\vspace{-1em}
\paragraph{The reverse dominated consequence effect and the strict dominance effect:}
Consider next the algorithmically generated anomaly in Table \ref{tab: generated anomalies, binary lotteries, reverse dominated consequence, main text}
The individual selects lottery A0. 
The lotteries in menu B again have the same payoffs as those in menu A, though the probabilities associated with the highest payoffs increase in both lotteries across these menus.  
This leads to an apparent choice reversal as the individual now selects lottery B1. 
These menus are again engineered such that the choice predictions of expected utility theory do not change across them. 

\begin{table}[htbp!]
\caption{An algorithmically generated anomaly for expected utility theory illustrating the reverse dominated consequence effect.}
\begin{subtable}{0.45\linewidth}
\begin{tabular}{c c c}
 & \multicolumn{1}{l}{Menu A $(x_A, y_A^*)$}\\
 \multicolumn{1}{c|}{\cellcolor{green!25}Lottery 0} & \$2.59 & \$8.87 \\
 \multicolumn{1}{c|}{} & 88\% & 12\% \\
 \hline
 \multicolumn{1}{c|}{Lottery 1} & \$3.51 & \$8.65 \\
 \multicolumn{1}{c|}{} & 99\% & 1\% \\
 \hline
 \end{tabular}
\end{subtable} %
\begin{subtable}{0.45\linewidth}
\begin{tabular}{c c c}
 & \multicolumn{1}{l}{Menu B $(x_B, y_B^*)$} \\
 \multicolumn{1}{c|}{Lottery 0} & \$2.59 & \$8.87 \\
 \multicolumn{1}{c|}{} & 49\% & 51\% \\
 \hline
 \multicolumn{1}{c|}{\cellcolor{green!25} Lottery 1} & \$3.51 & \$8.65 \\
 \multicolumn{1}{c|}{} &  65\% & 35\% \\
 \hline
 \end{tabular}
\end{subtable}
\floatfoot{\textit{Notes}: 
In each menu, we color the lottery that is with probability at least 0.50 in green. 
This anomaly was produced by the example morphing algorithm applied to the choice probability function with probability weighting parameter values $(\delta, \gamma) = (0.726, 0.309)$.
We round each payoff to the nearest cent and each probability to the nearest percentage.
See Section \ref{section: anomalies, categorization, binary payoffs}.}
\label{tab: generated anomalies, binary lotteries, reverse dominated consequence, main text}
\end{table}

Each anomaly in the second row of Table \ref{tab: anomaly categorization for binary payoffs} share a common structure. 
We now define the appropriate pair of lotteries as $\ell_0 = (p_0, z_0)$, $\ell_1 = (p_1, z_1)$ with $z_0 = (z_{0,1}, z_{0,2})$, $z_1 = (z_{1,1}, z_{1,2})$ and $\overline{z}_0 := \max_{j \in \{1, 2\}} z_{0j} < \max_{j \in \{1, 2\}} z_{1j} := \overline{z}_1$. 
Each of these anomalies can be summarized as: 
for some $\alpha_1 \leq \alpha_0$, one menu consists of the choice between lottery $\ell_0$ and lottery $\ell_1$, and the other menu consists of the choice between the compound lotteries $\alpha_0 \ell_0 + (1 - \alpha_0) \delta_{\overline{z}_0}$ and $\alpha_1 \ell_1 + (1 - \alpha_1) \delta_{\overline{z}_1}$.
Since the other menu mixes lotteries $\ell_0$ and $\ell_1$ with their maximal payoffs, selecting $\ell_1$ over $\ell_0$ must imply the individual also prefers $\alpha_1 \ell_1 + (1 - \alpha_1) \delta_{\overline{z}_1}$ over $\alpha_0 \ell_0 + (1 - \alpha_0) \delta_{\overline{z}_0}$ under expected utility theory.
The menus highlight a violation of expected utility theory based on mixing each lottery with dominating certain consequences; we therefore refer to this category as a ``reverse dominated consequence effect'' (Appendix Table \ref{tab: generated anomalies, binary lotteries, reverse dominated consequence} provides additional examples).

Finally, all anomalies in the third row of Table \ref{tab: anomaly categorization for binary payoffs} exhibit what we call a ``strict dominance effect.'' 
We provide an illustrative example in Table \ref{tab: generated anomalies, binary lotteries, strict dominance, main text} (see Appendix Table \ref{tab: generated anomalies, binary lotteries, strict dominance effect} for additional examples).
In this case, the algorithmically generated anomaly shares a similar intuition as the original Allais paradox. 
The individual selects lottery A1 despite it having the lower expected payoff in the menu, demonstrating a degree of risk aversion.
The individual selects lottery B0 in menu B which is the higher expected payoff, and it again appears that their risk attitudes reversed.
The reversal here is particularly transparent.
Lottery B0 raises the probability of the lowest payoff in lottery A0, whereas lottery B1 raises the probability of the highest payoff in lottery A1. 
In this sense, the pair of menus highlights a violation of expected utility theory based on mixing lottery A1 with a certain consequence that strictly dominates the certain consequence mixed with lottery A0.
See Appendix \ref{section: simulations, binary payoffs, proofs of anomalies} for further discussion.

\begin{table}[htbp!]
\caption{An algorithmically generated anomaly for expected utility theory illustrating the strict dominance effect.}
\begin{subtable}{.45\linewidth}
\begin{tabular}{c c c}
& \multicolumn{1}{l}{Menu A $(x_A, y_A^*)$} \\
\multicolumn{1}{c|}{Lottery 0} & \$6.71 & \$8.98 \\
\multicolumn{1}{c|}{} & 22\% & 78\% \\
\hline
\multicolumn{1}{c|}{\cellcolor{green!25} Lottery 1} & \$7.17 & \$8.04 \\
\multicolumn{1}{c|}{} & 100\% & 0\% \\
\hline
\end{tabular}
\end{subtable} %
\begin{subtable}{.45\linewidth}
\begin{tabular}{c c c}
& \multicolumn{1}{l}{Menu B $(x_B, y_B^*)$} \\
\multicolumn{1}{c|}{\cellcolor{green!25} Lottery 0} & \$6.71 & \$8.98 \\
\multicolumn{1}{c|}{} & 49\% & 51\% \\
\hline
\multicolumn{1}{c|}{Lottery 1} & \$7.17 & \$8.04 \\
\multicolumn{1}{c|}{} &  45\% & 55\% \\     
\hline   
\end{tabular}
\end{subtable}
\floatfoot{\textit{Notes}: 
In each menu, we color the lottery that is selected with probability at least 0.50 in green. 
This anomaly was produced by the example morphing algorithm applied to the choice probability function with probability weighting parameter values $(\delta, \gamma) = (0.726, 0.309)$.
We round each payoff to the nearest cent and each probability to the nearest percentage.
See Section \ref{section: anomalies, categorization, binary payoffs}.}
\label{tab: generated anomalies, binary lotteries, strict dominance, main text}
\end{table}

While sharing similar intuitions, these final two categories are formally different than both the Common consequence effect and Common ratio effect.
These categories highlight violations of expected utility theory while using only two distinct payoffs in each lottery and mixing each lottery with particular certain consequences.
Our procedures generated categories of anomalies for expected utility theory that are implied by the probability weighting function, but to our knowledge had not been noticed before.

\vspace{-1em}
\paragraph{First-order stochastic dominance violations:}
All anomalies in the last row of Table \ref{tab: anomaly categorization for binary payoffs} are menus of lotteries in which the individual selects lotteries that are first-order stochastically dominated. 
See Appendix Table \ref{tab: generated anomalies, FOSD anomalies for subcertainty, binary payoffs} for specific examples.
Such first-order stochastic dominance violations were generally viewed as an undesirable ``bug'' in probability weighting that would be eliminated through an ``editing phase'' \citep[][]{KahnemanTversky(79)}.
What is intriguing is that our procedures generate first-order stochastic dominance violations on their own.

\section{Generating anomalies using real lottery choice data}\label{section: anomalies for choice under risk, choices13k}

We next apply our anomaly generation procedures to a publicly available dataset on real lottery choices. 
We construct a predictive algorithm that accurately predicts choice rates, and we then generate anomalies for expected utility theory implied by this estimated predictive algorithm. 
We produce 881 anomalies over menus of two-payoff lotteries and 1{,}996 anomalies over menus of three-payoff lotteries. 

\subsection{Constructing the predictive algorithm in real lottery choice data}

We revisit the Choices13K dataset collected in \cite{PetersonEtAl(21)-MLforChoiceTheory}.
Choices13K consists of over 1 million decisions made over 13,006 randomly sampled menus of lotteries. 
We focus on the 9,831 lottery menus in Choices13K that do not involve ambiguity nor correlation in the payoff structure between lotteries in a menu. 
For each menu $i$, we observe the fraction of all choices that selected Lottery 1 denoted $Y_i \in [0, 1]$ and the menu $X_i = \left( p_{i,0}, z_{i,0}, p_{i,1}, z_{i,1} \right)$. 

We train a shallow neural network (2 hidden layers with 32 nodes per layer) to predict choice rates $Y_i$ based on the lotteries $X_i$. 
We randomly select 8,831 lottery menus for training the neural network, and evaluate the resulting predictive algorithm $\widehat{f}^*$ on the remaining 1,000 held out lottery menus. 
On the held out lottery menus, our constructed predictive algorithm $\widehat{f}^*$ achieves a mean square error of approximately 0.014, which meaningfully improves over expected utility theory and is competitive with the best architecture in \cite{PetersonEtAl(21)-MLforChoiceTheory}. 

What does the predictive algorithm $\widehat{f}^*$ imply about lottery choice behavior?
To answer this question, we generate anomalies for expected utility theory implied by $\widehat{f}^*$, following the same steps as in Section \ref{section: main text, simulation, implementation discussion and benchmark}. 
We generate anomalies for expected utility theory over the space of menus of two-payoff lotteries and menus of three-payoff lotteries, applying our adversarial procedure and example morphing procedure to the constructed black box predictive algorithm $\widehat{f}^*$ and randomly initialized menus of either two-payoff or three-payoff lotteries.
We numerically verify whether each returned menu is an anomaly for expected utility theory at any increasing utility function and without noisy choices.
We report all resulting numerically verified anomalies for expected utility theory.

\subsection{Anomalies for expected utility theory over two-payoff lotteries}\label{section: choices13k, two payoff anomalies}

Our procedures generate 881 anomalies for expected utility theory over menus of two-payoff lotteries implied by the predictive algorithm $\hat{f}^*$.
We apply the same categorization based on the particular violation of expected utility theory they highlight, which we developed in Section \ref{section: anomalies, categorization, binary payoffs}.
The results are summarized in Table \ref{tab: anomaly categorization for binary payoffs, choices 13K}, and there are two findings worth highlighting.

\begin{table}[htbp!]
    \centering
    \begin{tabular}{c c}
    & \multicolumn{1}{c}{Predictive algorithm $\widehat{f}^*$} \\
    \multicolumn{1}{r|}{Dominated Consequence Effect} & 36 \\
    \multicolumn{1}{r|}{Reverse Dominated Consequence Effect} & 20 \\
    \multicolumn{1}{r|}{Strict Dominance Effect} & 72 \\
    \multicolumn{1}{r|}{First Order Stochastic Dominance} & 753 \\
    \hline \hline 
    \multicolumn{1}{r|}{\# of Anomalies} & 881 
    \end{tabular}
    \caption{Algorithmically generated anomalies for expected utility theory over menus of two-payoff lotteries implied by the predictive algorithm $\widehat{f}^*$.}
    \floatfoot{\textit{Notes}: 
    This table summarizes the anomalies for expected utility theory over menus of two-payoff lotteries generated by our procedures applied to a neural network $\widehat{f}^*$ constructed on the Choices13K dataset \citep[][]{PetersonEtAl(21)-MLforChoiceTheory}.  
    See Section \ref{section: choices13k, two payoff anomalies} for discussion.}
    \label{tab: anomaly categorization for binary payoffs, choices 13K}
\end{table}

First, our procedures generate many first order stochastic dominance violations implied by the predictive algorithm. 
This is unsurprising; first order stochastic dominance violations are common in Choices13K and occur on 14\% of all lottery menus \citep[][]{PetersonEtAl(21)-MLforChoiceTheory}.
Our procedures appear to generate the frequency of first order stochastic dominance violations in the dataset.
Second, all other generated anomalies illustrate choice reversals across menus of two-payoff lotteries.
As discussed in Section \ref{section: anomalies, categorization, binary payoffs}, the dominated consequence effect is a generalization of the Common ratio effect. 
They additionally generate the new categories (i.e., the reverse dominated consequence effect and the strict dominance effect).
Using a publicly available dataset on real lottery choices, our algorithms successfully generate known anomalies and novel anomalies for expected utility theory over two-payoff lotteries.

\subsection{Anomalies for expected utility theory over three-payoff lotteries}\label{section: choices13k, three payoff anomalies}

Our procedures generated 1,996 anomalies for expected utility theory over menus of three-payoff lotteries implied by the predictive algorithm $\widehat{f}^*$.
We illustrate two strategies for interpreting these generated anomalies: first, a theory-driven numeric categorization as before; and second, an empirical clustering based on features of the anomalies.

\vspace{-1em}
\paragraph{Theory-driven categorization of anomalies:} We first categorize the anomalies over menus of three-payoff lotteries based on the violation of expected utility theory they highlight. 
It is simple to check numerically whether an anomaly in menus of three-payoff lotteries constitutes a first-order stochastic dominance (FOSD) violation.
We find that 1,640 of our generated anomalies are first order stochastic dominance violations. 
Given the frequency of first order stochastic dominance violations in Choices13K mentioned earlier, it is unsurprising that our procedures automatically generate a large number of them.

Another category that is simple to check numerically is whether any generated pair of menus can be represented as compound lotteries over the same underlying simpler, two-payoff lotteries. 
A predicted choice reversal across such a pair would indicate a possible violation of the independence axiom. 
Table \ref{tab: generated anomaly, ternary lotteries, choices 13k example} provides an example of algorithmically generated menus in this category.
The predictive algorithm $\widehat{f}$ predicts individuals would be likely to select lottery A1.
This could be plausibly driven by risk aversion, since lottery A1 appears to be the ``safer'' option relative to lottery A0. 
Yet the predictive algorithm $\widehat{f}$ predicts that individuals would also be likely to select lottery B0, which appears to be the riskier (higher variance) lottery in Menu B.

\begin{table}[htbp!]
\caption{Algorithmically generated anomaly for expected utility theory over lotteries on three monetary payoffs implied by the predictive algorithm $\widehat{f}(\cdot)$.}
\begin{subtable}{0.45\linewidth}
\begin{tabular}{c c c c}
 & \multicolumn{2}{l}{Menu A $(x_A, y_A^*)$}\\
 \multicolumn{1}{c|}{Lottery 0} & \$4.30 & \$6.17 & \$8.51 \\ 
 \multicolumn{1}{c|}{} & 15\% & 61\% & 24\% \\
 \hline
 \multicolumn{1}{c|}{\cellcolor{green!25} Lottery 1} & \$4.63 \\
 \multicolumn{1}{c|}{} & 100\% \\
 \hline
 \end{tabular}
\end{subtable} %
\begin{subtable}{0.45\linewidth}
\begin{tabular}{c c c c}
 & \multicolumn{2}{l}{Menu B $(x_B, y_B^*)$} \\
 \multicolumn{1}{c|}{\cellcolor{green!25} Lottery 0} & \$4.30 & \$6.17 & \$8.51 \\ 
 \multicolumn{1}{c|}{} & 36\% & 36\% & 28\% \\
 \hline
 \multicolumn{1}{c|}{Lottery 1} & \$4.63 & \$5.04 & \$5.81 \\
 \multicolumn{1}{c|}{} & 30\% & 67\% & 3\% \\
 \hline
 \end{tabular}
\end{subtable}
\floatfoot{\textit{Notes}: 
In each menu, we color the lottery that is predicted to be selected with probability at least 0.50 in green.
This anomaly was produced by the example morphing algorithm applied to a neural network $\widehat{f}$ constructed on the Choices13K dataset \citep[][]{PetersonEtAl(21)-MLforChoiceTheory}.
We round each payoff to the nearest cent and each probability to the nearest percentage.
See Section \ref{section: choices13k, three payoff anomalies} for discussion.}
\label{tab: generated anomaly, ternary lotteries, choices 13k example}
\end{table}

The lotteries in menu A and menu B can be re-written as compound lotteries over the same two-payoff lotteries. 
Figure \ref{figure: illustration of choices 13k ternary example as compound} illustrates the decomposition, where lotteries A0 and B0 are compound lotteries over the same two-payoff lotteries (with mixing probabilities $\alpha_{A0} > \alpha_{B0}$), and similarly lotteries A1 and B1 are compound lotteries over the same two-payoff lotteries (with mixing probabilities $\alpha_{A1} < \alpha_{B1}$). 
The oddity is now clear: Lottery B1 raises the mixing probability on the dominating two-payoff lottery relative to lottery A1, yet lottery B0 raises the mixing probability on the two-payoff lottery with a lower expected value relative to lottery A0. 
The change in the mixing probabilities from menu A to menu B would appear to have made lottery 1 better and lottery 0 worse, yet choices flip.

\begin{figure}[htbp!]
\begin{minipage}{0.4\textwidth}
\centering
\begin{tikzpicture}[
    level 1/.style={sibling distance=3cm},
    level 2/.style={sibling distance=2cm}, 
    edge from parent/.style={draw, ->, >=stealth, midway, auto, semithick}
]

\node {\textbf{Lottery 0}}
    child{node {$\bullet$} 
        child {node {\$6.17} edge from parent node[left] {79\%}}
        child {node {\$8.51} edge from parent node[right] {21\%}}
        edge from parent node[left] {$\alpha_{A0} > \alpha_{B0}$ $\mbox{ }$ }
    }
    child {node {$\bullet$}
        child {node {\$4.30} edge from parent node[left] {66\%}}
        child {node {\$8.51} edge from parent node[right] {34\%}}
        edge from parent node[right] {}
    };

\end{tikzpicture}
\end{minipage}
\begin{minipage}{0.4\textwidth}
\centering
\begin{tikzpicture}[
    level 1/.style={sibling distance=3cm},
    level 2/.style={sibling distance=2cm}, 
    edge from parent/.style={draw, ->, >=stealth, midway, auto, semithick}
]

\node {\textbf{Lottery 1}}
    child{node {$\bullet$} 
        child {node {\$5.04} edge from parent node[left] {96\%}}
        child {node {\$5.81} edge from parent node[right] {4\%}}
        edge from parent node[left] {$\alpha_{A1} < \alpha_{B1}$ $\mbox{ }$ }
    }
    child {node {$\bullet$}
        child {node {\$4.63} edge from parent node[right] {100\%}}
        edge from parent node[right] {}
    };
\end{tikzpicture}
\end{minipage}
\caption{Decomposition of the algorithmically generated anomaly in Table \ref{tab: generated anomaly, ternary lotteries, choices 13k example} into compound lotteries.}
\floatfoot{\textit{Notes}: We depict each compound lottery in extensive form, where the mixing probability $\alpha$ gives the probability of receiving the simple two-payoff lottery on the left hand side. 
The mixing probabilities are $\alpha_{A0} = 0.45, \alpha_{B0} = 0.77$ and $\alpha_{A1} = 0$, $\alpha_{B1} = 0.70$. 
We round each payoff to the nearest cent and each probability to the nearest percentage.
See Section \ref{section: choices13k, three payoff anomalies} for discussion.}
\label{figure: illustration of choices 13k ternary example as compound}
\end{figure}

The structure of this anomaly is also a more general pattern that we can search for.
We can check whether the lotteries in menu A and menu B can be rewritten as compound lotteries over the same two-payoff lotteries, where one of those two-payoff lotteries strictly dominates another.
Across the menus, we can then examine whether the predicted choice changes in the opposite direction relative to the change in the mixing probability placed on the dominating lottery (see Appendix \ref{section: categorization of three payoff anomalies} for discussion).
We generate 132 anomalies with this structure (Appendix Table \ref{tab: generated anomalies, additional examples ternary lotteries, choices 13k} provides additional examples).

\vspace{-1em}
\paragraph{Empirical clustering of anomalies:}
In addition to using our knowledge of expected utility for categorization, we further provide an empirical organization of our algorithmically generated anomalies. 
We focus on the 356 anomalies that do not violate first-order stochastic dominance. 
For each menu of two lotteries, we construct a feature vector that summarizes the comparison between the lottery predicted to be chosen and its alternative. 
The features include, for example, differences in expected value, payoff variance, payoff range, and probability range. 
We then cluster the anomalies using $K$-means, and we summarize the feature space by calculating principal components. 
Further details are provided in Appendix \ref{section: categorization of three payoff anomalies}.

\begin{figure}[htbp]
    \centering
    \begin{subfigure}[t]{0.48\textwidth}
        \centering
        \includegraphics[width=\linewidth]{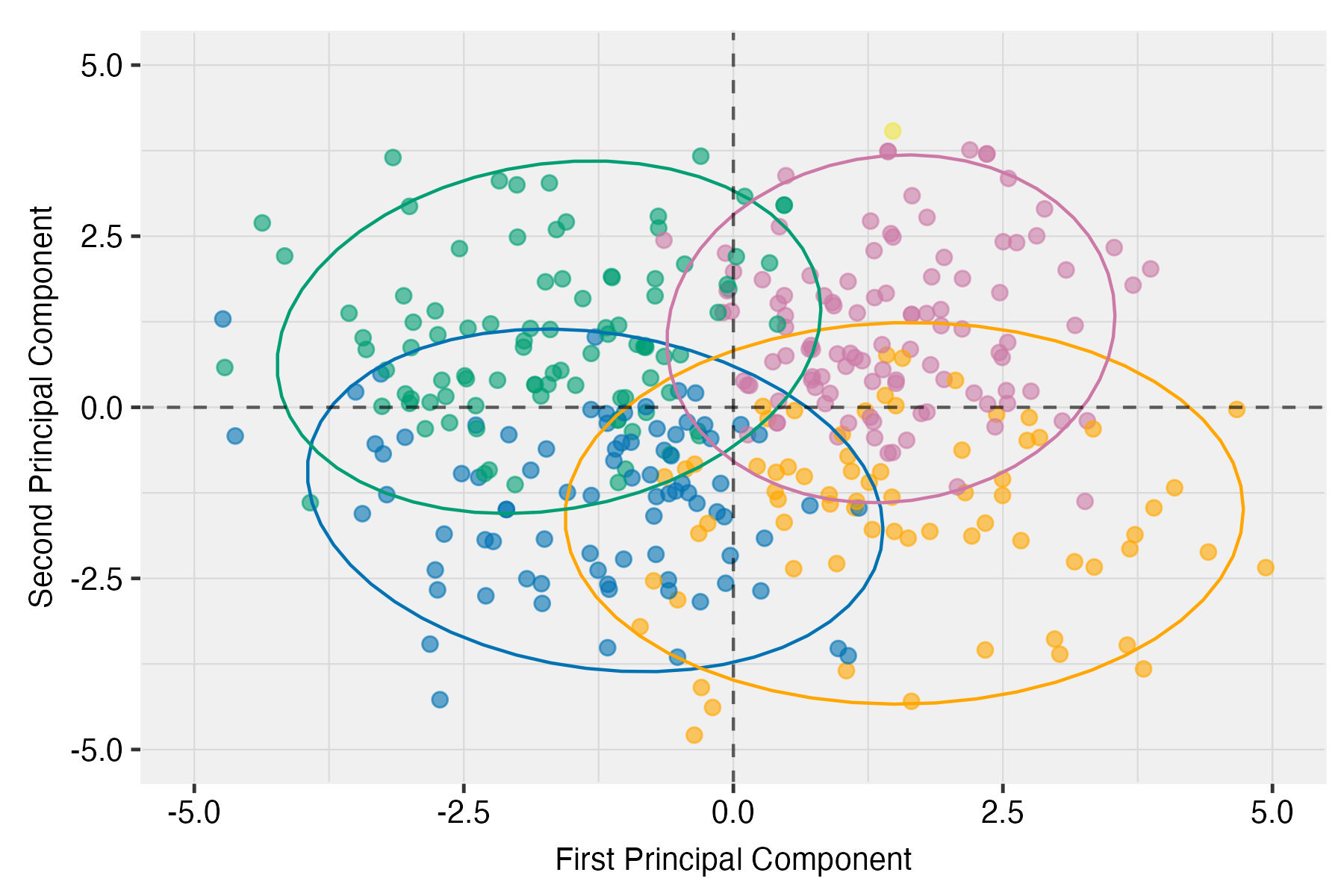}
        \caption{Clusters of anomalies in the space of the first two principal components.}
        \label{fig:cluster_visualization}
    \end{subfigure}
    \hfill
    \begin{subfigure}[t]{0.48\textwidth}
        \centering
        \includegraphics[width=\linewidth]{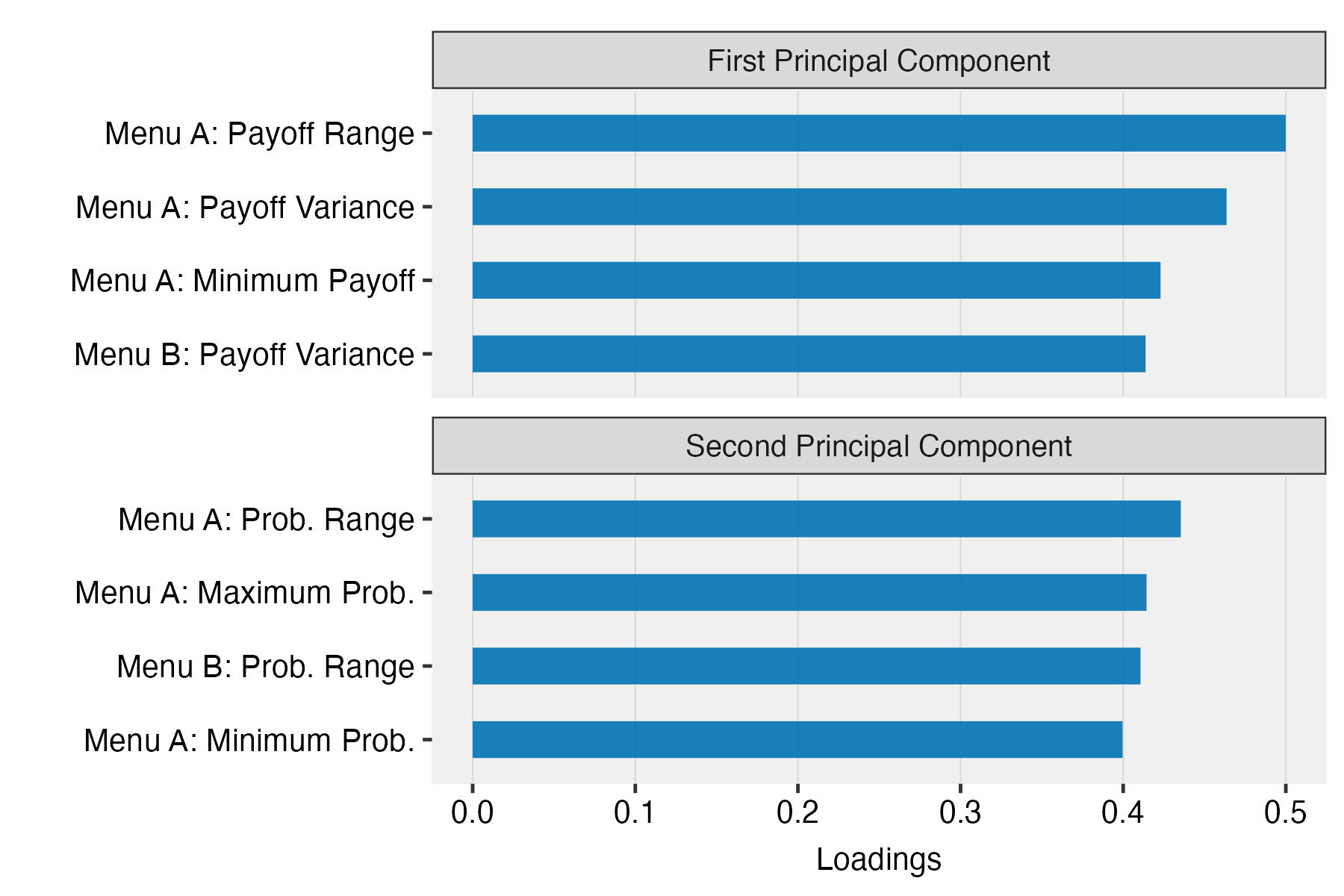}
        \caption{Top loadings on the first two principal components.}
        \label{fig:cluster_loadings}
    \end{subfigure}
    \caption{Empirical clustering of algorithmically generated anomalies over lotteries on three monetary payoffs implied by the predictive algorithm $\widehat{f}$.}
    \floatfoot{\textit{Notes}: We focus on the $356$ generated anomalies that are not first order stochastic dominance violations, and we construct clusters using $K$-means on features calculated for each menu.
    Panel (a) visualizes each anomaly in the space of the first two principal components of the anomaly features and overlays the cluster groups. 
    Panel (b) lists the features with the highest-magnitude loadings on the first two principal components. See Section~\ref{section: choices13k, three payoff anomalies} for discussion.}
    \label{fig:anomaly_clustering_results}
\end{figure}

Figure \ref{fig:cluster_visualization} visualizes each anomaly in the space spanned by the first two principal components and overlays the cluster groups. 
The clusters occupy the four quadrants with modest overlap, providing a coarse partition of the anomalies. 
To better interpret these axes, Figure \ref{fig:cluster_loadings} reports the largest loadings for each component by magnitude.
The first principal component loads heavily on the difference in payoff dispersion between lotteries --- especially the payoff variance. 
Variation along the first principal component suggests these are anomalies that are engineered to manipulate classic risk considerations.
The second principal component loads on the probability compositions of the lotteries -- particularly differences in the range and extreme probabilities across the menus. 
Variation along the second principal component suggests these are anomalies that manipulate the salience of probabilities that are not well captured by the expected value and variance of the payoffs.

\subsection{Experimental test of algorithmically generated anomalies}\label{subsection: experimental test of algorithmically generated anomalies}

When applied to a predictive algorithm, our procedures generate novel anomalies for expected utility theory over menus of two-payoff lotteries and three-payoff lotteries.
But do individuals violate expected utility theory as predicted?
Answering this question is where our procedures end, and the careful experimental work needed for the empirical testing of anomalies begins. 

We randomly selected 30 algorithmically generated anomalies for expected utility theory summarized in Table \ref{tab: anomaly categorization for binary payoffs, choices 13K} over lotteries with two monetary payoffs.
These anomalies were sampled to span the categories (i.e., the dominated consequence effect, the reverse dominated consequence effect, and the strict dominance effect).
We also randomly selected 30 algorithmically generated anomalies over lotteries with three-payoffs sharing the particular structure we highlighted in Section \ref{section: choices13k, three payoff anomalies}.
We recruited respondents on Prolific to make incentivized choices on these anomalies in surveys. 
See Appendix \ref{section: appendix, description of survey design} for the design of these online surveys.
Altogether, we recruited 258 and 266 respondents on our two surveys of anomalies over lotteries with two monetary payoffs, and 260 and 263 respondents on our two surveys of anomalies over three monetary payoffs.

\begin{figure}[htbp]
    \centering
    \begin{subfigure}[t]{0.48\textwidth}
        \centering
        \includegraphics[width=\linewidth]{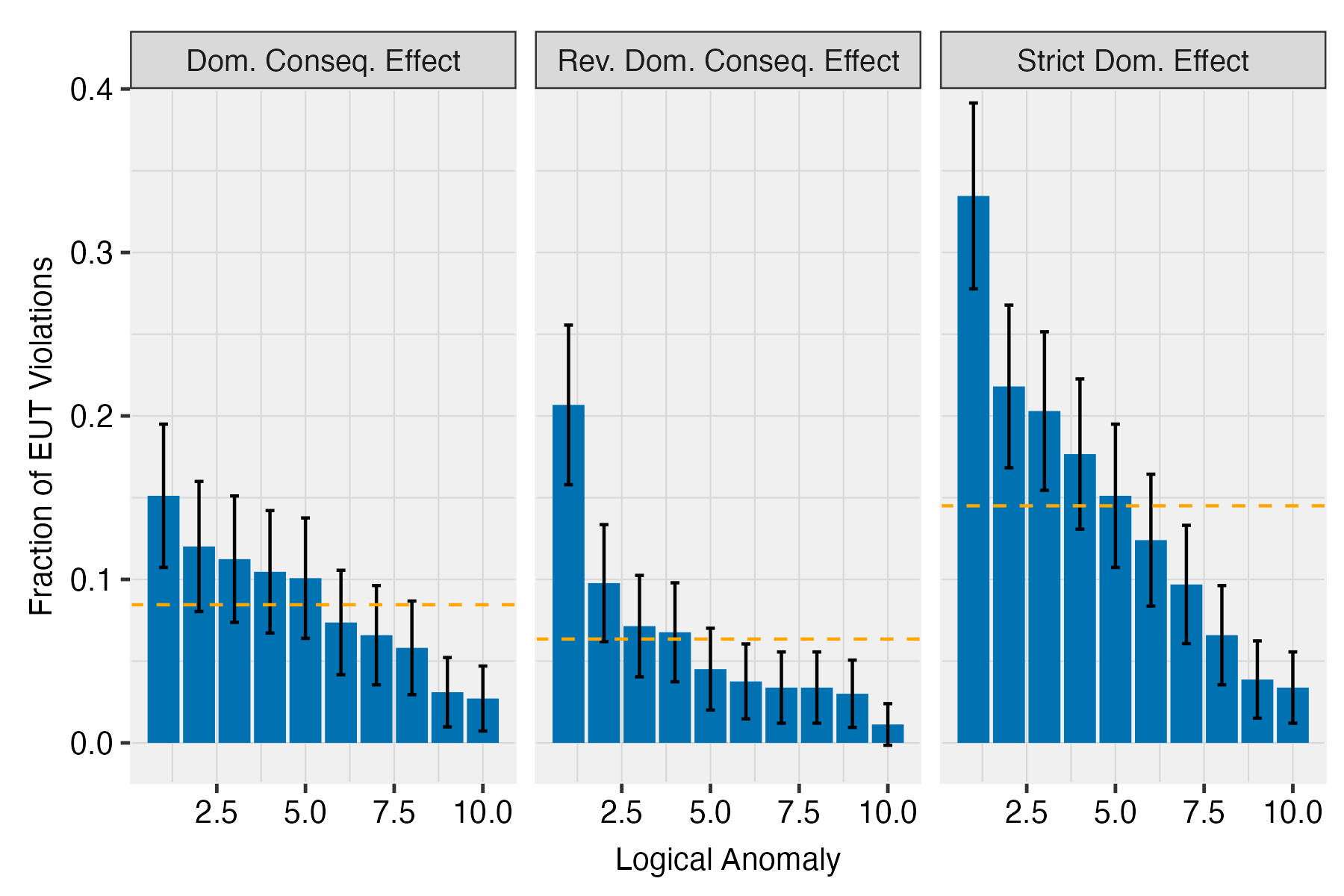}
        \caption{Two-payoff lotteries.}
        \label{fig:anomaly-fraction-binary}
    \end{subfigure}
    \hfill
    \begin{subfigure}[t]{0.48\textwidth}
        \centering
        \includegraphics[width=\linewidth]{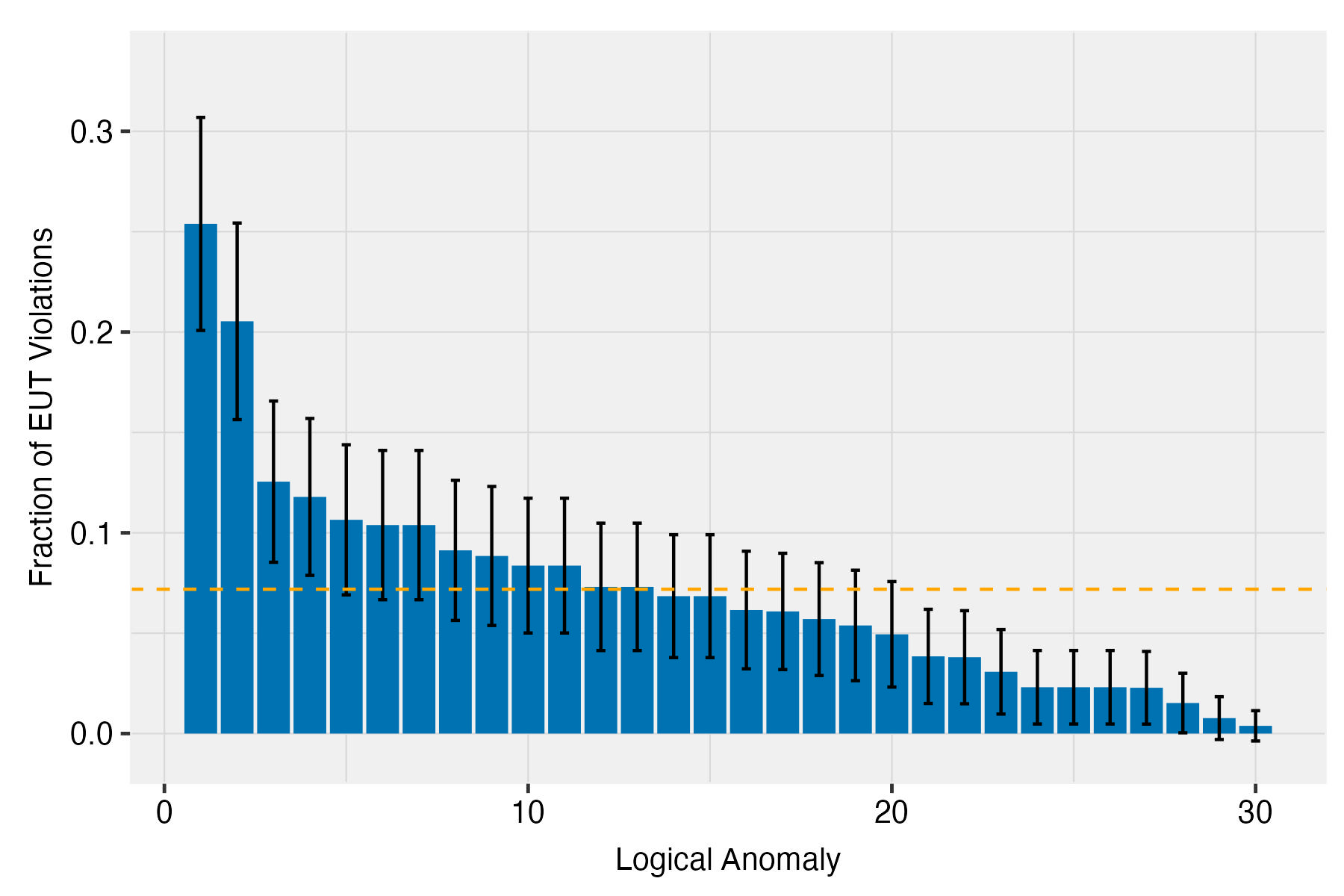}
        \caption{Three-payoff lotteries.}
        \label{fig:anomaly-fraction-ternary}
    \end{subfigure}
    \caption{Fraction of respondents whose choices violate expected utility on algorithmically generated anomalies.}
    \floatfoot{\textit{Notes}: Blue bars plot the fraction of respondents who violate expected utility on each algorithmically generated anomaly; black error bars are 95\% confidence intervals (standard errors clustered at the respondent level). The orange dashed line reports pooled violation rates: in panel~(a), pooling within the same category; in panel~(b), pooling across all anomalies. Within each panel, anomalies are sorted in decreasing order of violation rates and assigned numeric identifiers after sorting. See Section~\ref{subsection: experimental test of algorithmically generated anomalies} for discussion.}
    \label{fig:anomaly-fraction-bycat}
\end{figure}

We find strong evidence that the pooled respondents' choices are inconsistent with expected utility theory across our algorithmically generated anomalies.
Figure \ref{fig:anomaly-fraction-binary} and Figure \ref{fig:anomaly-fraction-ternary}
reports the fraction of respondents whose choices violate expected utility theory on our algorithmically generated  anomalies over two-payoff lotteries and three-payoff lotteries respectively.
Appendix Table \ref{tab: summary statistics, anomalous fraction, by anomaly category, binary and ternary payoffs} provides summary statistics on the expected utility theory violation rates, pooling across our algorithmically generated anomalies.
On our anomalies over two-payoff lotteries, the pooled expected utility theory violation rate is 11.4\% (p-value $<0.001$) on dominated consequence effect anomalies, 8.5\% (p-value $<0.001$) on reverse dominated consequence effect anomalies, and 12.7\% (p-value $<0.001$) on strict dominance effect anomalies.
On our anomalies over three-payoff lotteries, the pooled expected utility theory violation rate is 7.2\% (p-value $<0.001$).

These pooled estimates mask heterogeneity in the fraction of respondents violating expected utility theory across anomalies. 
More than 15\% of respondents' choices violate expected utility theory on several strict dominance effect anomalies over two-payoff lotteries as well as several of our anomalies over three-payoff lotteries.
Applying a conservative Bonferroni correction for multiple hypotheses across all anomalies over two-payoff lotteries in our surveys, the expected utility theory violation rate is statistically different than zero at the 5\% level for 26 out of 30.
On our anomalies over three monetary payoffs, the expected utility theory violation rate is statistically different than zero at the 5\% level for 22 out of 30 (after Bonferroni correction).

We compare our generated anomalies against known anomalies in the behavioral economics literature. 
Several recent papers provide meta-analyses of past experiments and conduct comprehensive experimental designs to systematically evaluate the Allais paradox, the common ratio effect, and the certainty effect.
\cite{BlavatskyyEtAl(22)-RobustnessOfAllais} find that 16\% of respondents' choices demonstrate the Allais paradox (``fanning out'' choices), and the median experiment with real financial incentives finds that 13.7\% of respondents' choices do so. 
\cite{McGranahanEtAl(23)-CommonRatio} find that 15.6\% of respondents' choices demonstrate the common ratio effect and 12.9\% demonstrate the reverse common ratio effect in experiments conducted on Prolific with real financial incentives.
\cite{JainNielsen(23)} find that 8.3\% of subjects display the certainty effect with financial incentives.
In this light, our algorithmically generated anomalies yield expected utility theory violation rates in line with known anomalies like the Allais paradox and the common ratio effect.

\section{Conclusion}\label{section: conclusion}

We develop procedures to generate anomalies for an existing theory from predictive algorithms. 
Our procedures contrast the theory with the predictive model, searching for minimal examples on which the theory cannot explain the black box's predictions.
Because supervised learners often uncover signals researchers miss, these anomalies are natural targets for new data collection and formal misspecification tests.
A two-step workflow recurs throughout the development of economic theory --- researchers first generate anomalies and then collect data to rigorously test them. 
Our framework hopefully illustrates that this first step of anomaly generation can be modeled, and as a result we can build procedures to accelerate this process.

We expect anomaly generation to be valuable when two conditions hold: there is ample data to form predictions about an observed outcome; and the theory is a tool for prediction about that same outcome.
Our illustration to choice under risk meets both these conditions: there exists large datasets of choices between risky lotteries, and expected utility theory makes sharp predictions about what choices are possible. 
Other important economic domains fit these conditions as well, such as discrete choice models, choice under ambiguity, and strategic interactions.
When these two conditions are met, it is natural to ask: what is implied by predictive algorithms that would conflict with the theory? 
Anomaly generation procedures are suited to help answer that question. 

There are many promising avenues for further work. We placed no constraints on the feature space over which our procedures optimized; adding restrictions (e.g., only search over round numbers) could yield more interpretable anomalies. 
As mentioned, an active literature in adversarial learning offers many gradient-based methods \citep[e.g.,][]{MokhtariEtAl(20), RazaviyaynEtAl(20)}, and specific techniques from interpretability and explainability may be adapted for anomaly generation \citep[][]{Molnar(25)}. 
Our evaluation metrics provide a common yardstick for future comparisons.
Finally, our empirical testing collects new data on algorithmically generated anomalies; this mirrors holdout evaluations in supervised machine learning. 
While feasible in many settings, it is worth exploring whether uncertainty quantification for the predictive algorithm $\widehat{f}^*$ can be directly translated into tests of the generated anomalies. 
More broadly, it is our hope to spur interest in this overlooked step in theory development.

\singlespacing
\bibliographystyle{aea}
\bibliography{Bibliography}

\clearpage
\newpage

\setcounter{page}{1}
\pagenumbering{arabic} 
\renewcommand{\thepage}{S-\arabic{page}}

\appendix
\singlespacing

\begin{center}
\vspace{-1.8cm}{\Large \textbf{From Predictive Algorithms to \\ Automatic Generation of Anomalies}} 
\bigskip \\
{\Large \textit{Online Appendix}}
\bigskip \\
\large Sendhil Mullainathan \& Ashesh Rambachan \medskip \\
\bigskip
\end{center}

\section{Omitted proofs}\label{section: proofs of main results}

\subsection{Proof of Proposition \ref{prop: plug in vs. pop max-min program}}

As a first step, we establish that the $\widehat{\cE}$ approximately solves the plug-in max-min optimization program up to the optimization errors associated with the approximate inner minimization and outer maximization routines.

\begin{lemma}\label{lemma: plug-in solution approx solves plug-in program}
Under the same conditions as Proposition \ref{prop: plug in vs. pop max-min program}, 
$$
    \left\| \widehat{\cE} - \max_{x_{1:n}} \min_{f(\cdot) \in \cF^{T}} n^{-1} \sum_{i=1}^{n} \ell\left( f(x_i), \widehat{f}^*(x_i) \right) \right\| \leq \delta + \nu.
$$
\begin{proof}
As notation, let $\widehat{f}^T(\cdot; x_{1:n})$ denote the optimal solution to $\min_{f(\cdot) \in \cF^{T}} n^{-1} \sum_{i=1}^{n} \ell\left( f(x_i), \widehat{f}^*(x_i) \right)$. 
Observe that 
$$
\left\| n^{-1} \sum_{i=1}^{n} \ell\left( \widetilde{f}(\widetilde{x}_i; \widetilde{x}_{1:n}), \widehat{f}^*(\widetilde{x}_i) \right) -  \max_{x_{1:n}} \min_{f(\cdot) \in \cF^{T}} n^{-1} \sum_{i=1}^{n} \ell\left( f(x_i), \widehat{f}^*(x_i) \right) \right\| \overset{(1)}{\leq}
$$
$$
\left\|  n^{-1} \sum_{i=1}^{n} \ell\left( \widetilde{f}(\widetilde{x}_i; \widetilde{x}_{1:n}), \widehat{f}^*(\widetilde{x}_i) \right) - \max_{x_{1:n}} n^{-1} \sum_{i=1}^{n} \ell\left( \widehat{f}^T(\cdot; x_{1:n}), \widehat{f}^*(x_i) \right) \right\| + 
$$
$$
\left\| \max_{x_{1:n}} n^{-1} \sum_{i=1}^{n} \ell\left( \widehat{f}^T(\cdot; x_{1:n}), \widehat{f}^*(x_i) \right) - \max_{x_{1:n}} \min_{f(\cdot) \in \cF^{T}} n^{-1} \sum_{i=1}^{n} \ell\left( f(x_i), \widehat{f}^*(x_i) \right) \right\| \overset{(2)}{\leq}
$$
$$
\nu + \left\| \max_{x_{1:n}} n^{-1} \sum_{i=1}^{n} \ell\left( \widehat{f}^T(\cdot; x_{1:n}), \widehat{f}^*(x_i) \right) - \max_{x_{1:n}} \min_{f(\cdot) \in \cF^{T}} n^{-1} \sum_{i=1}^{n} \ell\left( f(x_i), \widehat{f}^*(x_i) \right) \right\| \overset{(3)}{\leq}
$$
$$
\nu + \left\| \max_{x_{1:n}} \left\{ n^{-1} \sum_{i=1}^{n} \ell\left( \widehat{f}^T(\cdot; x_{1:n}), \widehat{f}^*(x_i) \right) - \min_{f(\cdot) \in \cF^{T}} n^{-1} \sum_{i=1}^{n} \ell\left( f(x_i), \widehat{f}^*(x_i) \right) \right\} \right\| \overset{(4)}{\leq} \nu + \delta
$$
where (1) adds/subtracts $\max_{x_{1:n}} \min_{f(\cdot) \in \cF^{T}} n^{-1} \sum_{i=1}^{n} \ell\left( f(x_i), \widehat{f}^*(x_i) \right)$ and applies the triangle inequality,  (2) follows from properties of the approximate outer maximization routine, (3) uses sub-additivity of the maximum, and (4) follows from the properties of the approximate inner minimization routine. 
\end{proof}
\end{lemma}

To analyze the convergence of the plug-in estimator, observe that 
$$
\left\| \widehat{\cE} - \cE \right\| \leq \left\| \widehat{\cE} - \max_{x_{1:n}} \min_{f(\cdot) \in \cF^{T}} n^{-1} \sum_{i=1}^{n} \ell\left( f(x_i), \widehat{f}^*(x_i) \right) \right\| + 
\left\| \max_{x_{1:n}} \min_{f(\cdot) \in \cF^{T}} n^{-1} \sum_{i=1}^{n} \ell\left( f(x_i), \widehat{f}^*(x_i) \right) - \cE \right\|.
$$
Lemma \ref{lemma: plug-in solution approx solves plug-in program} establishes that the first term is bounded by $\nu + \delta$. Therefore, we only need to establish a bound on the second term. 
Towards this, we rewrite the second term as 
$$
\left\| \max_{x_{1:n}} \min_{f(\cdot) \in \cF^{T}} n^{-1} \sum_{i=1}^{n} \ell\left( f(x_i), \widehat{f}^*(x_i) \right) - \cE \right\| \leq
$$
$$
\left\| \max_{x_{1:n}} \left\{ \min_{f(\cdot) \in \cF^{T}} n^{-1} \sum_{i=1}^{n} \ell\left( f(x_i), \widehat{f}^*(x_i) \right) - \min_{f(\cdot) \in \cF^{T}} n^{-1} \sum_{i=1}^{n} \ell\left( f(x_i), f^*(x_i) \right) \right\} \right\|.
$$
Defining $\widehat{f}^T(\cdot; x_{1:n})$ to be the minimizer for $\min_{f(\cdot) \in \cF^{T}} n^{-1} \sum_{i=1}^{n} \ell\left( f(x_i), \widehat{f}^*(x_i) \right)$ and $f^T(\cdot; x_{1:n})$ as the minimizer for $\min_{f(\cdot) \in \cF^{T}} n^{-1} \sum_{i=1}^{n} \ell\left( f(x_i), f^*(x_i) \right)$, we rewrite 
$$
\min_{f(\cdot) \in \cF^{T}} n^{-1} \sum_{i=1}^{n} \ell\left( f(x_i), \widehat{f}^*(x_i) \right) - \min_{f(\cdot) \in \cF^{T}} n^{-1} \sum_{i=1}^{n} \ell\left( f(x_i), f^*(x_i) \right) = 
$$
$$
n^{-1} \sum_{i=1}^{n} \ell\left( \widehat{f}^T(x_i; x_{1:n}), \widehat{f}^*(x_i) \right) - n^{-1} \sum_{i=1}^{n} \ell\left( f^T(x_i; x_{1:n}), f^*(x_i) \right) =
$$
$$
\underbrace{n^{-1} \sum_{i=1}^{n} \left\{ \ell\left( \widehat{f}^T(x_i; x_{1:n}), \widehat{f}^*(x_i) \right) - \ell\left( \widehat{f}^T(x_i; x_{1:n}), f^*(x_i) \right) \right\}}_{(a)} + 
$$
$$
\underbrace{n^{-1} \sum_{i=1}^{n} \left\{ \ell\left( \widehat{f}^T(x_i; x_{1:n}), f^*(x_i) \right) - \ell\left( f^T(x_i; x_{1:n}), f^*(x_i) \right) \right\}}_{(b)}.
$$
Consider (a). Since $\ell(\cdot, \cdot)$ is convex in its second argument, (a) is bounded above by
$$
n^{-1} \sum_{i=1}^{n} \left\{ \nabla_{2} \ell\left( \widehat{f}^T(x_i; x_{1:n}), \widehat{f}^*(x_i) \right) \left( \widehat{f}^*(x_i) - f^*(x_i) \right) \right\} \leq
$$
$$
n^{-1} K \| \widehat{f}^*(x_{1:n}) - f^*(x_{1:n}) \|_{1} \leq K \| \widehat{f}^*(x_{1:n}) - f^*(x_{1:n}) \|_{\infty}
$$
where we defined the shorthand notation $f(x_{1:n}) = \left( f(x_1), \hdots, f(x_n) \right)$, used that the loss function has bounded gradients, and the inequality $\|f(x_{1:n})\|_{1} \leq n \| f(x_{1:n}) \|_{\infty}$.
Next, we can rewrite (b) as being bounded by
$$
n^{-1} \sum_{i=1}^{n} \left\{ \ell\left( \widehat{f}^T(x_i; x_{1:n}), f^*(x_i) \right) - \ell\left( f^T(x_i; x_{1:n}), f^*(x_i) \right) \right\} = 
$$
$$
n^{-1} \sum_{i=1}^{n} \left\{ \ell\left( \widehat{f}^T(x_i; x_{1:n}), f^*(x_i) \right) - \ell\left( f^T(x_i; x_{1:n}), f^*(x_i) \right) \right\} - 
$$
$$
n^{-1} \sum_{i=1}^{n} \left\{ \ell\left( f^T(x_i; x_{1:n}), f^*(x_i) \right) - \ell\left( f^T(x_i; x_{1:n}), \widehat{f}^*(x_i) \right) \right\} \overset{(1)}{\leq}
$$
$$
n^{-1} \sum_{i=1}^{n} \left\{ \ell\left( \widehat{f}^{T}(x_i; x_{1:n}), f^*(x_i) \right) -  \ell\left( f^T(x_i; x_{1:n}), \widehat{f}^*(x_i) \right) \right\} -
$$
$$
n^{-1} \sum_{i=1}^{n} \left\{ \nabla_{2} \ell\left( f^T(x_i; x_{1:n}), \widehat{f}^*(x_i) \right) (f^*(x_i) - \widehat{f}^*(x_i)) \right\}  \overset{(2)}{\leq}
$$
$$
n^{-1} \sum_{i=1}^{n} \left\{ \ell\left( \widehat{f}^{T}(x_i; x_{1:n}), f^*(x_i) \right) -  \ell\left( \widehat{f}^T(x_i; x_{1:n}), \widehat{f}^*(x_i) \right) \right\} -
$$
$$
n^{-1} \sum_{i=1}^{n} \left\{ \nabla_{2} \ell\left( f^T(x_i; x_{1:n}), \widehat{f}^*(x_i) \right) (f^*(x_i) - \widehat{f}^*(x_i)) \right\} \overset{(3)}{\leq}
$$
$$
n^{-1} \sum_{i=1}^{n} \left\{ \nabla_2 \ell(\widehat{f}^T(x_i; x_{1:n}), \widehat{f}^*(x_i)) \left( \widehat{f}^*(x_i) - f^*(x_i) \right) \right\} -
$$
$$
n^{-1} \sum_{i=1}^{n} \left\{ \nabla_{2} \ell\left( f^T(x_i; x_{1:n}), \widehat{f}^*(x_i) \right) (f^*(x_i) - \widehat{f}^*(x_i))  \right\}
$$
where (1) uses that the loss is convex in its second argument, (2) uses $n^{-1} \sum_{i=1}^{n} \ell(f^T(x_i; x_{1:n}), \widehat{f}^*(x_i)) \geq n^{-1} \sum_{i=1}^{n} \ell(\widehat{f}^T(x_i; x_{1:n}), \widehat{f}^*(x_i))$, and (3) again uses that the loss is convex in it second argument.
By the same argument as before, it follows that this is bounded by 
$2 K \left\| \widehat{f}^*(x_{1:n}) - f^*(x_{1:n}) \right\|_{\infty}$.
Combining the bound on (a), (b) yields the desired result. $\Box$

\subsection{Proof of Proposition \ref{prop: descent direction for morphing}}

To prove the first result, let us define the shorthand notation $g^* = \nabla f^*(x)$, $g = \mbox{Proj}\left( \nabla f^*(x) \mid \cN(x) \right)$, and $g^{\perp} = g^* - g$. Observe that 
$$
\langle -\mbox{Proj}\left( \nabla f^*(x) \mid \cN(x) \right), \nabla f^*(x) \rangle = \langle -g, g^* \rangle = \langle -g, g^\perp + g \rangle = -\|g\|^2 \leq 0,
$$
and so $-\mbox{Proj}\left( \nabla f^*(x) \mid \cN(x) \right)$ is a descent direction for $f^*(\cdot)$. 

To prove the second result, let $\Omega$ to be the orthogonal projection matrix onto $\cN(x)$ and define $\widehat{g}^* = \nabla \widehat{f}^*(x)$, $\widehat{g} = \mbox{Proj}\left( \nabla \widehat{f}^*(x) \mid \cN(x) \right)$ and $\widehat{g}^{\perp} = \widehat{g}^* - \widehat{g}$. 
Observe that 
$$
\langle -\mbox{Proj}\left( \nabla \widehat{f}^*(x) \mid \cN(x)  \right), \nabla f^*(x) \rangle = \langle -\widehat{g}, g^* \rangle = \langle -\widehat{g}, g + g^\perp \rangle = \langle -\widehat{g}, g \rangle =
$$
$$
\langle -\widehat{g} + g - g, g \rangle = -\|g\|^2 + \langle g - \widehat{g}, g \rangle \leq -\|g\|^2 + \|g - \widehat{g} \| \|g \|,
$$
where the last inequality follows by the Cauchy-Schwarz inequality. The stated condition implies that 
$$
\|g - \widehat{g}\| \leq \|g\|
$$
since $\|g - \widehat{g}\| = \| \Omega (g^* - \widehat{g}^*)\| \leq \|\Omega\|_{op} \| g^* - \widehat{g}^* \|$ and $\|\Omega\|_{op} \leq 1$. But the previous display can be equivalently rewritten as 
$$
-\|g\|^2 + \|g - \widehat{g}\| \|g\| \leq 0
$$
thus proving the result. $\Box$

\section{Appendix figures and tables}\label{section: appendix tables}
\renewcommand{\thefigure}{A\arabic{figure}}
\renewcommand{\thetable}{A\arabic{table}}
\setcounter{figure}{0}
\setcounter{table}{0}
\begin{figure}[h!]
    \centering
    \includegraphics[width=0.6\textwidth]{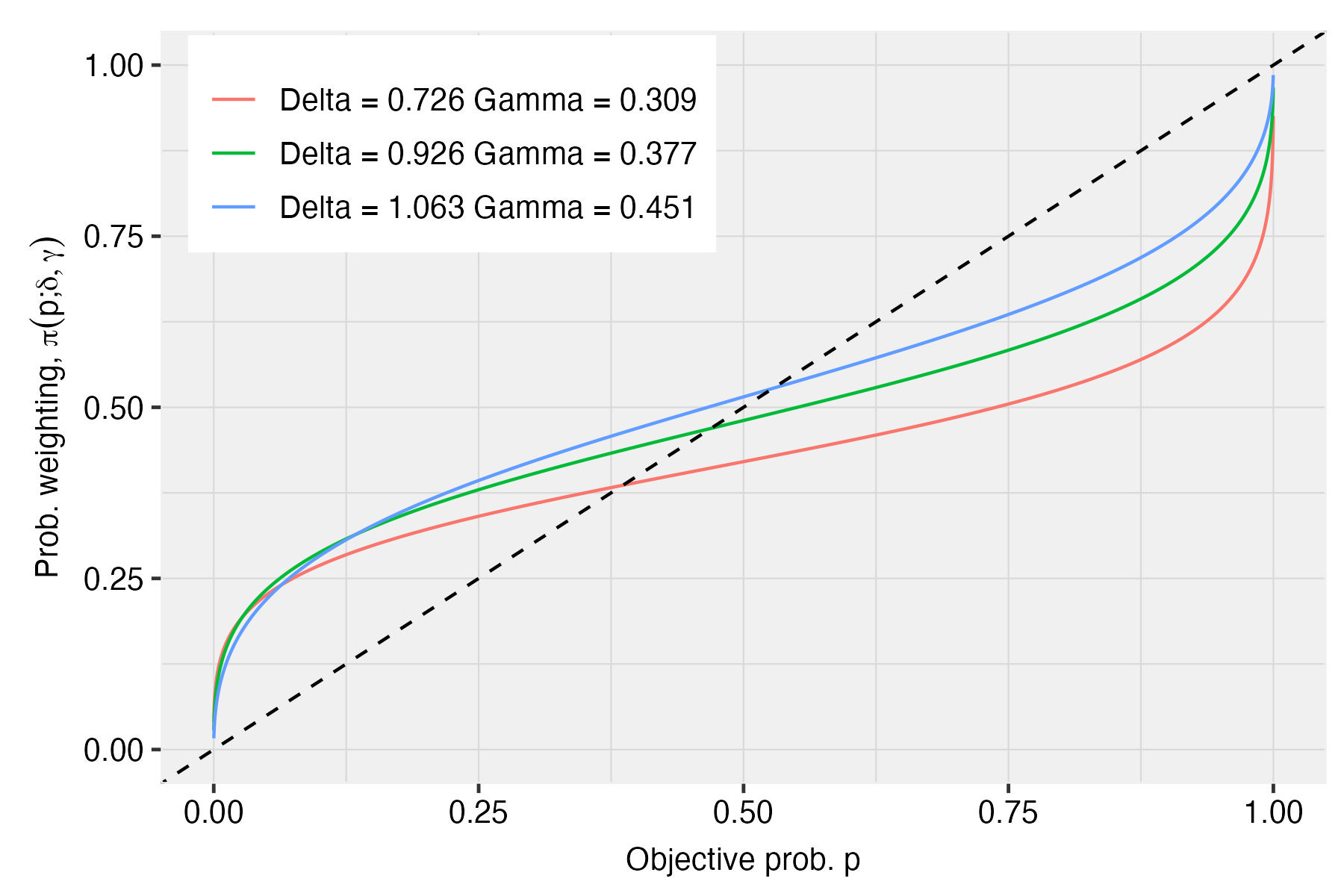}
    \caption{Probability weighting function for calibrated parameter values $(\delta, \gamma)$ in our numerical illustration based on prospect theory.}
    \floatfoot{\textit{Notes}: This figure plots the probability weighting function for the calibrated parameter values $(\delta, \gamma)$ used in our illustration to choice under risk. 
    We calibrate $(\delta, \gamma)$ to be equal to $(0.726, 0.309), (0.926, 0.377)$, and $(1.063, 0.451)$ using the pooled estimates based on the large-scale choice experiments in \citet{BruhinEtAl(10)} (reported in their Table V and Table IX). 
    See Section \ref{section: main text, simulation design} for discussion.
    } 
    \label{figure: probability weighting function, visualization}
\end{figure}

\begin{table}[htbp!]
\begin{subtable}{.4\linewidth}
 \centering
 \caption{Anomaly \#1}
 \begin{tabular}{c c c}
 & \multicolumn{1}{l}{Menu A $(x_A, y_A^*)$} \\
 \multicolumn{1}{c|}{Lottery 0} & 1.10 & 7.48 \\
 \multicolumn{1}{c|}{} & 15\% & 85\% \\
 \hline
 \multicolumn{1}{c|}{\cellcolor{green!25} Lottery 1} & 1.50 & 5.94 \\
 \multicolumn{1}{c|}{} & 1\% & 99\% \\
 \hline
 \end{tabular}
 \begin{tabular}{c c c}
 & \multicolumn{1}{l}{Menu B $(x_B, y_B^*)$} \\
 \multicolumn{1}{c|}{\cellcolor{green!25} Lottery 0} & 1.10 & 7.48 \\
 \multicolumn{1}{c|}{} & 45\% & 55\% \\
 \hline
 \multicolumn{1}{c|}{Lottery 1} & 1.50 & 5.94 \\
 \multicolumn{1}{c|}{} & 18\% & 82\% \\
 \hline
 \end{tabular}
\end{subtable} \hspace{1em}
\begin{subtable}{.4\linewidth}
 \centering
 \caption{Anomaly \#2}
 \begin{tabular}{c c c}
 & \multicolumn{1}{l}{Menu A $(x_A, y_A^*)$} \\
 \multicolumn{1}{c|}{Lottery 0} & 0.08 & 9.26 \\
 \multicolumn{1}{c|}{} & 34\% & 66\% \\
 \hline
 \multicolumn{1}{c|}{\cellcolor{green!25} Lottery 1} & 0.76 & 5.54 \\
 \multicolumn{1}{c|}{} & 0\% & 100\% \\
 \hline
 \end{tabular}
 \begin{tabular}{c c c}
 & \multicolumn{1}{l}{Menu B $(x_B, y_B^*)$} \\
 \multicolumn{1}{c|}{\cellcolor{green!25} Lottery 0} & 0.08 & 9.26 \\
 \multicolumn{1}{c|}{} & 63\% & 37\% \\
 \hline
 \multicolumn{1}{c|}{Lottery 1} & 0.76 & 5.54 \\
 \multicolumn{1}{c|}{} & 13\% & 87\% \\
 \hline
 \end{tabular}
\end{subtable} \\
\begin{subtable}{.4\linewidth}
 \centering
 \caption{Anomaly \#3}
 \begin{tabular}{c c c}
 & \multicolumn{1}{l}{Menu A $(x_A, y_A^*)$} \\
 \multicolumn{1}{c|}{\cellcolor{green!25} Lottery 0} & 2.52 & 7.64 \\
 \multicolumn{1}{c|}{} & 39\% & 61\% \\
 \hline
 \multicolumn{1}{c|}{Lottery 1} & 3.10 & 5.78 \\
 \multicolumn{1}{c|}{} & 21\% & 79\% \\
 \hline
 \end{tabular}
 \begin{tabular}{c c c}
 & \multicolumn{1}{l}{Menu B $(x_B, y_B^*)$} \\
 \multicolumn{1}{c|}{Lottery 0} & 2.52 & 7.64 \\
 \multicolumn{1}{c|}{} & 7\% & 93\% \\
 \hline
 \multicolumn{1}{c|}{\cellcolor{green!25} Lottery 1} & 3.10 & 5.78 \\
 \multicolumn{1}{c|}{} & 0\% & 100\% \\
 \hline
 \end{tabular}
\end{subtable} \hspace{1em}
\begin{subtable}{.4\linewidth}
 \centering
 \caption{Anomaly \#4}
 \begin{tabular}{c c c}
 & \multicolumn{1}{l}{Menu A $(x_A, y_A^*)$} \\
 \multicolumn{1}{c|}{Lottery 0} & 6.17 & 7.60 \\
 \multicolumn{1}{c|}{} & 9\% & 91\% \\
 \hline
 \multicolumn{1}{c|}{\cellcolor{green!25} Lottery 1} & 5.72 & 8.61 \\
 \multicolumn{1}{c|}{} & 27\% & 73 \\
 \hline
 \end{tabular}
 \begin{tabular}{c c c}
 & \multicolumn{1}{l}{Menu B $(x_B, y_B^*)$} \\
 \multicolumn{1}{c|}{\cellcolor{green!25} Lottery 0} & 6.17 & 7.60 \\
 \multicolumn{1}{c|}{} & 0\% & 100\% \\
 \hline
 \multicolumn{1}{c|}{Lottery 1} & 5.72 & 8.61 \\
 \multicolumn{1}{c|}{} & 0\% & 92\% \\
 \hline
 \end{tabular}
\end{subtable} \\
\begin{subtable}{.4\linewidth}
 \centering
 \caption{Anomaly \#5}
 \begin{tabular}{c c c}
 & \multicolumn{1}{l}{Menu A $(x_A, y_A^*)$} \\
 \multicolumn{1}{c|}{\cellcolor{green!25} Lottery 0} & 1.74 & 5.21 \\
 \multicolumn{1}{c|}{} & 10\% & 90\% \\
 \hline
 \multicolumn{1}{c|}{Lottery 1} & 1.83 & 4.71 \\
 \multicolumn{1}{c|}{} & 7\% & 93\% \\
 \hline
 \end{tabular}
 \begin{tabular}{c c c}
 & \multicolumn{1}{l}{Menu B $(x_B, y_B^*)$} \\
 \multicolumn{1}{c|}{Lottery 0} & 0.70 & 5.96 \\
 \multicolumn{1}{c|}{} & 2\% & 98\% \\
 \hline
 \multicolumn{1}{c|}{\cellcolor{green!25} Lottery 1} & 0.23 & 7.48 \\
 \multicolumn{1}{c|}{} & 0\% & 100\% \\
 \hline
 \end{tabular}
\end{subtable} \hspace{1em}
\begin{subtable}{.4\linewidth}
 \centering
 \caption{Anomaly \#6}
 \begin{tabular}{c c c}
 & \multicolumn{1}{l}{Menu A $(x_A, y_A^*)$} \\
 \multicolumn{1}{c|}{\cellcolor{green!25} Lottery 0} & 1.98 & 9.21 \\
 \multicolumn{1}{c|}{} & 48\% & 52\% \\
 \hline
 \multicolumn{1}{c|}{Lottery 1} & 2.49 & 7.69 \\
 \multicolumn{1}{c|}{} & 34\% & 66\% \\
 \hline
 \end{tabular}
 \begin{tabular}{c c c}
 & \multicolumn{1}{l}{Menu B $(x_B, y_B^*)$} \\
 \multicolumn{1}{c|}{Lottery 0} & 1.98 & 9.21 \\
 \multicolumn{1}{c|}{} & 5\% & 95\% \\
 \hline
 \multicolumn{1}{c|}{\cellcolor{green!25} Lottery 1} & 2.49 & 7.69 \\
 \multicolumn{1}{c|}{} & 0\% & 100\% \\
 \hline
 \end{tabular}
\end{subtable}
\caption{Examples of algorithmically generated anomalies for expected utility theory illustrating the dominated consequence effect.}
\floatfoot{\textit{Notes}: 
In each menu, we color the lottery that is selected with probability at least 0.50 in green. 
All payoffs are denominated in dollars.
We round each payoff to the nearest cent and each probability to the nearest percentage.
See Section \ref{section: anomalies, categorization, binary payoffs} and Appendix \ref{section: simulations, binary payoffs, proofs of anomalies} for discussion.}
\label{tab: generated anomalies, binary lotteries, dominated consequence}
\end{table} 

\begin{table}[htbp!]
\begin{subtable}{.4\linewidth}
 \centering
 \caption{Anomaly \#1}
 \begin{tabular}{c c c}
 & \multicolumn{1}{l}{Menu A $(x_A, y_A^*)$} \\
 \multicolumn{1}{c|}{Lottery 0} & 4.44 & 7.76 \\
 \multicolumn{1}{c|}{} & 100\% & 0\% \\
 \hline
 \multicolumn{1}{c|}{\cellcolor{green!25} Lottery 1} & 3.65 & 7.83 \\
 \multicolumn{1}{c|}{} & 95\% & 5\% \\
 \hline
 \end{tabular}
 \begin{tabular}{c c c}
 & \multicolumn{1}{l}{Menu B $(x_B, y_B^*)$} \\
 \multicolumn{1}{c|}{\cellcolor{green!25} Lottery 0} & 4.44 & 7.76 \\
 \multicolumn{1}{c|}{} & 36\% & 64\% \\    
 \hline
 \multicolumn{1}{c|}{Lottery 1} & 3.65 & 7.83 \\
 \multicolumn{1}{c|}{} & 23\% & 77\% \\
 \hline
 \end{tabular}
\end{subtable} \hspace{5em}
\begin{subtable}{.4\linewidth}
 \centering
 \caption{Anomaly \#2}
 \begin{tabular}{c c c}
 & \multicolumn{1}{l}{Menu A $(x_A, y_A^*)$} \\
 \multicolumn{1}{c|}{Lottery 0} & 1.36 & 5.91 \\
 \multicolumn{1}{c|}{} & 100\% & 0\% \\
 \hline
 \multicolumn{1}{c|}{\cellcolor{green!25} Lottery 1} & 0.05 & 6.05 \\
 \multicolumn{1}{c|}{} & 0.93\% & 7\% \\
 \hline
 \end{tabular}
 \begin{tabular}{c c c}
 & \multicolumn{1}{l}{Menu B $(x_B, y_B^*)$} \\
 \multicolumn{1}{c|}{\cellcolor{green!25} Lottery 0} & 1.36 & 5.91 \\
 \multicolumn{1}{c|}{} & 68\% & 32\% \\
 \hline
 \multicolumn{1}{c|}{Lottery 1} & 0.05 & 6.05 \\
 \multicolumn{1}{c|}{} & 56\% & 44\% \\
 \hline
 \end{tabular}
\end{subtable} \\
\begin{subtable}{.4\linewidth}
 \centering
 \caption{Anomaly \#3}
 \begin{tabular}{c c c}
 & \multicolumn{1}{l}{Menu A $(x_A, y_A^*)$} \\
 \multicolumn{1}{c|}{\cellcolor{green!25}  Lottery 0} & 2.23 & 7.69 \\
 \multicolumn{1}{c|}{} & 62\% & 38\% \\
 \hline
 \multicolumn{1}{c|}{Lottery 1} & 0.75 & 7.77 \\
 \multicolumn{1}{c|}{} & 38\% & 62\% \\
 \hline
 \end{tabular}
 \begin{tabular}{c c c}
 & \multicolumn{1}{l}{Menu B $(x_B, y_B^*)$} \\
 \multicolumn{1}{c|}{Lottery 0} & 2.23 & 7.69 \\
 \multicolumn{1}{c|}{} & 99\% & 1\% \\
 \hline
 \multicolumn{1}{c|}{\cellcolor{green!25} Lottery 1} & 0.75 & 7.77 \\
 \multicolumn{1}{c|}{} & 83\% & 17\% \\
 \hline
 \end{tabular}
\end{subtable} \hspace{4em}
\begin{subtable}{.4\linewidth}
 \centering
 \caption{Anomaly \#4}
 \begin{tabular}{c c c}
 & \multicolumn{1}{l}{Menu A $(x_A, y_A^*)$} \\
 \multicolumn{1}{c|}{\cellcolor{green!25}  Lottery 0} & 3.02 & 8.12 \\
 \multicolumn{1}{c|}{} & 80\% & 20\% \\
 \hline
 \multicolumn{1}{c|}{Lottery 1} & 0.29 & 9.43 \\
 \multicolumn{1}{c|}{} & 49\% & 51\% \\
 \hline
 \end{tabular}
 \begin{tabular}{c c c}
 & \multicolumn{1}{l}{Menu B $(x_B, y_B^*)$} \\
 \multicolumn{1}{c|}{Lottery 0} & 3.02 & 8.12 \\
 \multicolumn{1}{c|}{} & 100\% & 0\% \\
 \hline
 \multicolumn{1}{c|}{\cellcolor{green!25} Lottery 1} & 0.29 & 9.43 \\
 \multicolumn{1}{c|}{} & 79\% & 21\% \\
 \hline
 \end{tabular}
\end{subtable} \\
\begin{subtable}{.4\linewidth}
 \centering
 \caption{Anomaly \#5}
 \begin{tabular}{c c c}
 & \multicolumn{1}{l}{Menu A $(x_A, y_A^*)$} \\
 \multicolumn{1}{c|}{Lottery 0} & 0.84 & 9.88 \\
 \multicolumn{1}{c|}{} & 51\% & 49\% \\
 \hline
 \multicolumn{1}{c|}{\cellcolor{green!25} Lottery 1} & 3.32 & 9.25 \\
 \multicolumn{1}{c|}{} & 76\% & 24\% \\
 \hline
 \end{tabular}
 \begin{tabular}{c c c}
 & \multicolumn{1}{l}{Menu B $(x_B, y_B^*)$} \\
 \multicolumn{1}{c|}{\cellcolor{green!25}Lottery 0} & 0.84 & 9.88 \\
 \multicolumn{1}{c|}{} & 85\% & 15\% \\
 \hline
 \multicolumn{1}{c|}{Lottery 1} & 3.32 & 9.25 \\
 \multicolumn{1}{c|}{} & 100\% & 0\% \\
 \hline
 \end{tabular}
\end{subtable} \hspace{4em}
\begin{subtable}{.4\linewidth}
 \centering
 \caption{Anomaly \#6}
 \begin{tabular}{c c c}
 & \multicolumn{1}{l}{Menu A $(x_A, y_A^*)$} \\
 \multicolumn{1}{c|}{Lottery 0} & 0.93 & 6.82 \\
 \multicolumn{1}{c|}{} & 18\% & 82\% \\
 \hline
 \multicolumn{1}{c|}{\cellcolor{green!25} Lottery 1} & 2.02 & 6.78 \\
 \multicolumn{1}{c|}{} & 28\% & 72\% \\
 \hline
 \end{tabular}
 \begin{tabular}{c c c}
 & \multicolumn{1}{l}{Menu B $(x_B, y_B^*)$} \\
 \multicolumn{1}{c|}{\cellcolor{green!25} Lottery 0} & 0.93 & 6.82 \\
 \multicolumn{1}{c|}{} & 95\% & 5\% \\
 \hline
 \multicolumn{1}{c|}{Lottery 1} & 2.02 & 6.78 \\
 \multicolumn{1}{c|}{} & 100\% & 0\% \\
 \hline
 \end{tabular}
\end{subtable}
\caption{Examples of algorithmically generated anomalies for expected utility theory illustrating the reverse dominated consequence effect.}
\floatfoot{\textit{Notes}: 
In each menu, we color the lottery in the menu that is selected with probability at least 0.50 in green. 
All payoffs are denominated in dollars.
We round each payoff to the nearest cent and each probability to the nearest percentage.
See Section \ref{section: anomalies, categorization, binary payoffs} and Appendix \ref{section: simulations, binary payoffs, proofs of anomalies} for discussion.}
\label{tab: generated anomalies, binary lotteries, reverse dominated consequence}
\end{table} 

\begin{table}[htbp!]
\begin{subtable}{.4\linewidth}
 \centering
 \caption{Anomaly \#1}
 \begin{tabular}{c c c}
 & \multicolumn{1}{l}{Menu A $(x_A, y_A^*)$} \\
 \multicolumn{1}{c|}{Lottery 0} & 6.28 & 6.91 \\
 \multicolumn{1}{c|}{} & 65\% & 35\% \\
 \hline
 \multicolumn{1}{c|}{\cellcolor{green!25} Lottery 1} & 5.94 & 7.77 \\
 \multicolumn{1}{c|}{} & 53\% & 47\% \\
 \hline
 \end{tabular}
 \begin{tabular}{c c c}
 & \multicolumn{1}{l}{Menu B $(x_B, y_B^*)$} \\
 \multicolumn{1}{c|}{\cellcolor{green!25} Lottery 0} & 6.28 & 6.91 \\
 \multicolumn{1}{c|}{} & 100\% & 0\% \\
 \hline
 \multicolumn{1}{c|}{Lottery 1} & 5.94 & 7.77 \\
 \multicolumn{1}{c|}{} & 24\% & 76\% \\
 \hline
 \end{tabular}
\end{subtable} \hspace{5em}
\begin{subtable}{.4\linewidth}
 \centering
 \caption{Anomaly \#2}
 \begin{tabular}{c c c}
 & \multicolumn{1}{l}{Menu A $(x_A, y_A^*)$} \\
 \multicolumn{1}{c|}{Lottery 0} & 3.93 & 7.26 \\
 \multicolumn{1}{c|}{} & 39\% & 61\% \\
 \hline
 \multicolumn{1}{c|}{\cellcolor{green!25} Lottery 1} & 5.02 & 5.71 \\
 \multicolumn{1}{c|}{} & 100\% & 0\% \\
 \hline
 \end{tabular}
 \begin{tabular}{c c c}
 & \multicolumn{1}{l}{Menu B $(x_B, y_B^*)$} \\
 \multicolumn{1}{c|}{\cellcolor{green!25} Lottery 0} & 3.93 & 7.26 \\
 \multicolumn{1}{c|}{} & 41\% & 59\% \\
 \hline
 \multicolumn{1}{c|}{Lottery 1} & 5.02 & 5.71 \\
 \multicolumn{1}{c|}{} & 98\% & 2\% \\
 \hline
 \end{tabular}
\end{subtable} \\
\begin{subtable}{.4\linewidth}
 \centering
 \caption{Anomaly \#3}
 \begin{tabular}{c c c}
 & \multicolumn{1}{l}{Menu A $(x_A, y_A^*)$} \\
 \multicolumn{1}{c|}{Lottery 0} & 3.63 & 4.32 \\
 \multicolumn{1}{c|}{} & 61\% 39\% \\
 \hline
 \multicolumn{1}{c|}{\cellcolor{green!25} Lottery 1} & 3.72 & 4.21 \\
 \multicolumn{1}{c|}{} & 51\% & 49\% \\
 \hline
 \end{tabular}
 \begin{tabular}{c c c}
 & \multicolumn{1}{l}{Menu B $(x_B, y_B^*)$} \\
 \multicolumn{1}{c|}{\cellcolor{green!25} Lottery 0} & 3.63 & 4.32 \\
 \multicolumn{1}{c|}{} & 100\% & 0\% \\
 \hline
 \multicolumn{1}{c|}{Lottery 1} & 3.72 & 4.21 \\
 \multicolumn{1}{c|}{} & 13\% & 87\% \\
 \hline
 \end{tabular}
\end{subtable} \hspace{4em}
\begin{subtable}{.4\linewidth}
 \centering
 \caption{Anomaly \#4}
 \begin{tabular}{c c c}
 & \multicolumn{1}{l}{Menu A $(x_A, y_A^*)$} \\
 \multicolumn{1}{c|}{\cellcolor{green!25} Lottery 0} & 6.89 & 8.24 \\
 \multicolumn{1}{c|}{} & 64\% & 36\% \\
 \hline
 \multicolumn{1}{c|}{Lottery 1} & 7.01 & 7.18 \\
 \multicolumn{1}{c|}{} & 46\% & 54\% \\
 \hline
 \end{tabular}
 \begin{tabular}{c c c}
 & \multicolumn{1}{l}{Menu B $(x_B, y_B^*)$} \\
 \multicolumn{1}{c|}{Lottery 0} & 6.89 & 8.24 \\
 \multicolumn{1}{c|}{} & 34\% & 66\% \\
 \hline
 \multicolumn{1}{c|}{\cellcolor{green!25} Lottery 1} & 7.01 & 7.18 \\
 \multicolumn{1}{c|}{} & 100\% & 0\% \\
 \hline
 \end{tabular}
\end{subtable} \\
\begin{subtable}{.4\linewidth}
 \centering
 \caption{Anomaly \#5}
 \begin{tabular}{c c c}
 & \multicolumn{1}{l}{Menu A $(x_A, y_A^*)$} \\
 \multicolumn{1}{c|}{Lottery 0} & 6.03 & 6.31 \\
 \multicolumn{1}{c|}{} & 41\% & 59\% \\
 \hline
 \multicolumn{1}{c|}{\cellcolor{green!25} Lottery 1} & 3.49 & 8.99 \\
 \multicolumn{1}{c|}{} & 35\% & 65\% \\
 \hline
 \end{tabular}
 \begin{tabular}{c c c}
 & \multicolumn{1}{l}{Menu B $(x_B, y_B^*)$} \\
 \multicolumn{1}{c|}{\cellcolor{green!25} Lottery 0} &  6.03 & 6.31 \\
 \multicolumn{1}{c|}{} & 100\% & 00\% \\
 \hline
 \multicolumn{1}{c|}{Lottery 1} & 3.49 & 8.99 \\
 \multicolumn{1}{c|}{} & 34\% & 66\% \\
 \hline
 \end{tabular}
\end{subtable} \hspace{4em}
\begin{subtable}{.4\linewidth}
 \centering
 \caption{Anomaly \#6}
 \begin{tabular}{c c c}
 & \multicolumn{1}{l}{Menu A $(x_A, y_A^*)$} \\
 \multicolumn{1}{c|}{Lottery 0} & 6.33 & 6.51 \\
 \multicolumn{1}{c|}{} & 42\% & 58\% \\
 \hline
 \multicolumn{1}{c|}{\cellcolor{green!25} Lottery 1} & 6.31 & 6.61 \\
 \multicolumn{1}{c|}{} & 52\% & 48\% \\
 \hline
 \end{tabular}
 \begin{tabular}{c c c}
 & \multicolumn{1}{l}{Menu B $(x_B, y_B^*)$} \\
 \multicolumn{1}{c|}{\cellcolor{green!25} Lottery 0} & 6.33 & 6.51 \\
 \multicolumn{1}{c|}{} & 100\% & 00\% \\
 \hline
 \multicolumn{1}{c|}{Lottery 1} & 6.31 & 6.61 \\
 \multicolumn{1}{c|}{} & 24\% & 76\% \\
 \hline
 \end{tabular}
\end{subtable}
\caption{Examples of algorithmically generated anomalies for expected utility theory illustrating the strict dominance effect.}
\floatfoot{\textit{Notes}: 
We color the lottery in the menu that is selected with probability at least 0.50 in green. 
All payoffs are denominated in dollars.
For simplicity, we round each payoff to the nearest cent and each probability to the nearest percentage.
See Section \ref{section: anomalies, categorization, binary payoffs} and Appendix \ref{section: simulations, binary payoffs, proofs of anomalies} for discussion.}
\label{tab: generated anomalies, binary lotteries, strict dominance effect}
\end{table}

\begin{table}[!htb]
    \begin{subtable}{.33\linewidth}
      \centering
        \caption{Anomaly \#1}
        \begin{tabular}{c | c c}
            Lottery 0 & 5.72 & 6.19 \\
            & 19\% & 81\% \\
            \hline 
            \cellcolor{green!25}{Lottery 1} & 5.26 & \\
            & 100\% & 
        \end{tabular}
    \end{subtable}%
    \begin{subtable}{.33\linewidth}
      \centering
        \caption{Anomaly \#2}
        \begin{tabular}{c | c c}
            \cellcolor{green!25}{Lottery 0} & 8.17 & \\
            & 100\% & \\
            \hline 
            Lottery 1 & 9.03 & 9.70 \\
            & 23\% & 77\%
        \end{tabular}
    \end{subtable} %
    \begin{subtable}{.33\linewidth}
      \centering
        \caption{Anomaly \#3}
        \begin{tabular}{c | c c}
            \cellcolor{green!25}{Lottery 0} & 7.97 & \\
            & 100\% & \\
            \hline 
            Lottery 1 & 8.85 & 9.88 \\
            & 59\% & 41\%
        \end{tabular}
    \end{subtable} %
    \begin{subtable}{.33\linewidth}
      \centering
        \caption{Anomaly \#4}
        \begin{tabular}{c | c c}
            Lottery 0 & 7.20 & 7.61 \\
            & 33\% & 67\% \\
            \hline 
            \cellcolor{green!25}{Lottery 1} & 6.99 & 7.50 \\
            & 98\% & 2\%
        \end{tabular}
    \end{subtable} %
    \begin{subtable}{.32\linewidth}
      \centering
        \caption{Anomaly \#5}
        \begin{tabular}{c | c c}
            Lottery 0 & 8.07 & 9.05 \\
            & 21\% & 79\% \\
            \hline 
            \cellcolor{green!25}{Lottery 1} & 7.84 & \\
            & 100\%
        \end{tabular}
    \end{subtable} %
    \begin{subtable}{.32\linewidth}
      \centering
        \caption{Anomaly \#6}
        \begin{tabular}{c | c c}
            Lottery 0 & 6.89 & 8.88 \\
            & 46\% & 54\% \\
            \hline 
            \cellcolor{green!25}{Lottery 1} & 6.87 & \\
            & 100\%
        \end{tabular}
    \end{subtable} \\
    \begin{subtable}{.33\linewidth}
      \centering
        \caption{Anomaly \#7}
        \begin{tabular}{c | c c}
            Lottery 0 & 6.30 & 6.85 \\
            & 18\% & 82\% \\
            \hline 
            \cellcolor{green!25}{Lottery 1} & 6.09 \\
            & 100\%
        \end{tabular}
    \end{subtable} %
    \begin{subtable}{.32\linewidth}
      \centering
        \caption{Anomaly \#8}
        \begin{tabular}{c | c c}
            \cellcolor{green!25}{Lottery 0} & 4.90 \\
            & 100\% \\
            \hline 
            Lottery 1 & 5.04 & 5.27 \\
            & 31\% & 69\% 
        \end{tabular}
    \end{subtable} %
    \begin{subtable}{.32\linewidth}
      \centering
        \caption{Anomaly \#9}
        \begin{tabular}{c | c c}
            \cellcolor{green!25}{Lottery 0} & 7.67 \\
            & 100\% \\
            \hline 
            Lottery 1 & 8.31 & 8.57 \\
            & 43\% & 57\% 
        \end{tabular}
    \end{subtable} \\
    \caption{Examples of algorithmically generated anomalies for expected utility theory illustrating first-order stochastic dominance violations.}
    \floatfoot{\textit{Notes}: 
    In each menu, we color the lottery that is selected with probability at least 0.50 in green.
    Each generated first-order stochastic dominance violation presented here $(x, y^*)$ is based on the probability weighting function $\pi(p; \delta, \gamma)$ with $(\delta, \gamma) = (0.726, 0.309)$. 
    All payoffs are denominated in dollars.
    We round each payoff to the nearest cent and each probability to the nearest percentage point.
    See Section \ref{section: anomalies, categorization, binary payoffs} for discussion.}
    \label{tab: generated anomalies, FOSD anomalies for subcertainty, binary payoffs}
\end{table}

\begin{table}[htbp!]
\begin{subtable}{.4\linewidth}
 \centering
 \caption{Anomaly \#1}
 \begin{tabular}{c c c c}
 & \multicolumn{2}{l}{Menu A $(x_A, y_A^*)$} \\
 \multicolumn{1}{c|}{\cellcolor{green!25} Lottery 0} & 4.23 & 4.31 & 9.15 \\
 \multicolumn{1}{c|}{} & 70\% & 2\% & 28\% \\
 \hline
 \multicolumn{1}{c|}{ Lottery 1} & 6.22 & 7.17 & 8.51 \\
 \multicolumn{1}{c|}{} & 52\% & 9\% & 39\% \\
 \hline
 \end{tabular}
 \begin{tabular}{c c c c}
 & \multicolumn{2}{l}{Menu B $(x_B, y_B^*)$} \\
 \multicolumn{1}{c|}{Lottery 0} & 4.23 & 4.31 & 9.15 \\
 \multicolumn{1}{c|}{} & 31\% & 32\% & 37\% \\
 \hline
 \multicolumn{1}{c|}{\cellcolor{green!25} Lottery 1} & 6.22 & 7.17 & 8.51 \\
 \multicolumn{1}{c|}{} & 54\% & 5\% & 41\% \\
 \hline
 \end{tabular}
\end{subtable} \hspace{5em}
\begin{subtable}{.4\linewidth}
 \centering
 \caption{Anomaly \#2}
 \begin{tabular}{c c c c}
 & \multicolumn{1}{l}{Menu A $(x_A, y_A^*)$} \\
 \multicolumn{1}{c|}{Lottery 0} & 2.82 & 4.35 & 9.98 \\
 \multicolumn{1}{c|}{} & 8\% & 80\% & 12\% \\
 \hline
 \multicolumn{1}{c|}{\cellcolor{green!25} Lottery 1} & 3.34 & 4.01 & 5.36 \\
 \multicolumn{1}{c|}{} & 89\% & 2\% & 9\% \\
 \hline
 \end{tabular}
 \begin{tabular}{c c c c}
 & \multicolumn{1}{l}{Menu B $(x_B, y_B^*)$} \\
 \multicolumn{1}{c|}{\cellcolor{green!25} Lottery 0} & 2.82 & 4.35 & 9.98 \\
 \multicolumn{1}{c|}{} & 51\% & 33\% & 16\% \\
 \hline
 \multicolumn{1}{c|}{ Lottery 1} & 3.34 & 4.01 & 5.36 \\
 \multicolumn{1}{c|}{} & 27\% & 37\% & 36\% \\
 \hline
 \end{tabular}
\end{subtable}        
\caption{Examples of algorithmically generated anomalies for expected utility theory over menus of lotteries on three monetary payoffs based on predictive algorithm $\widehat{f}(\cdot)$.}
\floatfoot{\textit{Notes}: 
In the menu, we color the lottery that is predicted to be selected with probability at least 0.50 in green.
Each algorithmically generated anomaly presented here is produced by our example morphing algorithm.
We round each payoff to the nearest cent and each probability to the nearest percentage.
See Section \ref{section: choices13k, three payoff anomalies} for discussion.}
\label{tab: generated anomalies, additional examples ternary lotteries, choices 13k}
\end{table} 

\begin{table}[!h]
\begin{tabular}{r | c c c c}
& \shortstack{Pooled \\ Average} & Median & \shortstack{First \\ Quartile} & \shortstack{Third \\ Quartile} \\ 
\hline \hline
\textbf{Two-Payoff Lotteries} \\
Dominated Consequence Effect & \shortstack{0.084 \\ (0.006)} & 0.087 & 0.060 & 0.110 \\ \\
Reverse Dominated Consequence Effect & \shortstack{0.063 \\ (0.005)} & 0.041 & 0.033 & 0.070 \\ \\
Strict Dominance Effect & \shortstack{0.145 \\ (0.007)} & 0.137 & 0.073 & 0.196 \\ \\
\textbf{Three-Payoff Lotteries} & \shortstack{0.071 \\ (0.003)} & 0.064 & 0.032 & 0.090
\end{tabular}
\caption{Summary statistics on the fraction of respondents whose choices violate expected utility theory on algorithmically generated anomalies.}
\floatfoot{\textit{Notes}: This table reports summary statistics on the fraction of respondents whose choices violate expected utility theory (``expected utility theory violation rate'') across the algorithmically generated anomalies tested in the online experiment. 
The ``pooled average'' column reports the expected utility theory violation rate, pooling together respondents' choices on all anomalies within the same category. 
Standard errors reported in parentheses are clustered at the respondent level. 
See Section \ref{subsection: experimental test of algorithmically generated anomalies} for  discussion.}
\label{tab: summary statistics, anomalous fraction, by anomaly category, binary and ternary payoffs}
\end{table}

\section{Additional theoretical discussion}\label{section: additional theoretical results}

\subsection{Additional examples of theories and anomalies}\label{section: additional examples}

In this appendix section, we first illustrate how additional examples map into our framework described in Section \ref{section: anomaly generation problem} of the main text.

\paragraph{Example: choice under risk}
As in the main text, consider individuals evaluating a lottery over $J > 1$ monetary payoffs. 
The features describe the lottery $x = (z, p)$, where $z \in \mathbb{R}^{J}$ is the lottery's payoffs and $p \in \Delta^{J-1}$ is the lottery's probabilities.
The modeled outcome is now the certainty equivalent $y^* \in \mathbb{R}$ for the lottery.
Expected utility theory searches for utility functions $u$ in some researcher-chosen class $\mathcal{U}$ that rationalize the certainty equivalents, yielding the allowable function class $\mathcal{F}^{T} = \{ f \colon \exists u \in \mathcal{U} \mbox{ s.t. } f(x) = u^{-1}\left( \sum_{j=1}^{J} p(j) u(z(j)) \right) \mbox{ for all } x \in \mathcal{X} \}$.
Given the collection $D$, expected utility theory returns utility functions $u \in \mathcal{U}$ satisfying $y^* = u^{-1}\left( \sum_{j=1}^{J} p(j) u(z(j)) \right)$ for all $(x, y^*) \in D$. 
On any new lottery $x$, expected utility theory returns predictions $T(x; D)$, where $y^* \in T(x; D)$ if and only if $y^* = u^{-1}\left( \sum_{j=1}^{J} p(j) u(z(j)) \right)$ for some utility function $u(\cdot)$ rationalizing $D$. 
Alternative behavioral models such as cumulative prospect theory yield alternative theories of of certainty equivalents in our framework. $\blacktriangle$

\paragraph{Example: multi-attribute discrete choice}
Consider individuals making choices from menus of $J$ items. 
The features are a complete description of each item $x = \left( z_1, p_1, \hdots, z_J, p_J \right)$, where $z_j$ are the attributes of item $j$ and $p_j$ is its price.
The features may also describe how items are presented in the menu or their ordering.
The modeled outcomes are menu choice probabilities $y^* \in \Delta^{J-1}$.

A popular class of parametric additive random utility models, such as the multinomial logit, specify the indirect utility of item $j$ as $v_j(x; \alpha, \beta) = z_j \beta - \alpha p_j + \epsilon_{j}$, where $(\alpha, \beta)$ are parameters and $\epsilon_{j}$ is a random taste shock with some known distribution.
Each choice of the parameters $(\alpha, \beta)$ then corresponds to a particular allowable function $f \in \mathcal{F}^T$.
Given a collection $D$, such a parametric additive random utility model searches for parameter values $(\alpha, \beta)$ that match the choice probabilities, meaning $y_j^* = P\left( j \in \arg \max_{\widehat{j}} \, v_{\widehat{j}}(x; \alpha, \beta) \right)$ for all $j = 1, \hdots, J$ and $(x, y^*) \in D$. 
On any new menu $x$, it returns $T(x; D)$, where $y^* \in T(x; D)$ if and only if $y_j^* = P\left( j \in \arg \max_{\widehat{j}} \, v_{\widehat{j}}(x; \alpha, \beta) \right)$ for some $(\alpha, \beta)$ that matches $D$. $\blacktriangle$

\medskip

We next illustrate the Certainty effect for expected utility theory and an anomaly for Nash equilibrium, following Definition \ref{defn: anomaly}.

\paragraph{Example: the Certainty effect} 
Consider the Certainty effect for expected utility theory in Table \ref{tab: the certainty effect} \citep{KahnemanTversky(79)}.
\begin{table}[!htb]
\caption{Menus of lotteries in the Certainty effect \citep{KahnemanTversky(79)}.}
\begin{subtable}{.5\linewidth}
    \centering
    \caption{Menu A}
    \begin{tabular}{c | c c c}
            Lottery 0 & \$4000 & \$0 \\
            & 80\% & 20\% \\
            \hline 
            \cellcolor{green!25}{Lottery 1} & \$3000 \\ 
            & 100\%
        \end{tabular}
    \end{subtable}%
    \begin{subtable}{.5\linewidth}
      \centering
        \caption{Menu B}
        \begin{tabular}{c | c c c}
            \cellcolor{green!25}{Lottery 0} & \$4000 & \$0 \\
            & 20\% & 80\% \\
            \hline 
            Lottery 1 & \$3000 & \$0 \\ 
            & \$25 & 75\%
        \end{tabular}
    \end{subtable} 
    \label{tab: the certainty effect}
    \floatfoot{\textit{Notes}: We highlight in green the conjectured choices on these two menus.} 
\end{table}
The Certainty effect is a pair of examples consisting of the menus $x_A, x_B$ and modeled outcomes $y_A^* = 0, y_B^* = 1$. 
Like the Allais paradox in the main text, the independence axiom implies that the choice on menu $x_A$ must be the same as the choice on menu $x_B$.
The Certainty effect is therefore inconsistent with expected utility theory, yet any single choice $(x_A, y_A^*)$ or $(x_B, y_B^*)$ is consistent with expected utility theory. 
The Certainty effect therefore satisfies Definition \ref{defn: anomaly}.$\blacktriangle$

\paragraph{Example: play in normal-form games} 
Consider the normal-form game in Table \ref{tab: anomaly for nash equilibrium}. In our framework, such a normal-form game is a particular feature $x \in \cX$.
Iterated elimination of strictly dominated strategies implies that $(Top, Left)$ is the unique Nash equilibrium of the game. 
Consequently, for any $D \in \mathcal{D}$, it must either be that $T(x; D) = \varnothing$ (i.e., $D$ contains pairs of payoff matrices and chosen strategies inconsistent with Nash equilibrium) or $T(x; D) = (1, 0, 0)$.
\begin{table}[htbp!]
    \centering
    \begin{tabular}{c || c c c}
    & Left & Center & Right \\ 
    \hline \hline
      Top & (10, 4) & (5,3) & (3,2) \\
      Middle & (0,1) & (4,6) & (6,0) \\
      Bottom & (2,1) & (3,5) & (2,8)
    \end{tabular}
    \caption{An example anomaly for Nash equilibrium based on Level-1 thinking.}
    \label{tab: anomaly for nash equilibrium}
\end{table}
Suppose instead the individual $m$ was a level-1 thinker. In this case, she would eliminate Bottom since it is strictly dominated but would fail to recognize the Right is now strictly dominated for her opponent by the iterated elimination of strictly dominated strategies. 
She would then play the game as if her opponent randomizes across all of her actions, and we may observe her strategy profile $y^*$ placing positive probability on both Top and Middle. 
By construction, a collection of examples that consisted of only this normal-form game and such a strategy profile would be an anomaly for Nash equilibrium (Definition \ref{defn: anomaly}). $\blacktriangle$

\subsection{Average anomalies across individuals}\label{section: average anomalies, appendix}

As mentioned in the main text, researchers have collected large experimental datasets of individuals making choices from menus of risky lotteries or selecting actions in normal-form games and fit black-box predictive algorithms to model the average behavior across individuals. 
Applied to such a predictive algorithm, our anomaly generation procedures search for anomalies that hold on average across individuals. 
We may instead be interested in generating anomalies that hold at the individual-level.
When does an anomaly that hold on average across individuals also hold at the individual-level?

More formally, suppose we observe we observe a random sample $(M_i, X_i, Y_i) \sim P(\cdot)$ for $i = 1, \hdots, N$, where $M_i$ is an individual-level identifier taking values $m \in \mathcal{M}$. 
Define $\bar{f}^*(x) := \mathbb{E}[Y_i \mid X_i = x]$ as the average relationship between the features and observed outcome across individuals.
Define $P(m \mid x) := P(M_i = m \mid X_i = x)$ and $f_m^*(x) := \mathbb{E}[g(Y_i) \mid M_i = m, X_i = x]$ for each individual $m \in \cM$. 
We assume throughout this discussion that $P(m \mid x) > 0$ for all $m \in \mathcal{M}$ and $x \in \mathcal{X}$.

\vspace{-1em}
\paragraph{Incompatible collections:}
A collection $x_{1:n}$ is an \textit{average} incompatible collection of examples if $D = \{ (x_1, \bar{f}^*(x_i)), \hdots, (x_n, \bar{f}^*(x_n)) \}$ is incompatible with theory $\mathcal{F}^{T}$. 
A collection $x_{1:n}$ is an incompatible collection \textit{for individual} $m$ if $D = \{ (x_1, f^*_m(x_i)), \hdots, (x_n, f^*_m(x_n)) \}$ is incompatible with theory $\mathcal{F}^{T}$.
Provided $\{ (x_1, \bar{f}^*(x_i)), \hdots, (x_n, \bar{f}^*(x_n)) \}$ is a ``systematically'' incompatible collection at the individual-level, then it is also an average incompatible collection.

\begin{proposition}\label{prop: average and systematic incompatible datasets}
Provided $x_{1:n}$ is an incompatible collection for some individual $m$ and satisfies
\begin{equation}\label{eqn: systematically incompatible}
    \sum_{m \neq \tilde{m}} \left( n^{-1} \sum_{i=1}^{n} P(m \mid x) P(\tilde{m} \mid x) \left( f_m^T(x_i) - f_m^*(x_i) \right) \left( f_{\tm}^T(x_i) - f_{\tm}^*(x_i) \right) \right) \geq 0,
\end{equation}
for all $f_{m}(\cdot), f_{\tm}(\cdot) \in \cF^{T}$, then $x_{1:n}$ is also an average incompatible collection.

\begin{proof}
To prove this result, it suffices to focus on squared loss $\ell(y, y^\prime) = \left( y - y^\prime \right)^2$.
Observe that 
$$
\min_{f(\cdot) \in \cF^{T}} n^{-1} \sum_{i=1}^{n} \left( f(x_i) - \bar{f}^*(x_i) \right)^2 \geq \min_{f_m(\cdot) \in \cF^T} n^{-1} \sum_{i=1}^{n} \left( \sum_{m \in \cM} P(m \mid x_i) ( f_m(x_i) - f_m^*(x_i) )  \right)^2,
$$
where 
$$
n^{-1} \sum_{i=1}^{n} \left( \sum_{m \in \cM} P(m \mid x_i) ( f_m(x_i) - f_m^*(x_i) )  \right)^2 =
$$
$$
n^{-1} \sum_{m \in \cM} \sum_{i=1}^{n} P(m \mid x_i)^2 (f_m(x_i) - f_m^*(x_i))^2 + 
n^{-1} \sum_{m \neq \tilde{m}} \sum_{i=1}^{n} P(m \mid x_i) P(\tm \mid x_i) (f_m(x_i) - f_m^*(x_i)) ( f_{\tm}(x_i) - f_{\tm}^*(x_i) ).
$$
Then, under Equation \eqref{eqn: systematically incompatible}, it follows that 
$$
\min_{f(\cdot) \in \cF^{T}} n^{-1} \sum_{i=1}^{n} \left( f(x_i) - \bar{f}^*(x_i) \right)^2 \geq \sum_{m \in \cM} \left\{ n^{-1} 
 \sum_{i=1}^{n} P(m \mid x_i)^2 (f_m(x_i) - f_m^*(x_i))^2 \right\}.
$$
The result follows since $x_{1:n}$ is incompatible collection for some individual $m$. 
\end{proof}
\end{proposition}

\noindent Equation \eqref{eqn: systematically incompatible} requires $x_{1:n}$ be ``systematically'' incompatible with $\mathcal{F}^{T}$ in the sense that the errors of the theory's best fitting allowable functions across individuals do not cancel out on average.

\vspace{-1em}
\paragraph{Representational anomalies:}
An \textit{average} representational anomaly is a pair of features $x_1, x_2$ such that $\bar{f}^*(x_1) \neq \bar{f}^*(x_2)$ but $f(x_1) = f(x_2)$ for all $f(\cdot) \in \mathcal{F}^{T}$.
If there are no compositional changes across these features, then $x_1, x_2$ is an average representational anomaly if and only if it is a representational anomaly for some individual $m$. 

\begin{proposition}\label{prop: avg representational anomalies and composition}
Consider features $x_1, x_2 \in \cX$ and suppose $P(m \mid x_1) = P(m \mid x_2)$ for all $m \in \cM$.
If $\{ (x_1, \bar{f}^*(x_1)), (x_2, \bar{f}^*(x_2)) \}$ is an average representational anomaly, then there exists some individual $m$ for whom $\{ (x_1, f_m^*(x_1), (x_2, f_m^*(x_2)) \}$ is a representational anomaly.
\begin{proof}
To prove this result, observe that 
$$
\bar{f}^*(x_1) - \bar{f}^*(x_2) = \sum_{m \in \cM} P(m \mid x_1) f_m^*(x_1) - \sum_{m \in \cM} P(m \mid x_2) f_m^*(x_2) 
$$
$$
= \sum_{m \in \cM} P(m \mid x_1) \left( f_m^*(x_1) - f_m^*(x_2) \right) + \sum_{m \in \cM} \left( P(m \mid x_1) - P(m \mid x_2) \right) f_m(x_2).
$$
Assuming that $P(m \mid x_1) = P(m \mid x_2)$ for all $m \in \cM$ implies that the second term in the previous display equals zero. The result is then immediate.
\end{proof}
\end{proposition}

\noindent The condition in Proposition \ref{prop: avg representational anomalies and composition} requires that there exists the same composition of individuals across features $x_1, x_2$. This is satisfied provided individuals are randomly assigned features in an experiment.

\subsection{Gradient descent ascent optimization over allowable functions}\label{section: GDA over allowable functions}

In Section \ref{section: main text, GDA} of the main text, we proposed a gradient descent ascent (GDA) procedure to optimize the plug-in max-min program \eqref{equation: plug-in max min}.
For some parametrization $\cF^{T} = \{ f_{\theta}(\cdot) \colon \theta \in \Theta \}$, initial feature values $x_{1:n}^0$, maximum number of iterations $S > 0$, and step size sequence $\{\eta_s\}_{s=0}^{S} > 0$, we iterate over $s = 0, \hdots, S$ and calculate
\begin{align*}
& \theta^{s+1} = \arg \min_{\theta \in \Theta} \ \widehat{\cE}(x_{1:n}^s; \theta) \\
& x_{1:n}^{s+1} = x_{1:n}^t + \eta_{t} \nabla \widehat{\cE}(x_{1:n}^s; \theta^{s+1})
\end{align*}
at each iteration, where $\widehat{\cE}(x_{1:n}, \theta) := n^{-1} \sum_{i=1}^{n} \ell\left( f_{\theta}(x_i), \widehat{f}^*(x_i)\right)$.

There is a vast literature studying max-min optimization using a variety of gradient based algorithms \citep[][]{RazaviyaynEtAl(20)}. 
As an illustration, we discuss how recent results from \cite{JinEtAl(19)} on non-convex/concave max-min optimization may applied to show this simple GDA procedure converges to an approximate stationary point of the falsifier's outer maximization problem.

Define $\bar{x}_{1:n}$ to be the random variable drawn uniformly over $\{ x_{1:n}^0, \hdots, x_{1:n}^S \}$ and define $\widehat{\cE}(x_{1:n}) = \min_{\theta \in \Theta} \widehat{\cE}(x_{1:n}, \theta)$. 
Define the \textit{Moreau envelope} of $\widehat{\cE}(x_{1:n})$ as
$$
\phi_{\lambda}(x_{1:n}) = \min_{x_{1:n}^\prime} \ \widehat{\cE}(x_{1:n}^\prime) + \frac{1}{2\lambda} \| x_{1:n} - x_{1:n}^\prime \|_2^2
$$
For non-convex functions, the Moreau envelope is a smooth, convex approximation that is often used to analyze the properties of gradient descent algorithms \citep[e.g, see][]{DavisDrusvyatskiy(18)}. 
Provided the loss function, predictive algorithm, and collection of parametrized allowable functions are sufficiently smooth, the simple GDA procedure provides a bound on the gradient of the Moreau envelope $\phi_{\lambda}(\cdot)$. 
Standard results in convex optimization establish that a bound on the gradient of the Moreau envelope implies a bound on the subdifferentials of $\widehat{\cE}(x_{1:n})$.

\begin{lemma}[Lemma 30 in \cite{JinEtAl(19)}]
Suppose $\widehat{\cE}(x_{1:n})$ is $b$-weakly convex. For an $\lambda < \frac{1}{b}$ and $\widetilde{x}_{1:n} = \arg \min_{x_{1:n}^\prime} \widehat{\cE}(x_{1:n}^\prime) + \frac{1}{2\lambda} \| x_{1:n} - x_{1:n}^\prime \|_2^2$, $\|\nabla \phi_{\lambda}(x_{1:n}) \| \leq \epsilon$ implies
$$
    \| \widetilde{x}_{1:n} - x_{1:n} \| = \lambda \epsilon \mbox{ and } \min_{g \in \partial \widehat{\cE}(\widetilde{x}_{1:n})} \|g \| \leq \epsilon,
$$
where $\partial$ denotes the subdifferential of a weakly convex function.
\end{lemma}

\begin{proposition}\label{proposition: GDA proposition}
Suppose $\ell(\cdot, \cdot), \widehat{f}^*(\cdot)$ and $\{f_{\theta}(\cdot) \colon \theta \in \Theta \}$ are $k$-times continuously differentiable with $K$-bounded gradients. 
Then, the output $\bar{x}_{1:n}$ of the gradient descent ascent algorithm with step size sequence given by $\eta_{s} = \eta_0 / \sqrt{S + 1}$ for some $\eta_0 > 0$ satisfies 
$$
\mathbb{E}\left[ \| \nabla \phi_{0.5 b}(\bar{x}_{1:n}) \|_2^2 \right] \leq 2 \frac{ \left( \phi_{0.5 b}(x_{1:n}^0) - \min_{x_{1:n}} \widehat{\cE}(x_{1:n}) \right) + b K^2 \eta_0^2 }{\eta_0 \sqrt{S+1}} + 4 b \delta,
$$
where $\delta \geq 0$ is the error associated with the approximate inner minimization oracle in Assumption \ref{assumption: optimization oracles}(i).
\begin{proof}
This result is an immediate consequence of Theorem 31 in \cite{JinEtAl(19)}.
\end{proof}
\end{proposition}

\noindent Our specific implementation of the adversarial anomaly generation algorithm in our choice under risk application satisfy the regularity conditions in Proposition \ref{proposition: GDA proposition}. 
Specifically, $\ell(\cdot, \cdot)$ is chosen to be the cross-entropy loss, $\widehat{f}^*(\cdot)$ is a parametrized neural network, and $\{ f_{\theta}(\cdot) \cdot \theta \in \Theta\}$ is a class of parametrized logit models. 
We search over lotteries with finite, bounded payoffs.

\section{Implementation details of anomaly generation procedures}\label{section: simulations, implementation details}

In this section, we describe the implementation details of our anomaly generation procedures in the illustration to choice under risk in Section \ref{section: anomalies for choice under risk, simulation} and Section \ref{section: anomalies for choice under risk, choices13k}. 

For both algorithms, we randomly initialize menus of lotteries in the following manner.
We sample each payoff in a lottery independently from a uniform distribution on $[0, 10]$. 
We simulate the probabilities in a lottery by drawing each lottery probability uniformly from the unit interval, and then normalizing the draws so they lie on the unit simplex.

\subsection{Adversarial algorithm}\label{section: implementation details, adversarial algorithm}

We first specify a parametric basis for the allowable functions of expected utility theory. 
We parametrize the utility function $u_{\theta}(\cdot)$ as a linear combination of polynomials up to order $K$ or I-splines with some number of knot points $q$ and degree $K$ \citep[see][]{Ramsay(88)}. 
We experimented with both choices of basis functions, varying the maximal degree of the polynomial bases as well as the number of knot points and degree of the I-spline bases.
Since we found qualitatively similar results, we focus on presenting anomalies generated by a polynomial utility function basis with order $K = 6$.
We set the learning rate to be $\eta = 0.01$.

For any particular choice of utility function basis and learning rate, we ran the gradient descent ascent procedure for 25,000 randomly initialized menus $x^0$. 
We set the maximum number of iterations to be $S = 50$.
For a particular choice of utility basis functions, we solve the inner minimization problem \eqref{eqn: GDA inner minimization} by minimizing the cross-entropy loss between the true choice probabilities on the menus $f^*(x^s)$ and the implied expected utility theory choice probabilities $f_{\theta}(x^s) = P\left( \sum_{j=1}^{J} p_{1j}^s u_{\theta}(z_{1j}) - \sum_{j=1}^{J} p_{0j}^s u_{\theta}(z_{0j}) + \xi \right)$ for $\xi$ an i.i.d. logistic shock.
We implement the outer gradient ascent step \eqref{eqn: GDA outer maximization step} directly. 
We take gradient ascent steps on only the probabilities of the lotteries in the menu.
After each gradient ascent step, we project the updated lottery probability vectors back into the unit simplex. 
We collect and return the pairs $(x^0, x^S)$ produced across all runs of the adversarial procedure. 

A subtle numerical issue arises as the gradients of the cross-entropy loss $\widehat{\cE}(x^s; \theta^{s+1})$ vanish whenever expected utility theory can exactly match the choice probabilities. 
To avoid this vanishing gradients problem, we instead implement the outer gradient ascent step \eqref{eqn: GDA outer maximization step} by following the gradient of $ \log\left( \frac{f^*(x^s)}{1 - f^*(x^s)} \right) \left( \sum_{j=1}^{J} p_{1j}^s u_{\theta^{s+1}}(z_{1j}) - \sum_{j=1}^{J} p_{0j} u_{\theta^{s+1}}(z_{0j}) \right)$.
This alternative loss function for the gradient ascent step applies the logit transformation to the choice probabilities so that $\log\left( \frac{f^*(x^s)}{1 - f^*(x^s)} \right)$ is positive whenever $f^*(x^s) > 0.5$ and weakly negative otherwise. 
The overall loss function is therefore positive whenever the expected utility difference between the lotteries is positive but $f^*(x^s) < 0.5$ and vice versa.

\subsection{Example morphing algorithm}\label{section: implementation details, example morphing algorithm}
We again specify a parametric basis for the allowable functions of expected utility theory. 
Like the adversarial algorithm, we experimented with both polynomial bases up to order $K$ and I-spline bases varying the number of knot points $q$ and degree $K$.
Since we found qualitatively similar results, we focus on presenting anomalies generated by the I-spline basis with $q = 10$ knot points and degree $K = 3$.
We set the learning rate to be $\eta = 10$.

For any particular utility function basis and learning rate, we ran the example morphing procedure 15,000 randomly initialized menus $x^0$. 
We set the maximum number of iterations to be $S = 50$.
At each iteration $s$, we solve for the best-fitting allowable function $\theta^{s}$ and sample $\theta_b$ independent from a multivariate normal distribution with mean vector equal to $\bar{\theta}^{s} = \frac{1}{s} \sum_{s^\prime = 1}^{s} \theta^{s^\prime}$ and variance matrix equal to $\frac{1}{s-1} \sum_{s^\prime=1}^{s} (\theta^{s^\prime} - \bar{\theta}^{s}) (\theta^{s^\prime} - \bar{\theta}^{s})^\prime$ for $b = 1, \hdots, B$.
We set $B = 200,000$.
We take gradient steps on only the probabilities of the lotteries in the menu.
We collect and return the pairs $(x^0, x^S)$ produced across all runs of the example morphing procedure.

\subsection{Numerical verification of anomalies for expected utility theory}\label{section: simulations, numerical verification of EUT anomalies}

The returned menus of lotteries over two monetary payoffs by our anomaly generation procedures are anomalies for expected utility theory using our parametrization of the utility function $\{ u_{\theta}(\cdot) \colon \theta \in \Theta \}$ and model of noise. 
Given any such returned menus of lotteries over two monetary payoffs, we numerically verify whether the dataset of returned menus is an anomaly for expected utility theory at any increasing utility function and without noisy choices. 
In the main text, we report all resulting numerically verified anomalies for expected utility theory at any increasing utility function. 

Concretely, consider $\{ (x_0, f^*(x_0)), (x_1, f^*(x_1)) \}$ returned by our anomaly generation procedures, where $x_0 = \left( z_{0,0}, p_{0,0}, z_{0,1}, p_{0,1} \right)$ and $x_1 = \left( z_{0,0}, p_{1,0}, z_{0,1}, p_{1,1} \right)$. 
For ease of exposition, we assume the monetary payoffs are the same across the two menus. 
Define $y_0^* = 1\{ f^*(x_0) \geq 0.50\}$ and $y_1^* = 1\{f^*(x_1) \geq 0.50\}$, and the ordered monetary payoffs as 
$
    z_{(1)} < z_{(2)} < z_{(3)} < z_{(4)}.
$
We check whether there exists any increasing utility function $u(z)$ satisfying $u(z_{(1)}) < u(z_{(2)}) < u(z_{(3)}) < u(z_{(4)})$ that could rationalize the given configuration of binary choices $(y^*_0, y^*_1)$. 
Abusing notation, let us redefine $p_{0,0} \in \Delta^{4}$ as the vector of probabilities associated with the ordered monetary payoffs, and $p_{0,1}, p_{1,0}, p_{1,1}$ analogously. 
Let $u = \left(u_1, u_2, u_3, u_4\right)$ denote the vector of utility values assigned to the ordered monetary payoffs.
Checking whether there exists any increasing utility function that could rationalize the given configuration of choices is equivalent to checking whether there exists a solution to a system of linear inequalities.
In particular, if (i) $y_0^* = y_1^* = 0$, we check whether there exists any vector $u$ satisfying $(p_{0,0} - p_{0,1})^\prime u > 0$ and $(p_{1,0} - p_{1,1})^\prime u > 0$; (ii) $y_0^* = 1, y_1^* = 0$, we check whether there exists any vector $u$ satisfying $(p_{0,0} - p_{0,1})^\prime u < 0$ and $(p_{1,0} - p_{1,1})^\prime u > 0$; and so on.

\section{Additional details for illustration to expected utility theory}\label{section: additional illustration details}

\subsection{Numeric categorization of anomalies for expected utility theory over two-payoff lotteries}\label{section: categorization details, two payoff anomalies}

As mentioned in Section \ref{section: anomalies, categorization, binary payoffs} of the main text, after applying the numerical verification described Appendix \ref{section: simulations, numerical verification of EUT anomalies}, we categorize the generated anomalies over two-payoff lotteries in two steps based on the violation of expected utility theory they highlight.
We first check whether the generated anomaly is a first-order stochastic dominance violation. 
Second, we check whether the generated pair of menus can be represented as compound lotteries.
Since the space of lotteries over two payoffs is simple enough, we can enumerate compounding operations. 
We next describe these steps in more detail.

Our generated anomalies contain two menus of lotteries over same two monetary payoffs $(x^0, x^S)$ as discussed in Appendix \ref{section: implementation details, adversarial algorithm} and Appendix \ref{section: implementation details, example morphing algorithm}. 
Denote the lotteries in menu A as $\ell_0^{A}$ with payoffs $\underline{z}_0^{A} \leq \overline{z}_0^{A}$ and $\ell_1^{A}$ with payoffs $\underline{z}_1^{A} \leq \overline{z}_1^{A}$.
Let $p_0^{A}$ denote the probability of the low payoff in lottery $0$ and $p_1^{A}$ denote the probability of the low payoff in lottery 1. 
We analogously write the lotteries in menu B $\ell_0^B, \ell_1^B$.
The labeling of the menus and the lotteries within each menu is arbitrary. 
We will next introduce relationships that express the lotteries in menu B in terms of the lotteries in menu A and a compounding operation. 
We check whether these relationships hold for any labeling of the menus and  labeling of the lotteries within the menus. 

We then check whether one lottery first-order stochastically dominates another in menu A or menu B and the individual is predicted to select the dominated lottery. 
If so, we categorize this generated anomaly as a first-order stochastic dominance anomaly. 
We note that all of our first-order stochastic dominance anomalies are produced on the morphed menu $x^S$.

Among all anomalies that are not first-order stochastic dominance anomalies, we next check whether the lotteries in menu B can be expressed in terms of the lotteries in menu A using one of three compounding operations.
Let $\delta(z)$ denote the lottery that returns payoff $z$ with certainty.
First, we check whether there exists values $\alpha_0, \alpha_1 \in [0, 1]$ such that $\ell_0^{B} = \alpha_0 \ell_0^{A} + (1 - \alpha_0) \delta(\underline{z}_0^A)$ and $\ell_1^{B} = \alpha_1 \ell_1^{A} + (1 - \alpha_1) \delta(\underline{z}_1^A)$. 
Second, we check whether there exists (possibly different) values $\alpha_0, \alpha_1 \in [0, 1]$ such that $\ell_0^{B} = \alpha_0 \ell_0^{A} + (1 - \alpha_0) \delta(\overline{z}_0^A)$ and $\ell_1^{B} = \alpha_1 \ell_1^{A} + (1 - \alpha_1) \delta(\overline{z}_1^A)$.
Finally, we check whether there exists (again possibly different) values $\alpha_0, \alpha_1 \in [0, 1]$ such that $\ell_0^{B} = \alpha_0 \ell_0^{A} + (1 - \alpha_0) \delta(\underline{z}_0^A)$ and $\ell_1^{B} = \alpha_1 \ell_1^{A} + (1 - \alpha_1) \delta(\overline{z}_1^A)$.
The dominated consequence effect is associated with the first compounding operation, the reverse dominated consequence effect is associated with the second compounding operation, and the strict dominance effect is associated with the third compounding operation. See Section \ref{section: anomalies, categorization, binary payoffs} and Appendix \ref{section: simulations, binary payoffs, proofs of anomalies} for the interpretation of these categories.

\subsection{Proofs of anomalies for expected utility theory over two-payoff lotteries}\label{section: simulations, binary payoffs, proofs of anomalies}

We now prove that the dominated consequence effect, reverse dominated consequence effect, and strict dominance effect (formally described in Appendix \ref{section: categorization details, two payoff anomalies}) are anomalies for expected utility theory. 

\vspace{-1em}
\paragraph{The dominated consequence effect:}
Consider the first menu defined over the pair of lotteries $\ell_0 = (p_0, z_0)$, $\ell_1 = (p_1, z_1)$ and the second menu defined over the lotteries $\ell_0^\prime = (p_0^\prime, z_0)$, $\ell_1^\prime = (p_1^\prime, z_1)$. 
Let $\underline{z}_0 = \min_{j} z_0(j)$ and $\underline{z}_1 = \min_{j} z_1(j)$.
Suppose the lotteries in the second menu can be written as $\ell_0^\prime = \alpha_0 \ell_0 + (1 - \alpha_0) \delta_{\underline{z_0}}$ and $\ell_1^\prime = \alpha_1 \ell_1 + (1 - \alpha_1) \delta_{\underline{z_1}}$ for some $\alpha_0, \alpha_1 \in [0,1]$.
Further assume (i) $\underline{z}_0 < \underline{z}_1$, (ii) $\ell_1 \succ \ell_0$, (iii) $\ell_0^\prime \succ \ell_1^\prime$, and (iv) $\alpha_1 \geq \alpha_0$. 
Under expected utility theory, we observe that
\begin{align*}
    & \ell_1 \succ \ell_0 \overset{(1)}{\implies} \alpha_1 \ell_1 + (1 - \alpha_1) \delta_{\underline{z}_1} \succ \alpha_1 \ell_0 + (1 - \alpha_1) \delta_{\underline{z}_1}, \\
    & \alpha_1 \ell_0 + (1 - \alpha_1) \delta_{\underline{z}_1} \overset{(2)}{\succ} \alpha_1 \ell_0 + (1 - \alpha_1) \delta_{\underline{z}_0}, \\ 
    & \alpha_1 \ell_0 + (1 - \alpha_1) \delta_{\underline{z}_0} \overset{(3)}{\succ} \alpha_0 \ell_0 + (1 - \alpha_0) \delta_{\underline{z}_0} 
\end{align*}
where (1) follows by the independence axiom, (2) follows by utility must be increasing in monetary payoffs and the independence axiom, and (3) follows by preservation of first-order stochastic dominance.
Transitivity then delivers that $\ell_1$ being preferred to $\ell_0$ must imply  $\ell_1^\prime$ is preferred to $\ell_0^\prime$ under expected utility theory.
The collection $\{ ((\ell_0, \ell_1),  1), ((\ell_0^\prime, \ell_1^\prime), 0) \}$ is therefore an anomaly for expected utility theory. 

Appendix Table \ref{tab: generated anomalies, binary lotteries, dominated consequence} provides more examples of dominated consequence effect anomalies.
Each example can be mapped into the dominated consequence effect through an appropriate choice of $\ell_0, \ell_1$ and $\ell_0^\prime, \ell_1^\prime$ in the algorithmically generated menus.

\vspace{-1em}
\paragraph{The reverse dominated consequence effect:}
Consider the first menu defined over the pair of lotteries $\ell_0 = (p_0, z_0)$, $\ell_1 = (p_1, z_1)$ and the second menu defined over the pair of lotteries $\ell_0^\prime = (p_0^\prime, z_0)$, $\ell_1^\prime = (p_1^\prime, z_1)$. 
Let $\overline{z}_0 = \max_{j} z_0(j)$ and $\overline{z}_1 = \max_{j} z_1(j)$.
Suppose the lotteries in the second menu can be written as $\ell_0^\prime = \alpha_0 \ell_0 + (1 - \alpha_0) \delta_{\overline{z_0}}$ and $\ell_1^\prime = \alpha_1 \ell_1 + (1 - \alpha_1) \delta_{\overline{z_1}}$ for some $\alpha_0, \alpha_1 \in [0,1]$.
Further assume (i) $\overline{z}_1 > \overline{z}_0$, (ii) $\ell_1 \succ \ell_0$, (iii) $\ell_0^\prime \succ \ell_1^\prime$, and (iv) $\alpha_0 \geq \alpha_1$. 
Under expected utility theory, we observe that
\begin{align*}
    & \ell_1 \succ \ell_0 \overset{(1)}{\implies} \alpha_1 \ell_1 + (1 - \alpha_1) \delta_{\overline{z}_1} \succ \alpha_1 \ell_0 + (1 - \alpha_1) \delta_{\overline{z}_1}, \\
    & \alpha_1 \ell_0 + (1 - \alpha_1) \delta_{\overline{z}_1} \overset{(2)}{\succ} \alpha_1 \ell_0 + (1 - \alpha_1) \delta_{\overline{z}_0} \\ 
    & \alpha_1 \ell_0 + (1 - \alpha_1) \delta_{\overline{z}_0} \overset{(3)}{\succ} \alpha_0 \ell_0 + (1 - \alpha_0) \delta_{\overline{z}_0} 
\end{align*}
where (1) follows by the independence axiom, (2) follows by utility must be increasing in monetary payoffs and the independence axiom, and (3) follows by preservation of first-order stochastic dominance.
Transitivity then delivers $\ell_1$ being preferred to $\ell_0$ must imply that $\ell_1^\prime$ is preferred to $\ell_0^\prime$ under expected utility theory.
Therefore, the collection $\{ ((\ell_0, \ell_1),  1), ((\ell_0^\prime, \ell_1^\prime), 0) \}$ is an anomaly for expected utility theory. 

Appendix Table \ref{tab: generated anomalies, binary lotteries, reverse dominated consequence} provides more examples of reverse dominated consequence effect anomalies. 
Each example can be mapped into the reverse dominated consequence effect through an appropriate choice of $\ell_0, \ell_1$ and $\ell_0^\prime, \ell_1^\prime$ in the algorithmically generated menus.

\vspace{-1em}
\paragraph{The strict dominance effect:}
Consider the first menu defined over the pair of lotteries $\ell_0 = (p_0, z_0)$, $\ell_1 = (p_1, z_1)$ and the second menu defined over the pair of lotteries $\ell_0^\prime = (p_0^\prime, z_0)$, $\ell_1^\prime = (p_1^\prime, z_1)$. 
Let $\underline{z}_0 = \max_{j} z_0(j)$ and $\overline{z}_1 = \max_{j} z_1(j)$.
Suppose the lotteries in the second menu can be written as $\ell_0^\prime = \alpha_0 \ell_0 + (1 - \alpha_0) \delta_{\underline{z_0}}$ and $\ell_1^\prime = \alpha_1 \ell_1 + (1 - \alpha_1) \delta_{\overline{z_1}}$ for some $\alpha_0, \alpha_1 \in [0,1]$.
Further assume (i) $\ell_1 \succ \ell_0$, and (ii) $\ell_0^\prime \succ \ell_1^\prime$.
Under expected utility theory, we observe that $\ell_1^\prime \overset{(1)}{\succ} \ell_1$ and $\ell_0 \overset{(2)}{\succ} \ell_0^\prime$, where (1) and (2) follow by preservation of first-order stochastic dominance.
Transitivity then delivers that $\ell_1$ being preferred to $\ell_0$ must imply that $\ell_1^\prime$ is preferred to $\ell_0^\prime$ under expected utility theory.
Therefore, the collection $\{ ((\ell_0, \ell_1),  1), ((\ell_0^\prime, \ell_1^\prime), 0) \}$ is an anomaly for expected utility theory. 

Appendix Table \ref{tab: generated anomalies, binary lotteries, strict dominance effect} provides additional examples of strict dominance effect anomalies. 
Each example can be mapped into the strict dominance effect through an appropriate choice of $\ell_0, \ell_1$ and $\ell_0^\prime, \ell_1^\prime$ in the algorithmically generated menus.

\subsection{Anomaly generation from an estimated choice probability function}\label{section: simulations, estimated choice probabilities, binary payoffs}

In this section, we generate anomalies based on an estimated choice probability function $\widehat{f}(\cdot)$ using a random sample of binary choices from a representative agent.

For each calibrated parameter value $(\delta, \gamma)$, we simulate a dataset of menus of two lotteries over two monetary payoffs and binary choices on each menu.  
For $i = 1, \hdots, n$, we simulate menus of two lotteries over two monetary payoffs $X_i$ by drawing each payoff in the lotteries independently from a uniform distribution on $[0, 10]$, and simulating the probabilities in each lottery by drawing uniformly from the unit interval $[0, 1]$ and normalizing the draws so they lie on the unit simplex.
We draw the binary choice according to $Y_i \mid X_i \sim Bernoulli(f^*(X_i))$, yielding the dataset $\{ (X_i, Y_i) \}_{i=1}^{n}$.

We estimate the choice probability function $f^*(x) = P( CPT(p_1, z_1; \delta, \gamma) - CPT(p_0, z_0; \delta, \gamma) + \xi \geq 0 )$ in two ways.
First, we consider correctly specified choice probability functions, and estimate the parameter values $(\widehat{\delta}, \widehat{\gamma})$ that minimize the average cross-entropy loss
\begin{equation}\label{eqn: fitted prob weighting function}
    (\widehat{\delta}, \widehat{\gamma}) = \arg \min_{\tilde{\delta}, \tilde{\gamma}} \frac{1}{n} \sum_{i=1}^{n} -Y_i \log( f_{(\tilde{\delta}, \tilde{\gamma})}(X_i)) - (1 - Y_i) \log(1 - f_{(\tilde{\delta}, \tilde{\gamma})}(X_i))
\end{equation}
for $f_{(\tilde{\delta}, \tilde{\gamma})}(x) = \frac{e^{CPT(p_1, z_1; \tilde \delta, \tilde \gamma) - CPT(p_0, z_0; \tilde \delta, \tilde \gamma)}}{1 + e^{CPT(p_1, z_1; \tilde \delta, \tilde \gamma) - CPT(p_0, z_0; \tilde \delta, \tilde \gamma)}}$. 
This yields the estimated choice probability function $\widehat{f}(\cdot) = f_{(\widehat{\delta}, \widehat{\gamma})}(\cdot)$.
Second, we consider the class of choice probability functions that can be characterized by deep neural networks. 
We specifically consider over-parametrized deep neural networks with four hidden layers and $500$ hidden nodes each with rectified linear unit (ReLU) activation functions. 
We minimize the average cross-entropy loss
\begin{equation}\label{eqn: fitted DNN}
    \widehat{f}^{DNN}(\cdot) = \arg \min_{\tilde{f} \in \cF^{DNN}} \frac{1}{n} \sum_{i=1}^{n} -Y_i \log( \tilde{f}(X_i)) - (1 - Y_i) \log(1 - \tilde{f}(X_i))
\end{equation}
using mini-batch gradient descent with a batch size of 256 observations over 2,000 epochs. 

For each calibrated parameter value $(\delta, \gamma)$, we simulate one binary choice dataset $\{(X_i, Y_i)\}_{i=1}^{n}$, and approximate the individual's true choice probability function $f^*(\cdot)$ using both the estimated probability weighting parameters \eqref{eqn: fitted prob weighting function} and the deep neural network \eqref{eqn: fitted DNN}.
As in Section \ref{section: main text, simulation, implementation discussion and benchmark} of the main text, we apply our anomaly generation procedures on the estimated choice probability function $\widehat{f}^*(\cdot)$.
Each returned pair of menus and the implied choices based on $\widehat{f}^*(\cdot)$ is an anomaly for expected utility theory at our particular parameterization of the utility function. 
We again numerically verify whether the returned menus and implied choices based on $\widehat{f}^*(\cdot)$ are anomalies for expected utility theory at any increasing utility function and without noisy choices, as discussed in Appendix \ref{section: simulations, numerical verification of EUT anomalies}.

Appendix Tables \ref{tab: anomaly categorization for binary payoffs, finite data, estimated CPT}-\ref{tab: anomaly categorization for binary payoffs, finite data, deep network} summarize the generated anomalies for expected utility theory using the estimated probability weighting parameters and the deep neural network respectively. 
We vary the size of the simulated dataset over $n = 1,000$, $5,000$, $10,000$ and $25,000$. 
As in the main text, we again logically verify the generated anomalies in two ways. 
First, our procedures continue to generate many anomalies for the parametric utility function classes -- for example, with $n = 1,000$, $763$, $1,133$, and $1,772$ are generated using the estimated probability weighting parameters and $1,963$, $2015$, and $1,869$ are generated using the estimated neural networks respectively across calibrated parameter values $(\delta, \gamma)$.
Second, our procedures also generate many anomalies for expected utility theory with any increasing utility function --- for example, with $n = 1,000$, $149$ are generated using the estimated probability weighting parameters and $282$ are generated using the estimated neural networks.
On estimated choice probability functions, we find similar performance of our optimization routines.
We further apply the same categorization as in Section \ref{section: anomalies, categorization, binary payoffs} of the main text, and we find that our procedures generate the same categories of anomalies for expected utility theory as we found in the main text.

\begin{table}[htbp!]
\begin{subtable}{\linewidth}
\centering
\caption{$(\delta, \gamma) = (0.726, 0.309)$}
\begin{tabular}{c c c c c c}
 & \multicolumn{4}{c}{Sample Size: $n$} & \multicolumn{1}{c}{True Choice Prob.} \\
 & 1,000 & 5,000 & 10,000 & 25,000 \\ 
 \multicolumn{1}{r|}{Dominated Consequence Effect} & 7 & 25 & 2 & 17 & 85 \\
 \multicolumn{1}{r|}{Reverse Dominated Consequence Effect} & 1 & 4 & 0 & 3 & 17 \\
 \multicolumn{1}{r|}{Strict Dominance Effect} & 10 & 77 & 9 & 57 & 45 \\
 \multicolumn{1}{r|}{First Order Stochastic Dominance} & 1 & 66 & 16 & 74 & 81 \\
 \multicolumn{1}{r|}{Other} & 1 & 4 & 0 & 3 & 3 \\
 \hline \hline 
 \multicolumn{1}{r|}{\# of Anomalies} & 20 & 176 & 27 & 154 & 231 \\
\end{tabular}
\end{subtable}
\begin{subtable}{\linewidth}
\centering
\caption{$(\delta, \gamma) = (0.926, 0.377)$}
\begin{tabular}{c c c c c c}
 & \multicolumn{4}{c}{Sample Size: $n$} & \multicolumn{1}{c}{True Choice Prob.} \\
 & 1,000 & 5,000 & 10,000 & 25,000 \\ 
 \multicolumn{1}{r|}{Dominated Consequence Effect} & 2 & 3 & 9 & 5 & 34 \\
 \multicolumn{1}{r|}{Reverse Dominated Consequence Effect} & 9 & 2 & 4 & 5 & 15 \\
 \multicolumn{1}{r|}{Strict Dominance Effect} & 17 & 5 & 1 & 1 & 1 \\
 \multicolumn{1}{r|}{First Order Stochastic Dominance} & 2 & 3 & 5 & 0 & 0 \\
 \multicolumn{1}{r|}{Other} & 2 & 2 & 0 & 0 & 1 \\
 \hline \hline 
 \multicolumn{1}{r|}{\# of Anomalies} & 32 & 15 & 19 & 11 & 51 \\
\end{tabular}  
\end{subtable}
\begin{subtable}{\linewidth}
\centering
\caption{$(\delta, \gamma) = (1.063, 0.451)$}
\begin{tabular}{c c c c c c}
 & \multicolumn{4}{c}{Sample Size: $n$} & \multicolumn{1}{c}{True Choice Prob.} \\
 & 1,000 & 5,000 & 10,000 & 25,000 \\ 
 \multicolumn{1}{r|}{Dominated Consequence Effect} & 5 & 7 & 2 & 0 & 10 \\
 \multicolumn{1}{r|}{Reverse Dominated Consequence Effect} & 13 & 4 & 3 & 5 & 14 \\
 \multicolumn{1}{r|}{Strict Dominance Effect} & 39 & 0 & 0 & 1 & 0 \\
 \multicolumn{1}{r|}{First Order Stochastic Dominance} & 33 & 0 & 0 & 1 & 2 \\
 \multicolumn{1}{r|}{Other} & 7 & 0 & 0 & 0 & 1 \\
 \hline \hline 
 \multicolumn{1}{r|}{\# of Anomalies} & 97 & 11 & 5 & 7 & 27
\end{tabular}
\end{subtable}
\caption{Anomalies for expected utility theory over lotteries on two payoffs, generated using $f_{(\widehat{\delta}, \widehat{\gamma})}(\cdot)$.}
\floatfoot{\textit{Notes}: For each calibrated parameter values $(\delta,\gamma)$, we estimate the parameter values $(\delta, \gamma)$ that minimize average cross-entropy loss \eqref{eqn: fitted prob weighting function}, varying the size of the binary choice dataset over $n = 1,000, 5,000, 10,000$ and $25,000$. 
For reference, the column ``True Choice Prob.'' reproduces Table \ref{tab: anomaly categorization for binary payoffs}, which generated anomalies using the true choice probability function $f^*(\cdot)$.
See Appendix \ref{section: simulations, estimated choice probabilities, binary payoffs} for further discussion.}
\label{tab: anomaly categorization for binary payoffs, finite data, estimated CPT}
\end{table}

\begin{table}[htbp!]
\begin{subtable}{\linewidth}
\centering
\caption{$(\delta, \gamma) = (0.726, 0.309)$}
\begin{tabular}{c c c c c c}
 & \multicolumn{4}{c}{Sample Size: $n$} & \multicolumn{1}{c}{True Choice Prob.} \\
 & 1,000 & 5,000 & 10,000 & 25,000 \\ 
 \multicolumn{1}{r|}{Dominated Consequence Effect} & 21 & 18 & 17 & 13 & 85 \\
 \multicolumn{1}{r|}{Reverse Dominated Consequence Effect}  & 14 & 3 & 3 & 0 & 17 \\
 \multicolumn{1}{r|}{Strict Dominance Effect} & 35 & 7 & 2 & 1 & 45 \\
 \multicolumn{1}{r|}{First Order Stochastic Dominance} & 45 & 16 & 27 & 13 & 81 \\
 \multicolumn{1}{r|}{Other} & 3 & 0 & 1 & 3 & 3 \\
 \hline \hline 
 \multicolumn{1}{r|}{\# of Anomalies} & 118 & 44 & 50 & 30 & 231
\end{tabular}
\end{subtable}
\begin{subtable}{\linewidth}
\centering
\caption{$(\delta, \gamma) = (0.926, 0.377)$}
\begin{tabular}{c c c c c c}
 & \multicolumn{4}{c}{Sample Size: $n$} & \multicolumn{1}{c}{True Choice Prob.} \\
 & 1,000 & 5,000 & 10,000 & 25,000 \\ 
 \multicolumn{1}{r|}{Dominated Consequence Effect} & 16 & 17 & 22 & 15 & 34 \\
 \multicolumn{1}{r|}{Reverse Dominated Consequence Effect} & 17 & 6 & 4 & 5 & 15 \\
 \multicolumn{1}{r|}{Strict Dominance Effect} & 33 & 5 & 1 & 0 & 1 \\
 \multicolumn{1}{r|}{First Order Stochastic Dominance} & 25 & 18 & 17 & 10 & 0 \\
 \multicolumn{1}{r|}{Other} & 1 & 2 & 2 & 3 & 1 \\
 \hline \hline 
 \multicolumn{1}{r|}{\# of Anomalies} & 92 & 48 & 46 & 33 & 51
\end{tabular}  
\end{subtable}
\begin{subtable}{\linewidth}
\centering
\caption{$(\delta, \gamma) = (1.063, 0.451)$}
\begin{tabular}{c c c c c c}
 & \multicolumn{4}{c}{Sample Size: $n$} & \multicolumn{1}{c}{True Choice Prob.} \\
 & 1,000 & 5,000 & 10,000 & 25,000 \\ 
 \multicolumn{1}{r|}{Dominated Consequence Effect} & 19 & 15 & 22 & 23 & 10 \\
 \multicolumn{1}{r|}{Reverse Dominated Consequence Effect} & 8 & 7 & 6 & 4 & 14 \\
 \multicolumn{1}{r|}{Strict Dominance Effect} & 26 & 2 & 3 & 0 & 0 \\
 \multicolumn{1}{r|}{First Order Stochastic Dominance} & 16 & 17 & 18 & 11 & 2 \\
 \multicolumn{1}{r|}{Other} & 3 & 0 & 3 & 4 & 1 \\
 \hline \hline 
 \multicolumn{1}{r|}{\# of Anomalies} & 72 & 41 & 52 & 42 & 27
\end{tabular}
\end{subtable}
\caption{Anomalies for expected utility theory over lotteries on two payoffs, generated using $\widehat{f}^{DNN}(\cdot)$.}
\floatfoot{\textit{Notes}: 
For each calibrated parameter values $(\delta,\gamma)$, we fit a deep neural network to minimize average cross-entropy loss \eqref{eqn: fitted DNN}, varying the size of the simulated binary choice dataset over $n = 1,000, 5,000, 10,000$ and $25,000$.
For reference, the column ``True Choice Prob.'' reproduces Table \ref{tab: anomaly categorization for binary payoffs}, which generated anomalies using the true choice probability function $f^*(\cdot)$.
See Appendix \ref{section: simulations, estimated choice probabilities, binary payoffs} for further discussion.}
\label{tab: anomaly categorization for binary payoffs, finite data, deep network}
\end{table}

\subsection{Categorization of anomalies for expected utility theory over three-payoff lotteries}\label{section: categorization of three payoff anomalies}

\subsubsection{Numeric categorization}

As mentioned in Section \ref{section: choices13k, three payoff anomalies} of the main text, after applying the numerical verification described Appendix \ref{section: simulations, numerical verification of EUT anomalies}, we categorize the generated anomalies over three-payoff lotteries in two steps based on the violation of expected utility theory they highlight. 
We first check whether the generated anomaly is a first-order stochastic dominance violation. 
Second, we check whether the generated pair of menus can be represented as a compound lotteries over the same two-payoff lotteries.
More formally, we check whether, for some two-payoff lotteries $\ell_{0}^\prime, \ell_{0}^{\prime \prime}$ and $\ell_1^\prime, \ell_1^{\prime \prime}$, we can write the lotteries in menu A as $\alpha_{A0} \ell_{0}^\prime + (1 - \alpha_{A0}) \ell_0^{\prime \prime}$ and $\alpha_{A1} \ell_{1}^\prime + (1 - \alpha_{A1}) \ell_1^{\prime \prime}$ and the lotteries in menu B as $\alpha_{B0} \ell_{0}^\prime + (1 - \alpha_{B0}) \ell_0^{\prime \prime}$ and $\alpha_{B1} \ell_{1}^\prime + (1 - \alpha_{B1}) \ell_1^{\prime \prime}$. If, for example, $\ell_0^\prime$ first order stochastically dominates $\ell_0^{\prime \prime}$, we can then check whether $\alpha_{A0} > \alpha_{B0}$ yet individuals are predicted to select lottery A1 and lottery B0.

\subsection{Empirical clustering}

We now describe in more detail the empirical clustering of algorithmically generated anomalies over three-payoff lotteries discussed in Section \ref{section: choices13k, three payoff anomalies}.
As mentioned in the main text, we focus on the 356 anomalies that do not violate first-order stochastic dominance. 

Each anomaly consists of two menus $m = \{A, B\}$. 
Each menu $m$ consists of two lotteries: the lottery predicted to be chosen by $\widehat{f}$, denoted by $\ell_1^m = (z_1^m, p_1^m)$ and the alternative lottery $\ell_m^0 = (z_0^m, p_0^m)$. 
For each menu $m$, we calculate summary statistics about the differences between the lottery predicted to be chosen $\ell_1^m$ and the alternative lottery $\ell_0^m$ in their expected payoffs, payoff variances, payoff skews, payoff range, minimum payoff, maximum payoff, probability range, minimum probability and maximum probability. This defines a feature vector $z^m$ for the menu, and the feature vector for the anomaly is then $z = (z^A, z^B)$.
To cluster and visualize them, we apply $K$-means to the standardized features of the anomalies and calculate the first two principle components of the standardized features. Figure \ref{fig:cluster_visualization} in the main text visualizes each anomaly in the space spanned by the first two principal components and overlays the cluster groups.
Figure \ref{fig:cluster_loadings} reports the top four features with the largest absolute loadings on each principal component. 

\section{Additional details for experimental test of algorithmically generated anomalies}

\subsection{Experimental design}\label{section: appendix, description of survey design}

As discussed in Section \ref{subsection: experimental test of algorithmically generated anomalies} of the main text, we randomly selected 30 algorithmically generated anomalies summarized in Table \ref{tab: anomaly categorization for binary payoffs, choices 13K} over lotteries with two monetary payoffs, which were chosen to span the dominated consequence effect, the reverse dominated consequence effect, and the strict dominance effect. 
We also randomly selected 30 anomalies over lotteries with three-payoffs sharing the structure we highlighted in Section \ref{section: choices13k, three payoff anomalies}.

We split these anomalies into four surveys, each containing 15 anomalies.
We present each anomaly as two separate choices of menus, and so each survey consists of 30 questions altogether. 
For a particular menu, we display the written probabilities and payoffs for each lottery in the menu, and we additionally depict each lottery as a color-coded pie chart.
Each survey randomizes the order of questions and the left-right positioning of lotteries in a menu across respondents.

We recruited respondents for all surveys on Prolific. 
Each respondent received a base payment of \$4 for completing a survey. 
We screened out inattentive respondents through comprehension questions and attention checks throughout the surveys.
We include screenshots of the instructions, comprehension checks, attention checks, and main survey questions in Appendix \ref{section: prolific survey screenshots}.
Respondents that successfully completed a survey without failing comprehension and attention checks were eligible for a bonus payment based on a ``random payment selection'' mechanism \citep[e.g.,][]{AzrieliEtAl(18)}. 
We determined the bonus by randomly selecting a lottery that was chosen by a respondent on the survey. 
The respondent was paid the realization of the randomly selected lottery.
On our surveys of lotteries with two monetary payoffs, the average bonus payments were \$4.82 and \$4.81 respectively. On our surveys of lotteries with three monetary payoffs, the average bonus payments \$4.98 and \$3.81 respectively. 
Since respondents completed a survey in roughly 15 minutes on average, the implied hourly wage was quite generous relative to Prolific standards.
We recruited 258 and 266 respondents on our two surveys of lotteries with two monetary payoffs, and 260 and 263 respondents on our two surveys of lotteries with three monetary payoffs.

\subsection{Could the algorithmically generated anomalies be explained by noisy choices?}\label{section: appendix, anomalies and noisy choices}

Recent experimental work suggests a simple extension to expected utility theory that could resolve many known anomalies: specifically, incorporating noise in individuals' choices \citep[e.g.,][]{McGranahanEtAl(23)-CommonRatio}.
We examine whether small amounts of noise in individuals' choices could explain the empirical findings on our algorithmically generated anomalies.
Following \cite{HarlessCamerer(94)}, we consider expected utility theory with idiosyncratic errors, in which individuals mistakenly select the wrong lottery with some probability $\epsilon \in [0, 0.5]$. 
We estimate the idiosyncratic error rate $\epsilon$ from preferences consistent with expected utility theory needed explain the observed choices of respondents on each of our algorithmically generated anomalies via minimum distance.
Appendix Figure \ref{figure: idiosyncratic error rates, by anomaly category, binary payoffs} reports the resulting estimated $\widehat{\epsilon}$ for each algorithmically generated anomaly over two-payoff lotteries separately, and Appendix Figure \ref{figure: idiosyncratic error rates, by anomaly category, ternary payoffs} reports the same quantity for each algorithmically generated anomaly over three-payoff lotteries.
On anomalies over two-payoff lotteries, the median estimated idiosyncratic error rate $\widehat{\epsilon}$ across algorithmically generated anomalies is 8.1\% for dominated consequence effect anomalies, 4.5\% for reverse dominated consequence effect anomalies, and 13.7\% for strict dominance effect anomalies. 
On three-payoff lotteries, the median estimated idiosyncratic error rate is 6.1\%.
There again exists heterogeneity in these estimates across anomalies. 
Explaining the observed choice fractions on several specific anomalies across categories would require that respondents erroneously deviate from their true preferences nearly 20\% of the time.

\begin{figure}[h!]
\centering
\includegraphics[width=.5\textwidth]{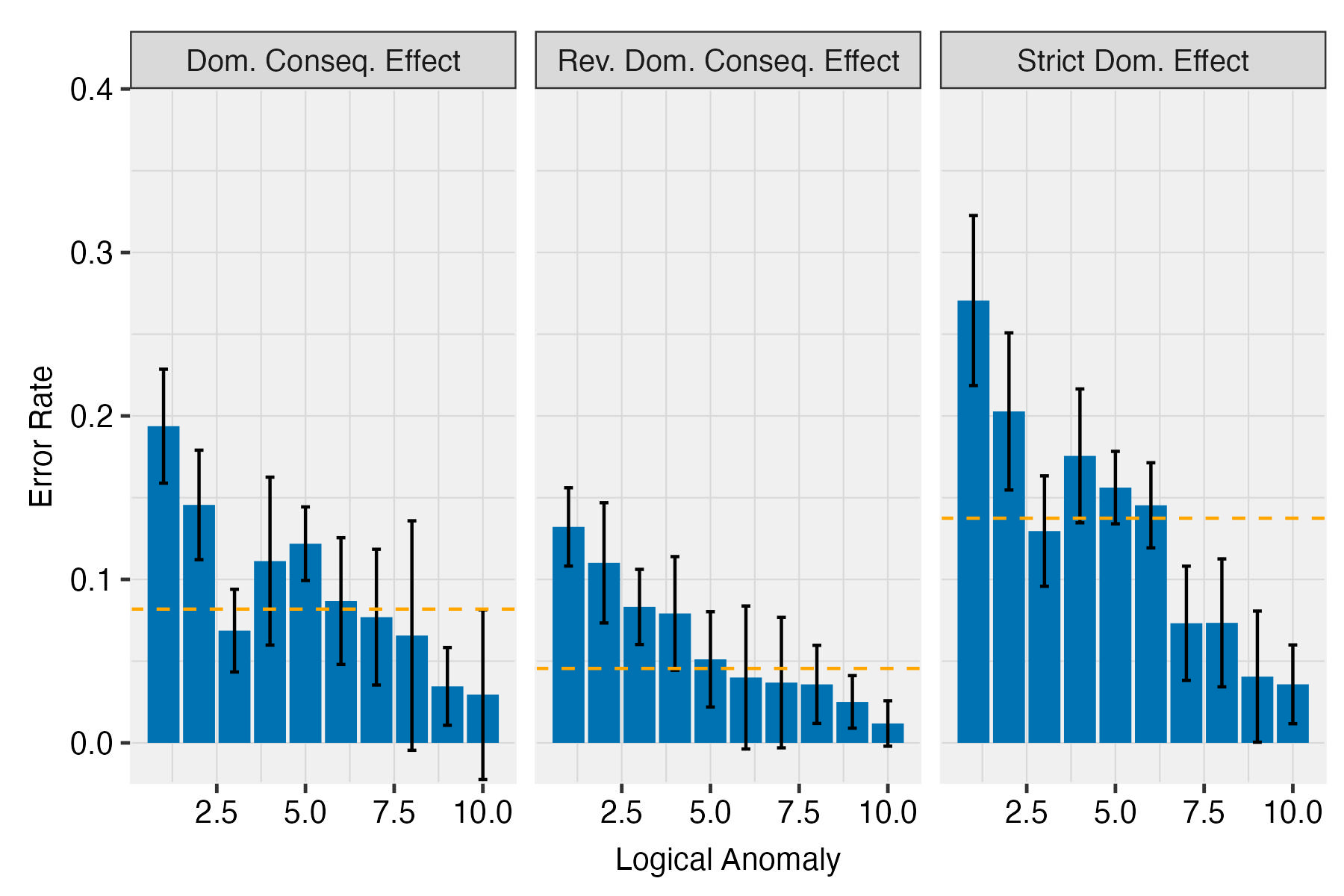}
\caption{Estimated idiosyncratic error rate $\widehat{\epsilon}$ on algorithmically generated anomalies over menus of two-payoff lotteries.}
\floatfoot{\textit{Notes}: The estimated idiosyncratic error rate $\widehat{\epsilon}$ are plotted in blue bars along with 95\% confidence intervals (black error bars; standard errors computed by the bootstrap). 
The orange dashed line reports the median estimated idiosyncratic error rate across all anomalies within the same category.
Within each category, we sort the anomalies and assign each anomaly an arbitrary numeric identifier in decreasing order based on the fraction of respondents whose choices violate expected utility theory. 
See Appendix \ref{section: appendix, anomalies and noisy choices} for further discussion.
}
\label{figure: idiosyncratic error rates, by anomaly category, binary payoffs}
\end{figure}

\begin{figure}[h!]
\centering
\includegraphics[width=.5\textwidth]{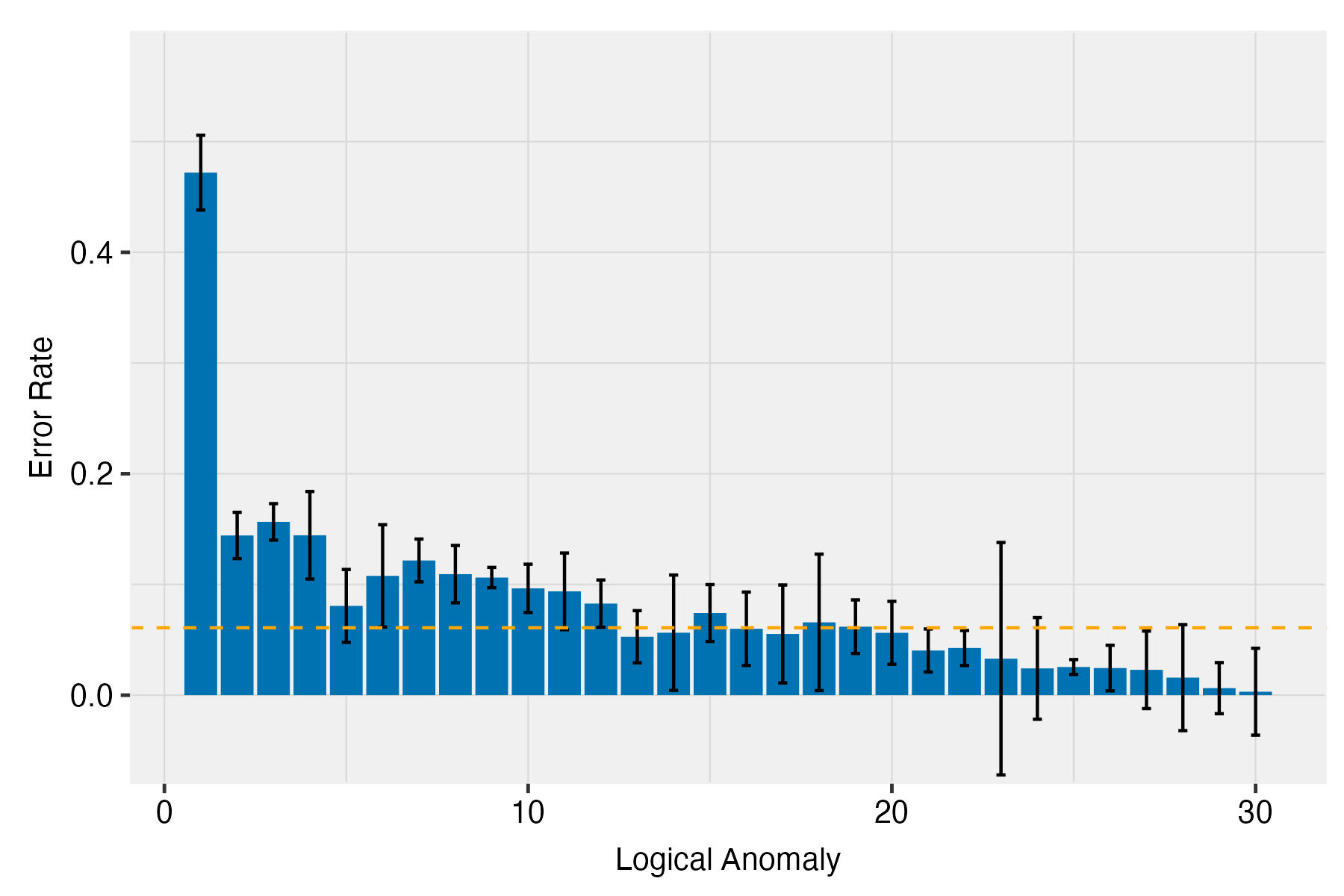}
\caption{Estimated idiosyncratic error rate $\widehat{\epsilon}$ on algorithmically generated anomalies over menus of three-payoff lotteries.}
\floatfoot{\textit{Notes}: The estimated idiosyncratic error rate $\widehat{\epsilon}$ are plotted in blue bars along with 95\% confidence intervals (black error bars; standard errors computed by the bootstrap).  
The orange dashed line reports the median estimated idiosyncratic error rate across all anomalies.
We sort the anomalies and assign each anomaly an arbitrary numeric identifier in decreasing order based on the fraction of respondents whose choices violate expected utility theory. 
See Appendix \ref{section: appendix, anomalies and noisy choices} for further discussion.
}
\label{figure: idiosyncratic error rates, by anomaly category, ternary payoffs}
\end{figure}

\subsection{Experimental instructions and questions for online surveys}\label{section: prolific survey screenshots}
\renewcommand{\thefigure}{F\arabic{figure}}
\renewcommand{\thetable}{F\arabic{table}}
\setcounter{figure}{0}
\setcounter{table}{0}

\begin{figure}[htbp!]
    \centering
    \begin{subfigure}[b]{0.33\textwidth}
         \centering
         \includegraphics[width=\textwidth]{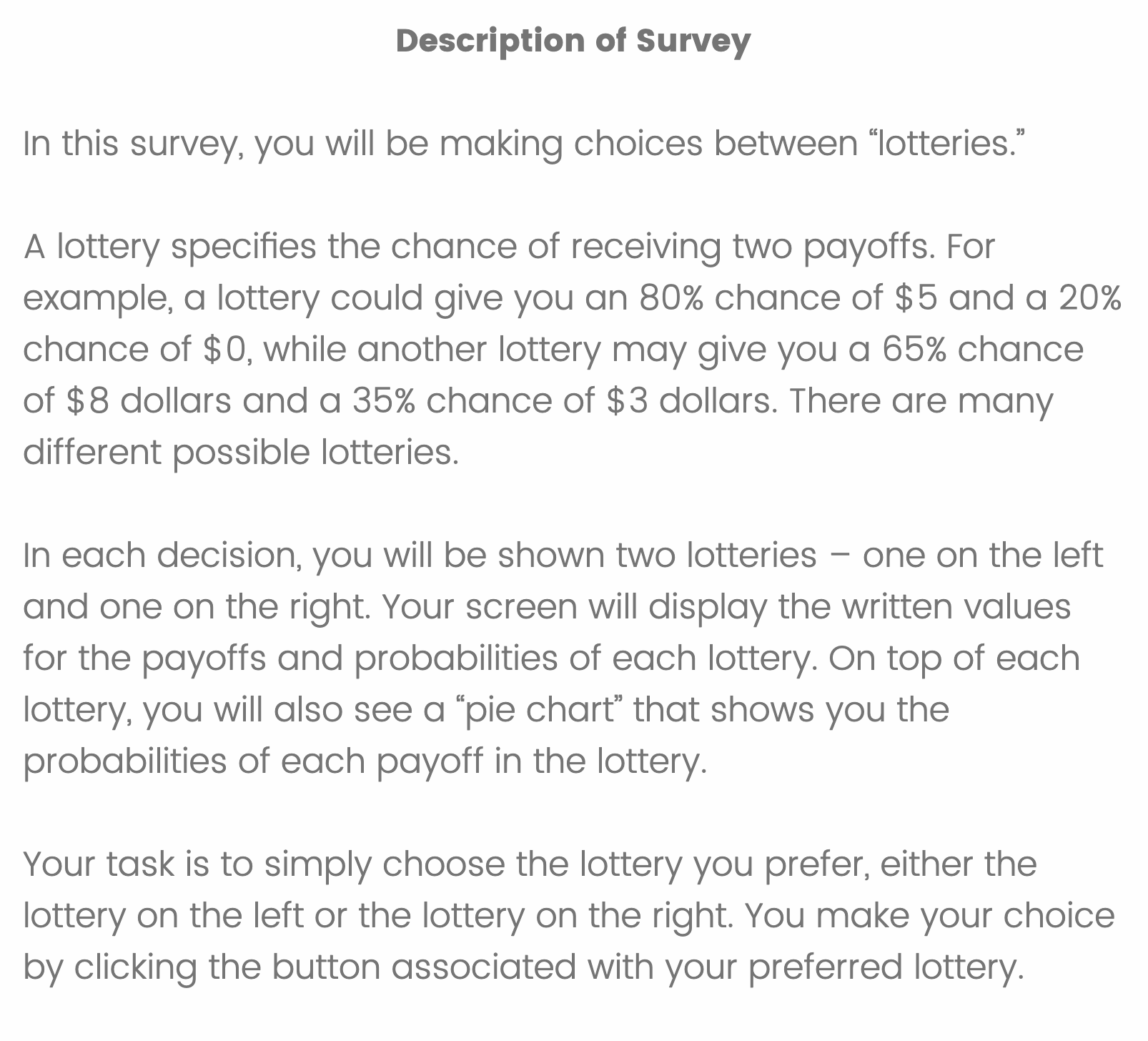}
     \end{subfigure}%
     \begin{subfigure}[b]{0.33\textwidth}
         \centering
         \includegraphics[width=\textwidth]{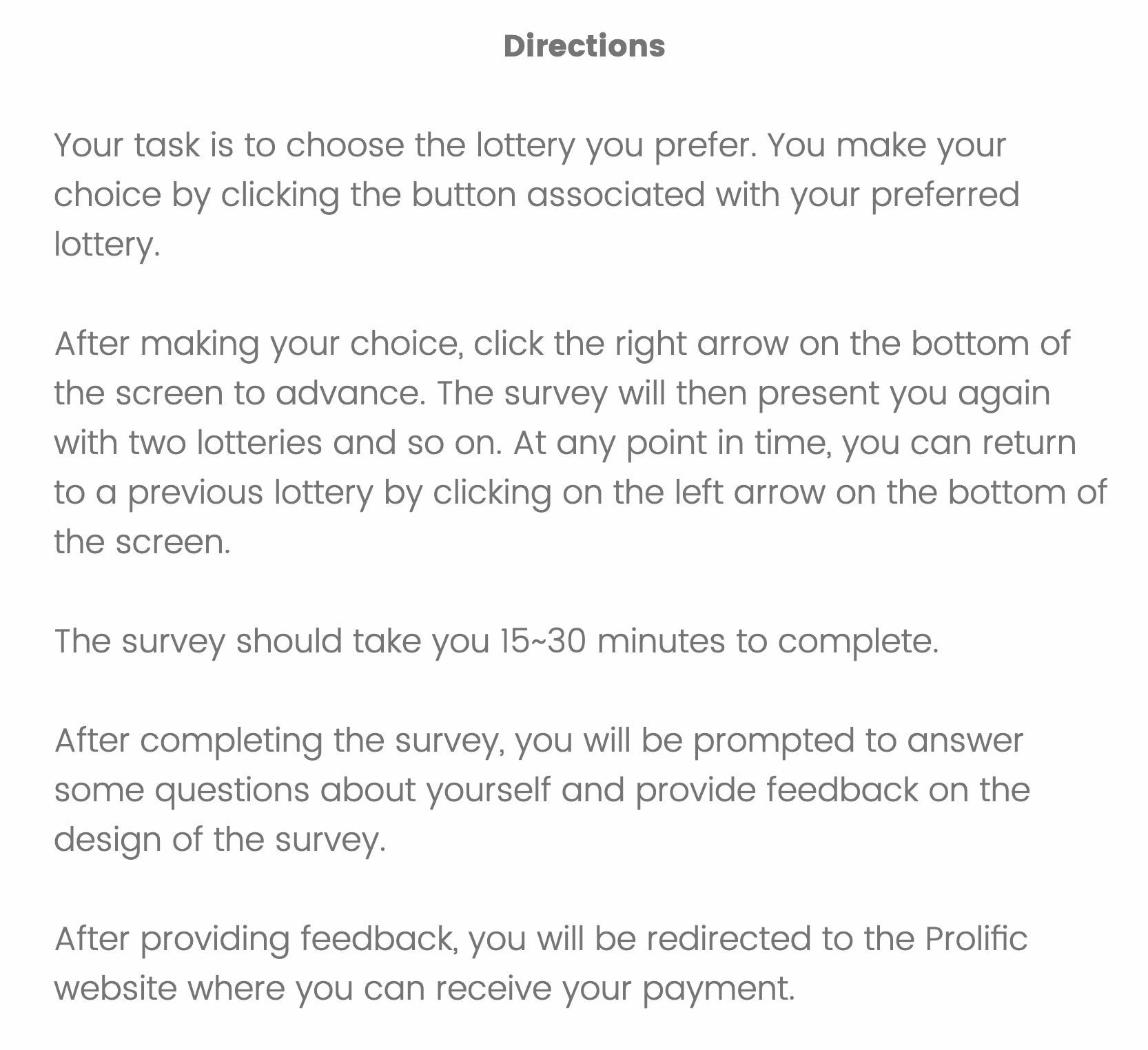}
     \end{subfigure}%
     \begin{subfigure}[b]{0.33\textwidth}
         \centering
         \includegraphics[width=\textwidth]{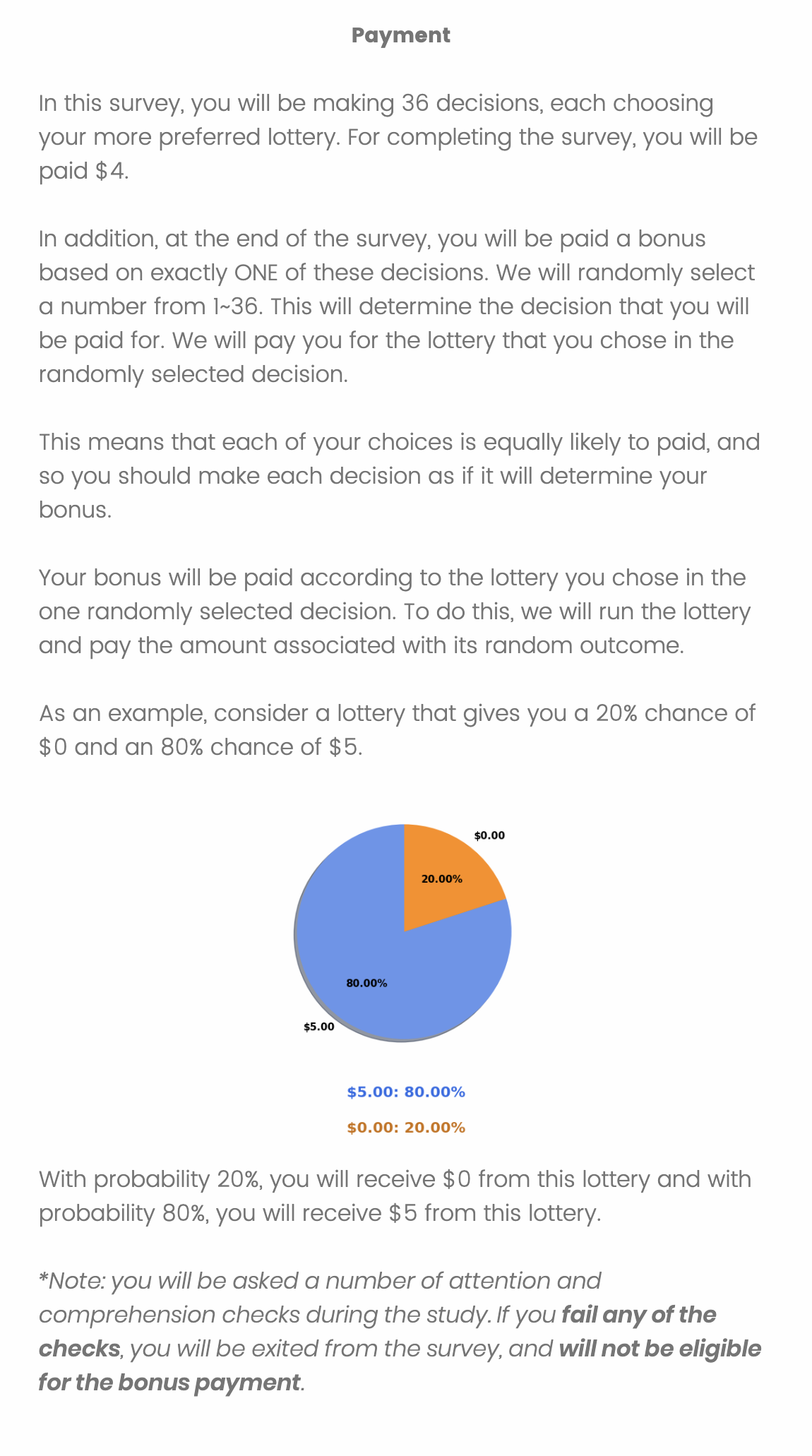}
     \end{subfigure}%
    \caption{Screenshots of directions for the online surveys on choices from menus of lotteries. See Section \ref{subsection: experimental test of algorithmically generated anomalies} and Appendix \ref{section: appendix, description of survey design} for further discussion.}
\end{figure}

\begin{figure}[htbp!]
    \centering
    \begin{subfigure}[b]{0.3\textwidth}
         \centering
         \includegraphics[width=\textwidth]{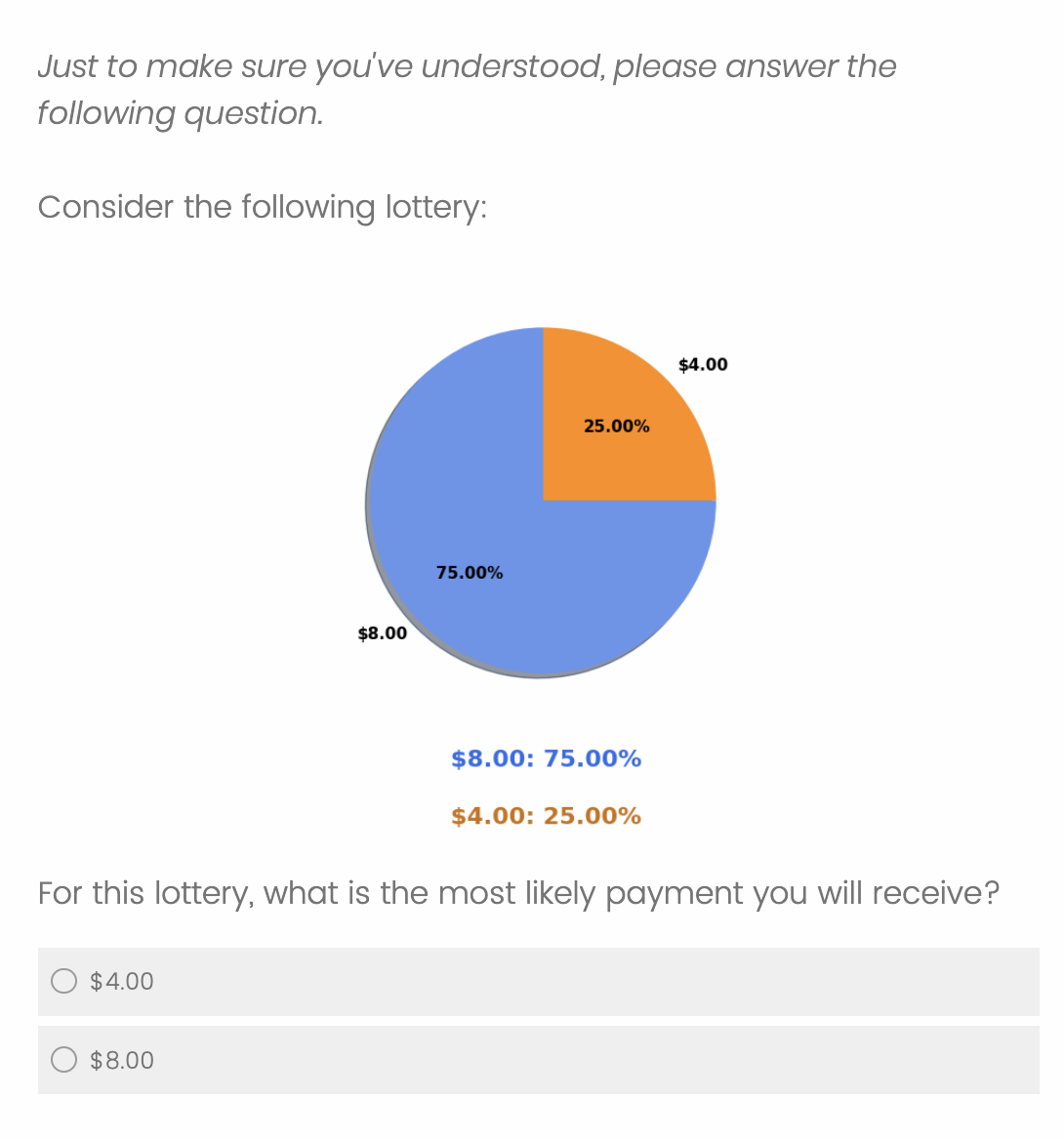}
     \end{subfigure}%
     \begin{subfigure}[b]{0.3\textwidth}
         \centering
         \includegraphics[width=\textwidth]{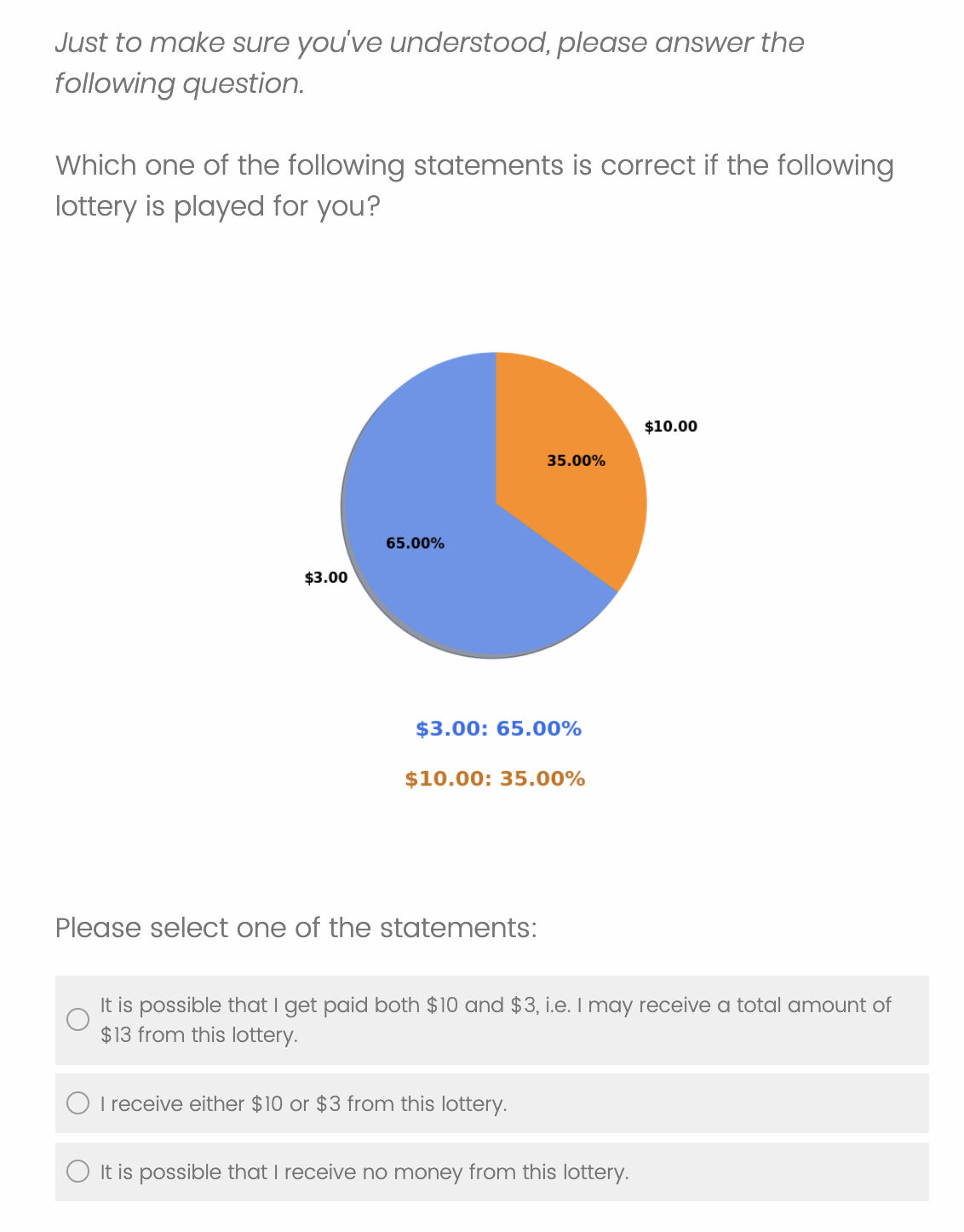}
     \end{subfigure}%
     \begin{subfigure}[b]{0.3\textwidth}
         \centering
         \includegraphics[width=\textwidth]{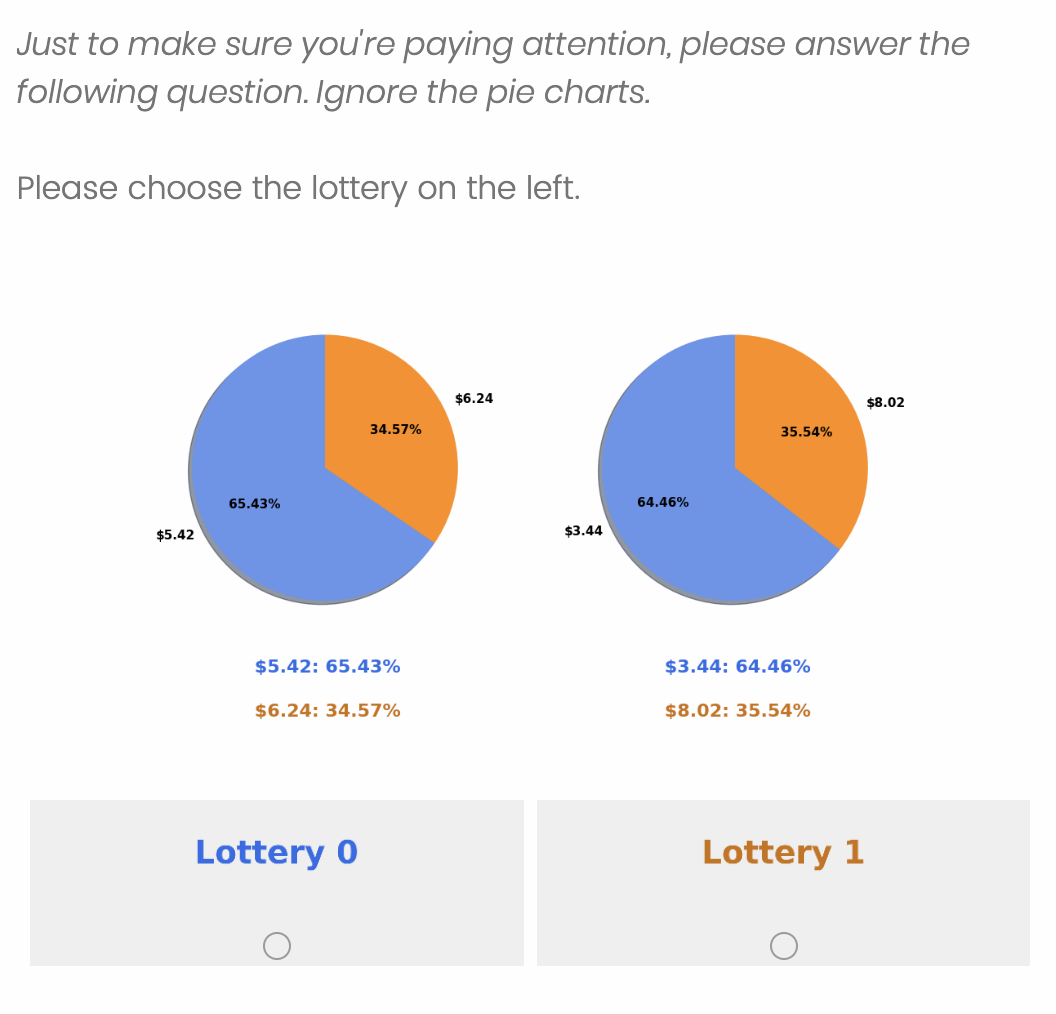}
     \end{subfigure}
    \caption{Screenshots of comprehension and attention checks for the online surveys on choices from menus of lotteries. See Section \ref{subsection: experimental test of algorithmically generated anomalies} and Appendix \ref{section: appendix, description of survey design} for further discussion.}
\end{figure}

\begin{figure}[htbp!]
    \centering
    \begin{subfigure}[b]{0.33\textwidth}
         \centering
         \includegraphics[width=\textwidth]{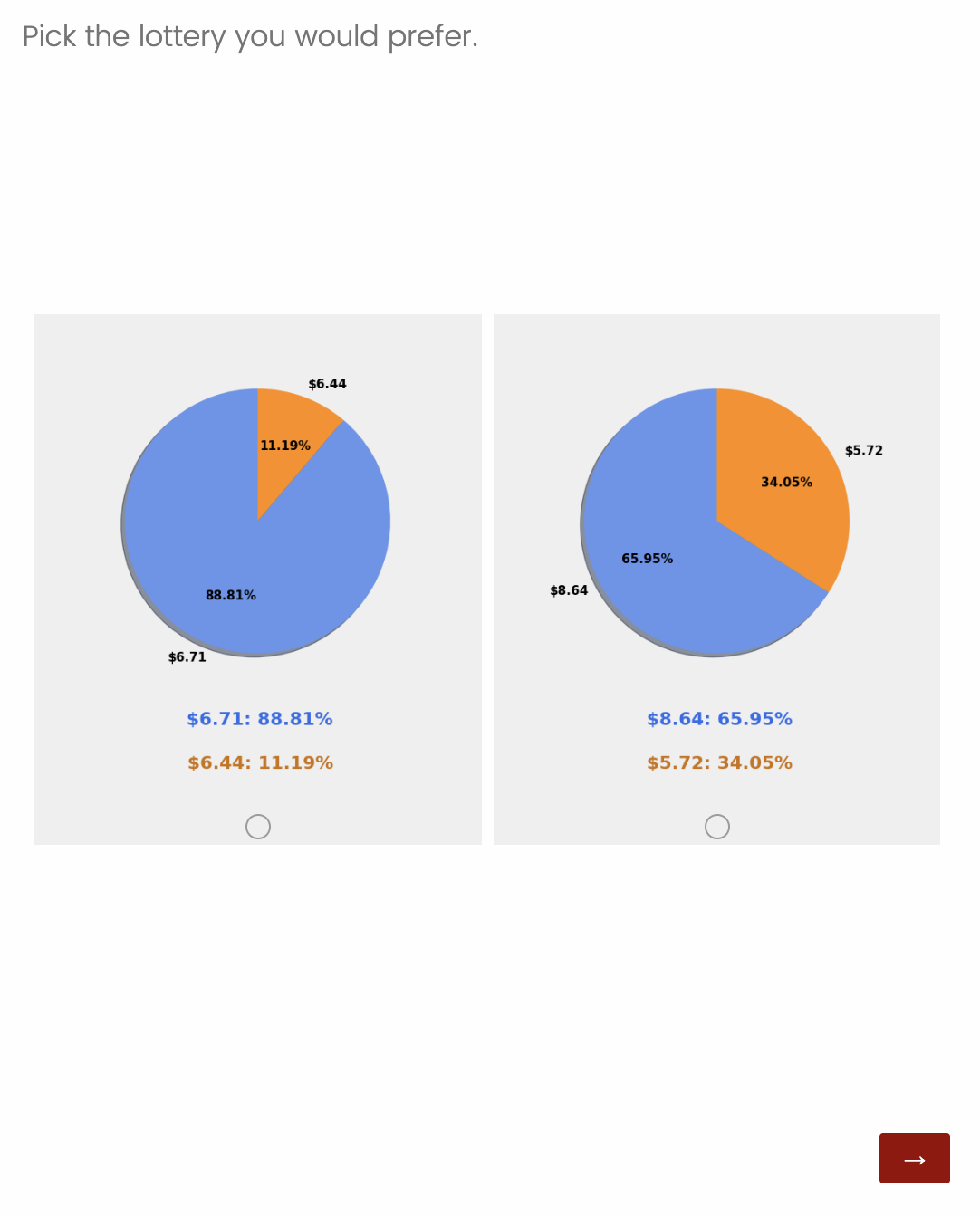}
     \end{subfigure}%
     \begin{subfigure}[b]{0.33\textwidth}
         \centering
         \includegraphics[width=\textwidth]{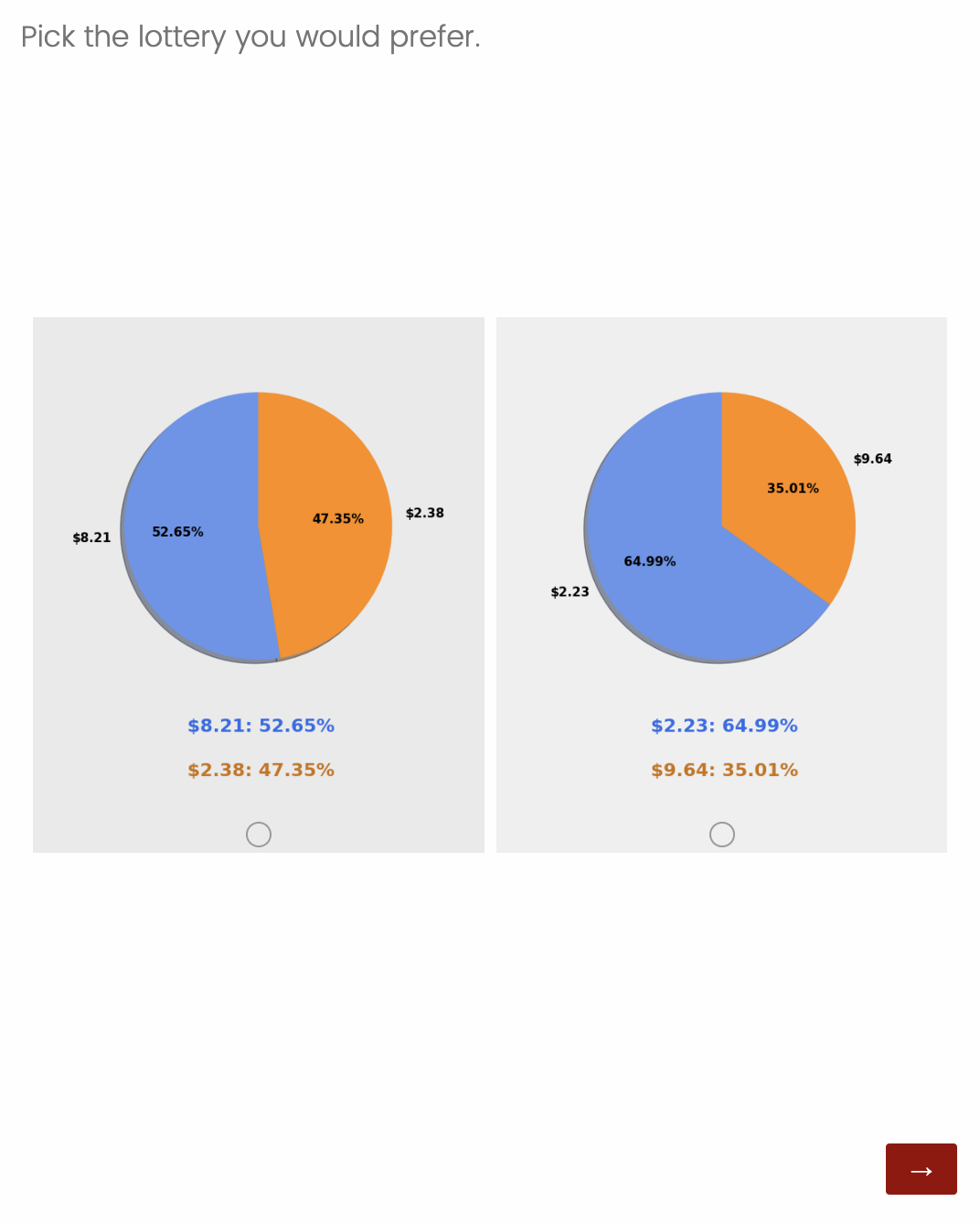}
     \end{subfigure}%
    \caption{Screenshots of two main survey questions for the online surveys on choices from menus of two lotteries with two monetary payoffs. See Section \ref{subsection: experimental test of algorithmically generated anomalies} and Appendix \ref{section: appendix, description of survey design} for further discussion.}
\end{figure}

\begin{figure}[htbp!]
    \centering
    \begin{subfigure}[b]{0.33\textwidth}
         \centering
         \includegraphics[width=\textwidth]{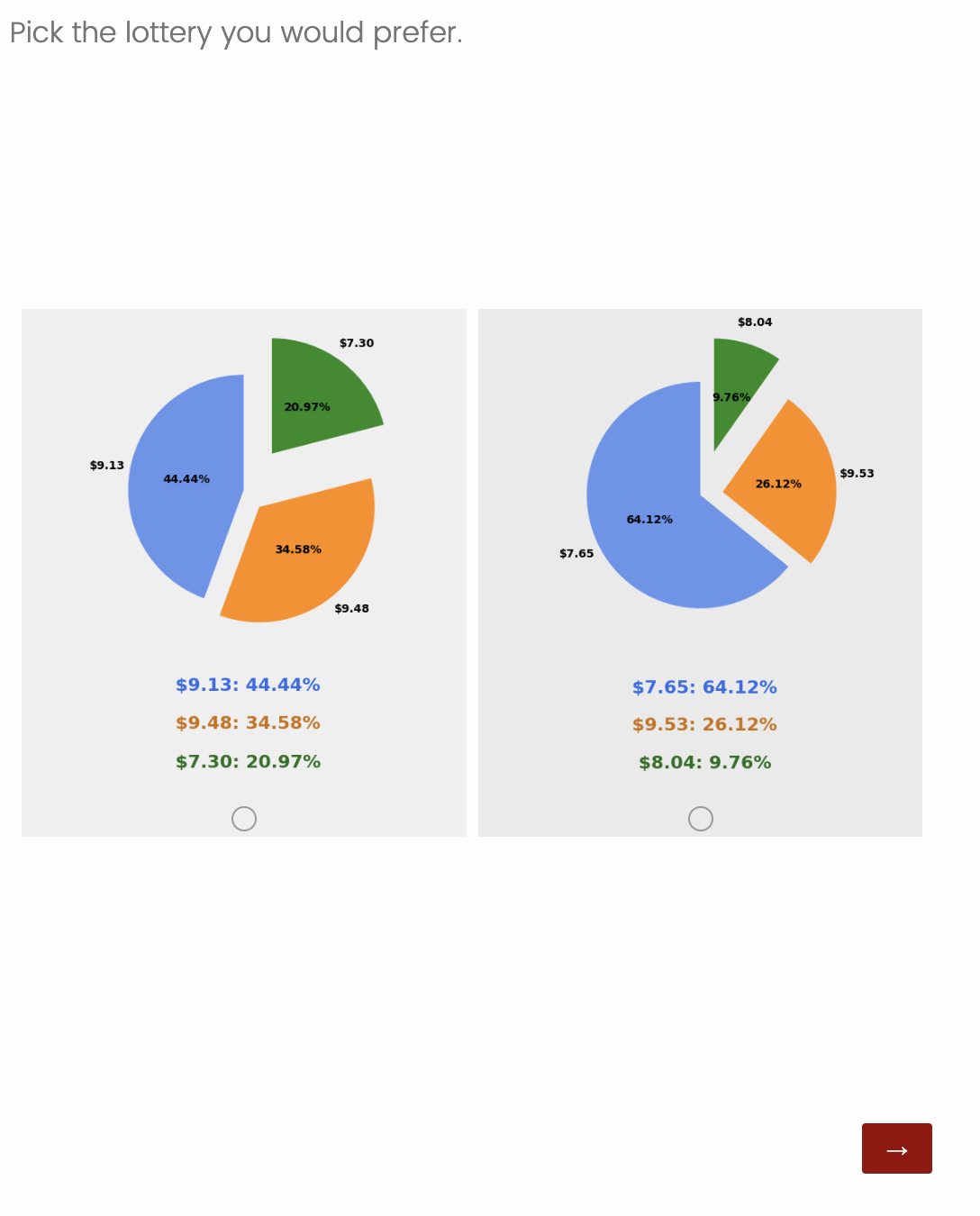}
     \end{subfigure}%
     \begin{subfigure}[b]{0.33\textwidth}
         \centering
         \includegraphics[width=\textwidth]{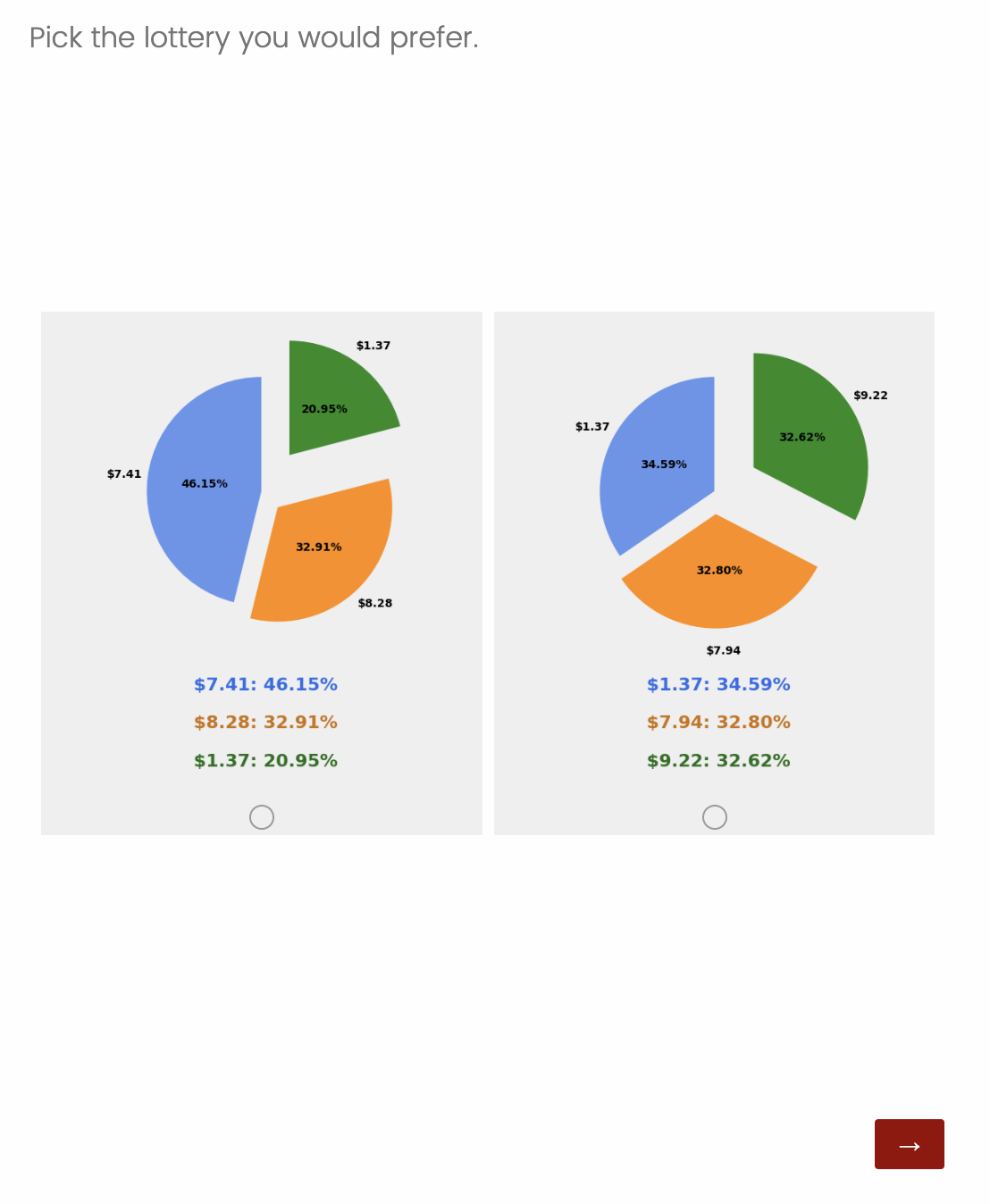}
     \end{subfigure}%
    \caption{Screenshots of two main survey questions for the online surveys on choices from menus of two lotteries over three monetary payoffs. See Section \ref{subsection: experimental test of algorithmically generated anomalies} and Appendix \ref{section: appendix, description of survey design} for further discussion.}
\end{figure}

\end{document}